\documentclass[prd,aps,preprintnumbers,superscriptaddress]{revtex4}
\usepackage{epsfig,subfigure}
\usepackage{amsmath,latexsym,amsfonts,mathrsfs,slashed}

\usepackage{epstopdf}

\newcommand{\F}{{\cal I}}
\newcommand{\W}{{\cal W}}
\newcommand{\nn}{\nonumber}
\newcommand{\vep}{\varepsilon}
\newcommand{\ep}{\epsilon}
\newcommand{\as}{\alpha_s}

\newcommand{\bbeta}{\bar\beta}

\def \ind{{\rm enc}}

\def\as{\ensuremath{\alpha_{s}}}
\def\a0{\alpha_0}

\def\vep{\varepsilon}

\def \ep{\epsilon}

\def \nn {\nonumber}

\def \J {{\cal J}}

\def\bea {\begin{eqnarray}}
\def\eea {\end{eqnarray}}

\def\be {\begin{equation}}
\def\ee {\end{equation}}
\def\bi {\begin{itemize}}
\def\ei {\end{itemize}}

\begin{document}

\baselineskip 18pt

\preprint{YITP-SB-14-15}
\preprint{Cavendish-HEP-14/12}

\title{Ultraviolet divergences and factorization for coordinate-space amplitudes}

\author{Ozan~Erdo\u{g}an}
\affiliation{Cavendish~Laboratory, University~of~Cambridge, Cambridge~CB3~0HE, United Kingdom}

\author{George~Sterman}
\affiliation{C.N.\ Yang Institute for Theoretical Physics and
   Department of Physics and Astronomy, 
 Stony Brook University, Stony Brook, New York 11794--3840, USA}

\date{\today}

\begin{abstract}
 We consider the coordinate-space matrix elements that correspond to fixed-angle scattering amplitudes involving partons and Wilson lines in coordinate space, working in Feynman gauge.   In coordinate space, both collinear and short-distance limits produce ultraviolet divergences.  We classify singularities in coordinate space, and identify neighborhoods  associated unambiguously with individual subspaces (pinch surfaces) where the integrals are singular. The set of such regions is finite for any diagram.    Within each of these regions, coordinate-space soft-collinear and hard-collinear approximations reproduce singular behavior.     Based on this classification of regions and approximations, we develop a series of nested subtraction approximations by analogy to the formalism in momentum space.  This enables us to rewrite each amplitude as a sum of terms to  which gauge-theory Ward identities can be applied, factorizing them into hard, jet and soft factors, and to confirm the multiplicative renormalizability of products of lightlike Wilson lines.   We study in some detail the simplest case, the color-singlet cusp linking two Wilson lines, and show that the logarithm of this amplitude, which is a sum of diagrams known as webs, is closely related to the corresponding subtracted amplitude order by order in perturbation theory.   This enables us to confirm that the logarithm of the cusp can be written as the integral of an ultraviolet-finite function over a surface.  We study to what extent this result generalizes to amplitudes involving multiple Wilson lines.  

\end{abstract}
\maketitle

\section{Introduction}

For many purposes, scattering amplitudes and the expectation values of gauge-theory Wilson lines may be studied in momentum space or in coordinate space, although most fixed-order computations are carried out in momentum space.   At the same time, a coordinate-space perspective may serve as a bridge between scattering amplitudes and certain observables, often those involving jets \cite{Sveshnikov:1995vi,Belitsky:2013xxa}.  Similarly, analyses in coordinate space have played a central role in correspondences between gauge theories and gravity~\cite{Alday:2008yw}, and dual conformal symmetries for select supersymmetric theories make a direct correspondence between choices of momenta for amplitudes and assignments of vertices for certain polygonal Wilson loops \cite{Drummond:2007aua,Elvang:2013cua}.    These considerations suggest that it may be useful to reexamine some of the all-orders properties of perturbative scattering amplitudes and cross sections that have been derived primarily from momentum-space analyses \cite{Sterman:1995fz,Collinsbook}  in terms of coordinate-space integrals.  In this spirit, we argued in Ref.\ \cite{Erdogan:2011yc} that the cusp formed by two Wilson lines can be written in a geometrical form to all orders in perturbation theory, as a surface integral over an ultraviolet-finite  function of the running coupling, whose scale varies with position on the surface.   The surface integrand itself is found from the web diagrams of the cusp \cite{Sterman:1981jc,Gatheral:1983cz,Frenkel:1984pz},  which will play a role in our discussion below.   A more general analysis of partonic amplitudes was undertaken in Ref.\ \cite{Erdogan:2013bga}, which examined the structure of coordinate-space singularities in massless gauge theories, by analyzing the pinch singularities of Feynman integrals in coordinate space \cite{Date:1982un} and developing a power-counting procedure to identify leading and nonleading behavior. 

In this paper we will apply and extend the results of Ref.\ ~\cite{Erdogan:2013bga}, where it was found, for example, that in renormalized matrix elements of the form
\be 
G^{\nu}(x_1,x_2) \ = \ \left\langle
  0\left|T\left(\phi(x_2)\,J^{\nu}(0)\,\phi^\dagger(x_1)\right)
  \right|0\right\rangle \, ,
   \label{eq:fvertexfnc}
\ee 
singularities occur only when the external points are on the light cone with respect to the current $J^\nu(0)$, that is, only at $x^2_I=0$, $I=1,2$, and that divergences in coordinate-space integrals are logarithmic, relative to tree level.    It was also argued that integrals in such ``leading regions'' factorize into hard, soft, and jet functions, in much the same way as in the well-known factorizations of momentum space~\cite{Bodwin:1984hc,Collins:1989gx}.   

In coordinate space, collinear and short-distance divergences are both of ultraviolet nature \cite{Erdogan:2013bga}, requiring $D<4$ in dimensional regularization, while the factorized soft function is finite when the external points are kept at finite distances from each other.    In contrast to short-distance singularities, collinear ultraviolet divergences are by their very nature nonlocal, and are not removed by the standard renormalization procedures for quantum field theory.   It is natural, however, to expect that they may be treated by analogy to collinear singularities in momentum space, where they are infrared, requiring $D>4$, and are factorized into universal functions \cite{Collins:1981uw}.   To derive and interpret the corresponding factorization properties for coordinate-space amplitudes, we will introduce a subtraction procedure that is similar to constructions in momentum space \cite{Collinsbook,Bodwin:1984hc,Collins:1989gx}.   The subtractions will enable us to reorganize perturbative amplitudes for gauge theories in a manner that makes their singularity structure and factorization properties manifest, after using the Ward identities of the theory.  

We work in Feynman gauge, to preserve Lorentz invariance and causality in the physical spacetime structure of the amplitudes we study.   Our construction is for gauge-theory amplitudes with the geometry of fixed-angle scattering, and so must overcome the  complication that in gauge theories with massless particles almost any subdiagram may produce collinear singularities or take part in the underlying short-distance process, in different parts of the integration space.   This is in contrast to lowest-order electroweak processes like Drell-Yan, where the hard scattering is associated only with subdiagrams including a specific vertex.   
  
Building on the results of Ref.\ \cite{Erdogan:2013bga}, we will study the ultraviolet singularities of  multiparton coordinate-space Green functions in configurations related to fixed-angle scattering, 
\bea
G_N(x_1, \dots, x_N)\ =\ \langle 0| T\left( \phi_N(x_N)\, \cdots \ \phi_1(x_1) \right) |0\rangle\, .
\label{eq:fields}
\eea 
These Green functions, of course, are not gauge invariant, but as we will observe, their leading singularities in coordinate space have  the same gauge-invariance properties as S-matrix amplitudes, as a result of the same Ward identities.   

The arguments that we give below carry over almost without change from coordinate space to momentum space, and we provide in this way a new all-orders analysis of factorization for scattering amplitudes in massless QCD and related theories in Feynman gauge.   Our work thus complements the momentum-space analyses carried out in physical gauge long ago in Ref.~\cite{Sen:1982bt} for scattering amplitudes, and recently in Ref.\ \cite{Feige:2014wja}, which uses physical gauges to analyze a large set of amplitudes and observables involving outgoing jets.   Our analysis of field theory perturbative amplitudes, based on an all-orders subtraction procedure to isolate, organize and cancel singular behavior,  can also play a role in improving and extending existing factorization proofs for electroweak annihilation \cite{Bodwin:1984hc,Collins:1989gx}, jet and single-particle inclusive cross sections in hadron-hadron collisions \cite{Sachrajda:1977mb}.

We also study the closely related multieikonal  products of path-ordered exponentials or Wilson lines \cite{Bialynicki-Birula,Wilson:1974sk}, in representations $f$, 
\bea
\Phi^{[f]}_{\dot \xi_C}(\tau_f,\tau_i)
=
P\, \exp\, \left[\, -ig\, \int_{\tau_i}^{\tau_f} d\tau\, \dot \xi_C(\tau)\cdot A^{[f]}\left(\xi_C(\tau)\right)\, \right]\, .
\label{oe}
\eea
Wilson lines that correspond to partonic amplitudes have constant velocities, $\dot \xi_C=  d\xi_C/d\tau =\beta_C$.    A four-Wilson line multieikonal vertex, for example, is defined by a constant matrix, $c_M$ in color space that links the color indices of the ordered exponentials at a point \cite{Brandt:1981kf,Sen:1982bt,Kidonakis:1998nf},
\bea
\Gamma^{[\rm f]}_{4,M}{}_{\{r_k\}} \left(\Lambda_1\beta_1,\dots \Lambda_4\beta_4 \right)
&=&
\sum_{\{d_i\}}
\langle0|\, \Phi_{\beta_4}^{[f_4]}(\Lambda_4,0)_{r_4,d_4}\; 
\Phi_{\beta_3}^{[f_3]}(\Lambda_3,0)_{r_3,d_3}\cr
&\ & \hspace{15mm} \times
\left( c_M\right)_{d_4d_3,d_2d_1}\; 
\Phi_{\beta_2}^{[f_2]}(0,-\Lambda_2)_{d_2,r_2}\;
\Phi_{\beta_1}^{[f_1]}(0,-\Lambda_1)_{d_1,r_1}\, |0\rangle \, .
\label{eq:wivertex}
\eea
For the eikonal Wilson lines of this expression, which are joined at the origin, constant velocities $\beta_I$ label the curves, which we can choose to be $\xi_J(\tau_J) = \beta_J\tau_J$.   They arrive at the vertex from the past, or emerge from the vertex toward infinity in the future.   In momentum space, Wilson lines appear as linear, ``eikonal" propagators.  The corresponding coordinate-space propagators are simply step functions, ordering the connections of gluons to the exponential.   The exponentiation properties of these products have received extensive attention over the past few years \cite{Gardi:2011wa,Gardi:2009qi,Vladimirov:2014wga,Gardi:2014kpa,DelDuca:2011ae}, in no small part for their relevance to phenomenological applications of resummation \cite{Mitov:2009sv,Kidonakis:1997gm,Bauer:2008jx}.

We begin Sec.~\ref{co-subt} with a review of the sources of ultraviolet poles in the coordinate-space calculation of multieikonal and partonic amplitudes \cite{Date:1982un,Erdogan:2013bga}.   We go on to define a series of approximation operators~\cite{Collinsbook,Zimmermann:1969jj} adapted to coordinate integrals.   Using  these operators, we construct a set of nested subtractions.  This is followed by a proof of the cancellation of coordinate-space overlapping divergences that are analogous to those in momentum space.  In  Sec.~\ref{sec:renorm-fact}  we show how the approximation operators match and organize singularities, and enable the renormalization of multieikonal amplitudes like Eq.~(\ref{eq:wivertex}) and the factorization of partonic amplitudes like Eq.~(\ref{eq:fields}) in appropriate limits, to all orders in perturbation theory.   Section~\ref{eqwbsub} deals with the special case of the two-eikonal amplitude, the singlet ``cusp''.  We will relate the subtraction procedure of  Sec.\ \ref{co-subt} directly to the logarithm of the cusp, given by the so-called web prescription~\cite{Sterman:1981jc,Gatheral:1983cz,Frenkel:1984pz}. In this context, the ultraviolet finiteness of the web function, and its relation to a surface integral \cite{Erdogan:2011yc} are confirmed.  We then turn in Sec.\ \ref{sec:multi-eik}  to fixed-angle multieikonal amplitudes, and study their exponentiation properties and geometrical interpretation in the large-$N_c$ limit and the general case.   We conclude with a summary and brief comments on possible applications.

\section{The Regularization of Collinear Singularities in Coordinate Space}
\label{co-subt} 

We begin this section with a review of the results of Ref.\ \cite{Erdogan:2013bga} regarding the coordinate-space singularities of partonic and eikonal amplitudes that remain after standard perturbative renormalization.   We follow this with the construction of a set of nonlocal ultraviolet subtractions, adapted in analogy to the Bogoliubov-Parasiuk-Hepp-Zimmerman momentum-space renormalization procedure \cite{Zimmermann:1969jj}, and in the spirit of the all-orders, all-logs treatment of infrared divergences in momentum space in Ref.\ \cite{Collinsbook}.    In the subsequent sections we will relate this additive regularization to the renormalization and factorization properties of  eikonal and partonic amplitudes in coordinate space.

\subsection{Leading regions, ultraviolet divergences and gauge invariance}
\label{subset:leading-regions}

\begin{figure}[t]
\centering
\subfigure[]{
\includegraphics[height=4.8cm]{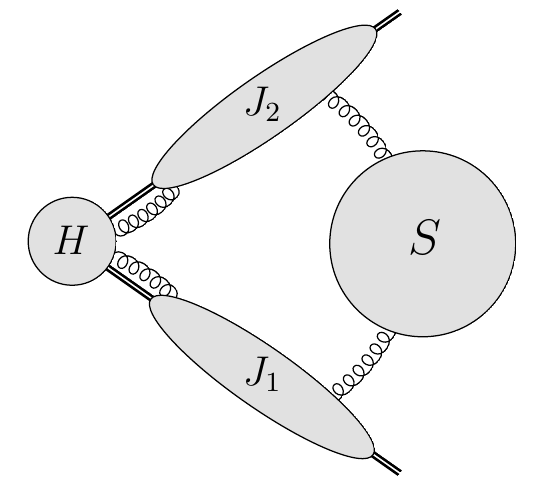}\label{fig:shjets-cusp}}
\hspace{20mm} 
\subfigure[]{
\includegraphics[height=4.8cm]{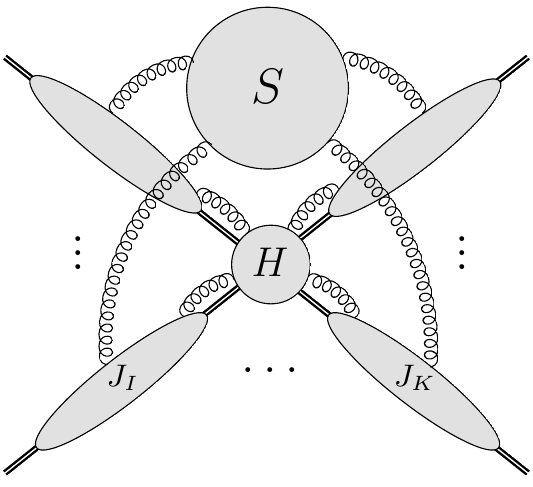}\label{fig:shjets-multi}}
\caption{Leading pinch surfaces represented by soft, jet and hard subdiagrams for
  \subref{fig:shjets-cusp} cusp and \subref{fig:shjets-multi} a
  typical multieikonal or multiparton amplitude. Gauge lines
  represent an arbitrary number of connections between the subdiagrams.  In
  \subref{fig:shjets-multi} the double line represents either Wilson
  lines or partonic propagators connected to the external vertices.}
\label{fig:shjets-first}
\end{figure}

In Ref.~\cite{Erdogan:2013bga}, the  most general regions from which divergences arise in coordinate-space integrals  were determined from their analytic structure  and a corresponding power-counting technique.   Divergences arise from pinches in the integrations over the positions of internal vertices considered as variables in complex coordinate space.  This is the direct analog of pinches in loop momenta  \cite{Collinsbook,Eden:1966,Sterman:1978bi}.    As in momentum space,  at each such leading region, the diagram describes a physical process with a fully consistent classical propagation for the set of lines that connect vertices that are lightlike separated.   We will refer to a manifold in coordinate space with a definite set of vertices pinched at lightlike or vanishing separations as a pinch surface (PS).   (We use this notation in the same sense as ``PSS'' in Ref.~\cite{Collinsbook}.)

In Ref.\ \cite{Erdogan:2013bga} it was shown that at such pinch surfaces, diagrams  are characterized by subdiagrams of soft, jet-like  and short-distance (hard) sets of lines, as depicted in Fig.~\ref{fig:shjets-first}, which is similar to the familiar structure of diagrams at pinch surfaces in momentum space both in direct QCD treatments \cite{Bodwin:1984hc,Collins:1989gx,Sterman:1978bi,Collinsbook} and in soft-collinear effective theory \cite{scet,Becher:2014oda}.    In the case of the massless cusp (Fig.~\ref{fig:shjets-cusp}), for example,   nonlocal ultraviolet subdivergences occur when subsets of vertices align along the Wilson lines, and these configurations define jet subdiagrams.  Other vertices remain at finite distances from both Wilson lines and the cusp in the soft subdiagram, while the remaining vertices move to the cusp and form the hard subdiagram~\cite{Erdogan:2011yc}\@.    

The same factorization into the same types of subdiagrams also occurs for multieikonal vertices with more Wilson lines and in partonic amplitudes in coordinate space whenever a single point in spacetime (the ``hard scattering")  is related to a set of external positions by lightlike distances, as illustrated in Fig.\ \ref{fig:shjets-multi}.    (We assume that no pair of external vertices is related by a lightlike distance.)   To anticipate, at these leading regions or PSs, one can make the coordinate-space soft-collinear and hard-collinear approximations, as defined in Ref.~\cite{Erdogan:2013bga}, which lead to the factorization of these subregions by the application of Ward identities in the same way as in momentum space~\cite{Bodwin:1984hc,Collins:1989gx}\@. We will give the expressions for these approximations for a leading PS below.  We use the term ``leading"  to denote an ultraviolet logarithm or a pole in the dimensionally regulated case,  and where necessary to distinguish PSs that produce such divergences from those that do not.   
\begin{figure}
\centering
\subfigure[]{
\includegraphics[height=4cm]{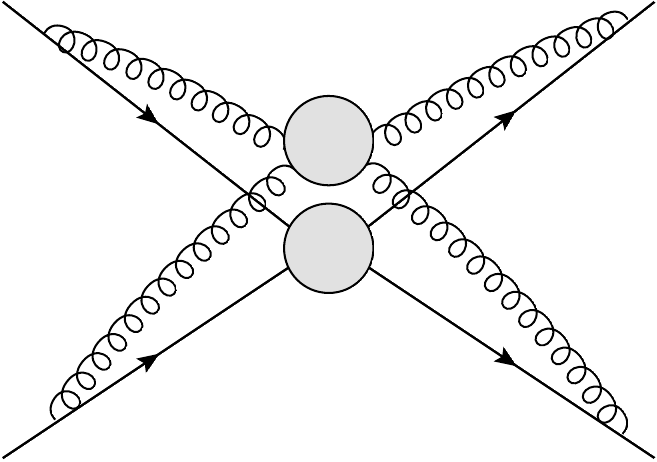}\label{fig:2hardsA}} \hspace{10mm}
\subfigure[]{
\includegraphics[height=4cm]{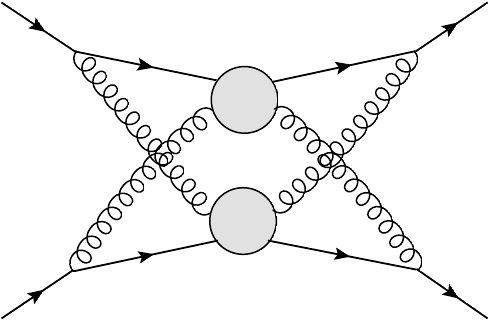}\label{fig:2hardsB}}
\caption{Examples of disconnected hard subdiagrams, representing:  \subref{fig:2hardsA} disconnected gluon-gluon and fermion-fermion scattering and \subref{fig:2hardsB} two disconnected gluon-fermion scattering subdiagrams.   The solid lines in \subref{fig:2hardsA} may also represent Wilson lines.}
\label{fig:2hards}
\end{figure}
As we quantify in the next subsection, for partonic amplitudes at leading PSs in Feynman and other covariant gauges, a single line from each jet carries a physical polarization (transverse for the gauge particle) into the hard subdiagram.    All other gauge lines attached to the hard scattering are scalar or longitudinally polarized  \cite{Erdogan:2013bga}.   We will use this result extensively below, and will assume that the external gauge fields of partonic amplitudes [Eq.\ (\ref{eq:fields})] are projected onto transverse polarizations.  We note that in physical gauges, only a single line connects each jet to the hard subdiagram \cite{Sterman:1978bi,Feige:2014wja}.

A complication for amplitudes involving physical processes with both incoming and outgoing external partons or Wilson lines is that PSs can have disconnected hard subdiagrams, as illustrated by the diagrams of Fig.~\ref{fig:2hards}.  We will confirm below, however, that these PSs are not associated with leading behavior.  As in momentum space \cite{Collinsbook}, their suppression follows from the Ward identities of the theory, which require that each jet subdiagram is connected to every connected component of the hard subdiagram by at least one line that is not a scalar-polarized gauge propagator.    Similar suppressions are described for cross sections in Refs.\ \cite{Labastida:1984gy} and \cite{Collins:2008sg}.    Leading pinch surfaces for partonic amplitudes involve at most a single, simply connected hard scattering. Similarly, for multieikonal amplitudes, the local multieikonal vertex must be part of every hard subdiagram.   Nevertheless, we will encounter diagrams like Fig.\ \ref{fig:2hards} in the full classification of leading regions and the elimination of double counting.

Before specifying the approximations, we pause to draw a few consequences of the observation that a physical picture associated with a pinch surface requires that the ``external'' propagators, beginning at the positions of fields, $x_I$ in Eq.\ (\ref{eq:fields}), be on the light cone with respect to  the position of the physical hard scattering.   The hard scattering may be mediated, for example, by exchange of a gluon in QCD  or by an electroweak current.    For multieikonal amplitudes, we can always set the vertex joining the eikonal lines to the origin.  In the case of partonic scattering, with external fields $\phi_I$ at points $x_I$, as in Eq.\ (\ref{eq:fields}), we consider $2\rightarrow N$ scattering, where $x_1^0,x_2^0$ are large and negative  and all $x_I^0$, $I>2$ are large and positive.   In this case, the requirement of a physical process allows hard scattering at a single, unique point, which, by translation invariance may also be taken as the origin.  A short proof is given in the Appendix.  In this coordinate system, all $x_I^2=0$ at the pinch surfaces, and we may identify velocity vectors by $\beta_I^\mu\sim x_I^\mu/x_I^0$ for each external field, with $\beta_I^2=0$.   These $\beta_I$ fix the directions of jets in the reduced diagrams of Fig. \ref{fig:shjets-multi}, for partonic scattering amplitudes, in the same way that Wilson lines fix jet directions for multieikonal amplitudes.  For each such line we introduce an additional, ``complement'' vector, $\bbeta_I$, $\bbeta_I^2=0$, normalized by $\beta_I\cdot\bbeta_I=1$.  The leading singularity of the diagram requires  that the light-cone singularity of each external propagator remains uncanceled.    We may think of this as the analog of the requirement that the S-matrix is the residue of the leading pole in every external line.    

The foregoing considerations on external propagators enable us to argue that the  leading behavior in coordinate space is gauge invariant, once external vector fields are projected onto transverse polarizations.    This follows the same way as in the diagrammatic proof of the gauge invariance of the S-matrix \cite{'tHooft:1972ue}.   In momentum space, an infinitesimal gauge transformation produces a sum of terms in which either external propagators are canceled, or vectors are projected onto scalar polarizations,   proportional to their own momenta.   The Fourier transformations of these relations are contributions in which an external propagator is replaced by a four-dimensional delta function, fixing its position at an internal vertex, or the divergence is taken of an external vector field, and hence a gradient of the external propagator.  The former case gives a suppression by $x_I^2$ relative to leading behavior, while the latter is eliminated by the same transverse projection that defines the S-matrix.

The general form of a coordinate amplitude can be written as
\bea
G_a(x_1,x_2, \dots, x_a)\ =\ \prod_{I=1}^a \int d^4 y_I\ G_2(x_I-y_I)\ \bar G_a(y_1, y_2, \dots, y_a)\, ,
\label{eq:gen-G}
\eea
where $\bar G_a$ is one-particle irreducible in each of the $x_I$ channels.   For much of the following analysis, we shall suppress the self-energies, which are factorized topologically, and whose renormalization is already included in the Lagrangian of the theory.   Except where indicated, therefore, our discussion will apply to the perturbative expansion of diagrams that contribute to $\bar G_a(y_1, \dots, y_a)$, in convolution with lowest-order propagators.  In the same way, for multieikonal amplitudes, our analysis will apply to single-eikonal-irreducible diagrams $\bar\Gamma_a$, related to the complete amplitudes by
\bea
\Gamma_a(\Lambda_1\beta_1,\Lambda_2\beta_2, \dots, \Lambda_a\beta_a)\ =\ \prod_{I=1}^a \int_0^{\Lambda_I} d \tau_I\ \Gamma_2((\Lambda_I-\tau_I)\beta_I)\ \bar \Gamma_a(\tau_1\beta_1, \tau_2\beta_2, \dots, \tau_a\beta_a)\, ,
\label{eq:gen-Gamma}
\eea
where $\tau_I\beta_I$ is the position of the outermost vertex on the $I$th Wilson line in $\bar \Gamma_a$, and where the $\Gamma_2$ represent self-energies of the Wilson lines.   Here all $\tau_I$ are taken positive, with the signs of the velocities $\beta_I^\mu$ adjusted as necessary for incoming lines.

\subsection{Variables, power counting and neighborhoods for pinch surfaces}
\label{subsec:power-ctg}

In the analysis of the pinch singularities of coordinate-space integrals, the soft, jet, and hard regions are specified by the identification of ``intrinsic'' and ``normal'' variables, which parametrize a pinch surface and its normal space, respectively~\cite{Sterman:1995fz,Erdogan:2013bga,Sterman:1978bi,Collinsbook}\@.     At a pinch surface, normal
variables are constructed to vanish as a distance scaling factor, $\lambda\rightarrow 0$ while intrinsic variables remain finite.   In the amplitudes that we discuss, at lowest order in normal variables, the propagator denominators of jet lines are linear in normal variables, those of the hard lines are quadratic in normal variables, and the soft lines are of zeroth order in normal variables.  (Our specific choices of normal variables for the amplitudes under consideration will be described shortly.)  Power counting can be performed by factoring out the lowest powers of $\lambda$ from each factor of the   integrand and the integration measure for each normal variable, $s_i$,
\bea
s_i\ \equiv\ \lambda s'_i\, .
\label{eq:sprime}
\eea
    Then, near a pinch surface, the integral for some quantity $g(x_I)$, depending on external parameters $x_I$  has the form~\cite{Erdogan:2013bga,Sterman:1978bi},
\be 
g(x_I) \sim \int_0^{d_0} d\lambda\, \lambda^{p-1}\int\prod_i ds'_i\;
\delta\left(1-\sum_i |s'_i|^2\right)\int\prod_j dr_j
\Big(\bar{I}_g(s'_i,r_j,x_I) + \mathcal{O}(\lambda^\eta) \Big) \ ,
\label{eq:scale-normal}
 \ee
where $\eta>0$.  The integrals over the intrinsic variables $r_j$ of the ``homogenous" integrand  $\bar{I}_g(s'_i,r_j,x_I)$, found by keeping only the lowest powers of $\lambda$ in each factor  \cite{Erdogan:2013bga,Sterman:1978bi}, will either be nonsingular or will have pinch surfaces generated when  subsets of the $s_i'$ vanish (in our case where a subset of vertices approaches the light cone or hard scattering faster than the others).   The scale $d_0$, which quantifies the maximum distance from the pinch surface, may be thought of as arbitrary at this point. The analysis of the homogeneous integrand determines the choice of normal variables near each PS \cite{Erdogan:2013bga}.   As found using the power counting developed in Ref.~\cite{Erdogan:2013bga}, the leading overall degree of divergence is $p=0$ for pinch surfaces of both eikonal and partonic amplitudes, relative to lowest order, indicating logarithmic divergences of their integrals in coordinate space.   We will review these results shortly, and only note here that  when $p>0$, the PS is integrable.

In these terms, leading regions are characterized by gauge vector propagators connecting the soft subdiagram to jet subdiagrams, with the following properties of the homogeneous integrals  \cite{Erdogan:2013bga}.
\begin{itemize}

\item The polarization tensors of all gauge vector propagators that attach the soft subdiagram to jet subdiagram $K$ are contracted only to the jet velocity vector, $\beta_K$.

\item The denominators of gauge propagators that attach the  soft subdiagram to jet subdiagram $K$ depend on the positions, $z^{(K)\mu}$ of the jet vertices to
which they attach only through a vector that depends on a single coordinate: $\bbeta_K\cdot z^{(K)} \beta_K^\mu$. 

\end{itemize}
Together, these properties specify the ``soft-collinear approximation", summarized by an operator, $sc(K)$, whose action is defined in Feynman gauge and dimensional regularization with $D=4-2\vep$, by
\bea 
sc(K)\, \left[ D^{\mu\nu}(x-z^{(K)}) \right] \ J_\nu(z^{(K)})    &=&
sc(K)\, \left[ \frac{-\ g^{\mu\nu} }{[-(x-z)^2+i\ep ]^{1-\vep}} \right] \ J_\nu(z^{(K)})
\nn\\
&=&
\beta^{\mu}_K\ \frac{-\ 1 }{[-2\,\bbeta_K\cdot (x - z^{(K)})\,\beta_K\cdot x +
  x^2_\perp +i\ep]^{1-\vep}} \bbeta^\nu_K \ J_\nu(z^{(K)})\, , 
  \label{eq:soft-co}
  \eea
where $x$ is the position of a soft vertex, or in the case of a gauge line exchanged between Wilson lines or jets,
a point on the other line or in the other jet.    The soft-collinear approximation drops terms that are
of order $\lambda^{1/2}$ near the pinch surface, where the denominator is finite.    It is then convenient to
 define coordinates that link the soft and jet  subdiagrams in
convolution for each vertex position, $z^{(K)}$,
\bea
d^Dz^{(K)}\ &\equiv& \ d\tau^{(K)}\ d^{D-1}z^{(K)}\, ,
\nonumber\\ 
\tau^{(K)}\ &=&\ \bbeta_K \cdot z^{(K)}\, .
\label{eq:tau-def}
\eea
Here, $\tau^{(K)}$ and the azimuthal angle of $z_\perp^{(K)}$ are intrinsic variables, while $\beta_K\cdot z^{(K)}$ and $z^{(K)2}_{\perp}/\bbeta_K\cdot z^{(K)}$ can be chosen as normal variables for this jet \cite{Erdogan:2013bga,Sterman:1978bi,Collins:1989gx}. 
For the special case of $z^{(K)}$ being a vertex on the $K$th Wilson line, 
we can identify $z^{(K)}{}^\mu=\tau^{(K)}\beta_K^\mu$.

In a similar fashion, we identify approximations that reproduce the homogeneous integral for lines that attach jet subdiagrams to the hard-scattering subdiagram at leading PSs \cite{Erdogan:2013bga}.

\begin{itemize}

\item Gauge field propagators that attach jet subdiagram $J_I$ to the hard subdiagram are either physically polarized or are contracted only to the complementary vector, $\bbeta_I^\mu$.

\item The denominators of  propagators attaching  jet subdiagram $I$ to the hard subdiagram depend on the coordinates $y^{(I)}$ of the hard vertices to
which they attach only through vectors $\beta_I\cdot y^{(I)} \bbeta_I^\mu$.

\end{itemize}
These conditions define the ``hard-collinear approximation", represented by an operator $hc(I)$, which acts on scalar-polarized gauge propagators as
\bea 
hc(I)\, \left[  D^{\mu\nu}(z-y^{(I)})  \right ]\ H_\nu(y^{(I)}) \ &=& \
hc(I)\, \left[ \frac{-\ g^{\mu\nu} }{[-(z-y^{(I)})^2+i\ep ]^{1-\vep}} \right] \ \ H_\nu(y^{(I)})
\nn\\
&=& \
\bbeta^{\mu}_I\, \frac{-\ 1}{[-2\,\beta_I\cdot (z - y^{(I)})\,\bbeta_I\cdot z +   z^2_\perp +i\ep]^{1-\vep}}\, \beta^\nu_I  \ \ H_\nu(y^{(I)})\, .
  \label{eq:co-hard}
  \eea
As we have noted above, in the case of partonic amplitudes, a leading PS requires that (exactly) one partonic propagator attaches each jet subdiagram to the hard scattering with a physical polarization for fermions or vectors.  For these propagators, the corresponding hard-collinear  approximation may be represented as
\bea 
hc(I) \left[ \Delta^{\omega\sigma}(z-y^{(I)}) \right ] \ H_\sigma(y^{(I)}) \ &=&\
 \Delta^{\omega\sigma}(z-\beta_I \cdot y^{(I)}\bbeta_I)\ {\cal T}_{\sigma}{}^{\sigma'} H_{\sigma'}(y^{(I)})\, ,
  \label{eq:co-hard-parton}
  \eea
   where ${\cal T}_{\sigma}{}^{\sigma'}$ is an appropriate projection for the leading physical polarizations, and $\Delta^{\omega\sigma}$ is the corresponding propagator, depending on the spin of the field.  The hard-collinear approximation drops terms that are of order $\lambda^{3/2}$ near the pinch surface in the denominators, whose leading behavior is order $\lambda$. 
Then, similarly to Eq.\ (\ref{eq:tau-def}) we define
\bea
d^Dy^{(I)}\ &\equiv& \ d\eta^{(I)}\ d^{D-1}y^{(I)}\, ,
\nonumber\\ 
\eta^{(I)}\ &=&\ \beta_I \cdot y^{(I)}\, .
\label{eq:zeta-def}
\eea
In the hard subdiagram, all components of the positions $y^\mu$ are normal variables.    In the generic case, where all components of $y^\mu$ appear linearly in the denominators of jet lines shown in Eq.\ (\ref{eq:co-hard}), all of these components are naturally taken to scale linearly in $\lambda$.   When there are precisely two incoming and two outgoing jets at the pinch surface hard scattering, however, one spacelike component of $y^\mu$, which we may call $y_{\rm out}$, does not appear in any factor $\beta_I\cdot y$, $I=1, \dots , 4$.    
Rather, it appears quadratically in every propagator attached to the vertex at $y^\mu$.  This coordinate defines the direction normal to the scattering plane in a center-of-momentum frame of the physical picture at the pinch surface.    In this case, the single variable $y_{\rm out}$ scales as $\lambda^{1/2}$, and the integral is correspondingly enhanced.   This enhancement is also a feature of the lowest-order, tree-level scattering, however, and does not change the logarithmic nature of radiative corrections \cite{Erdogan:2013bga}, which are the focus of our discussion.

In summary, for a partonic amplitude with hard scattering at the
origin and external points on the light cone $x_I^2\rightarrow 0$,
{\it all} pinch surfaces are specified by lists of vertices
$\{z_\mu^{(K)}\}$ that specify jet subdiagrams $J_K$,  and a list of
vertices $\{y_\mu\}$ that specifies the hard subdiagram $H$, while the
remaining vertices $\{x_\mu\}$ specify the soft subdiagram $S$. From these lists of vertices, we find
the normal variables of an arbitrary pinch surface $\rho$,
\bea
\{s_i^{(\rho)}\}\ =\ \left \{ \left \{\beta_K\cdot z^{(K)},\, \frac{z_\perp^{(K)2}}{\bbeta_K\cdot z^{(K)}} \right \}\, ,  \{y^\mu\} \right \}\, ,
\label{eq:normal-list}
\eea
that is, the opposite-moving and squared perpendicular components normalized by the longitudinal distance from the origin for each vertex in each jet, and all components of vertices in the hard subdiagram.     All other independent components are intrinsic variables,
\bea
\{ r_j^{(\rho)}\}\ =\ \left \{ \{x_i^\mu\}\, ,\ \{\bbeta_K\cdot z^{(K)},\phi^{(z_K)}\} \right \}\, ,
\label{eq:intrinsic-list}
\eea
with $\phi^{(z_K)}$ azimuthal angles for the transverse components of jet vertices.   We emphasize that the number of pinch surfaces is finite for any diagram of finite order, which are enumerated simply by the ways of assigning vertices to the jet, soft and hard subdiagrams.

\begin{figure}[t]
\centering
\includegraphics[height=5cm]{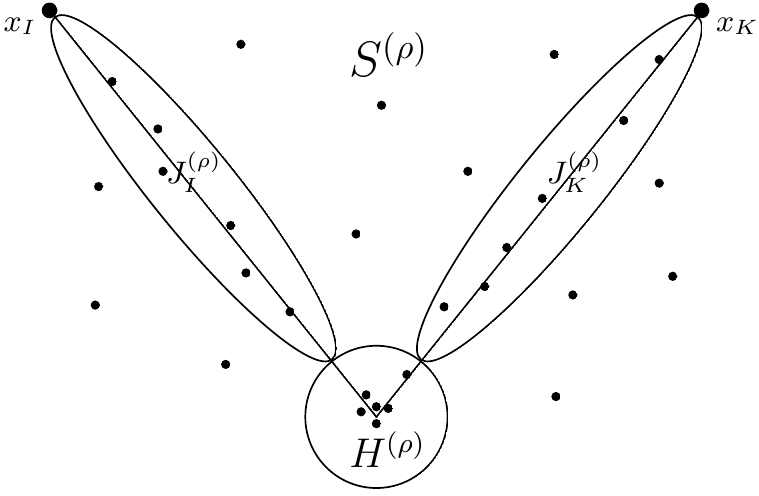}

\caption{Representation of the arrangement of vertices near a leading pinch surface $\rho$ directly in coordinate space and their assignments to jet, $J^{(\rho)}_I$, hard, $H^{(\rho)}$, and soft, $S^{(\rho)}$ subdiagrams. For every region, the direction of the jet $J^{(\rho)}_I$ is determined by the relative position of the external point $x_I$ with respect to the position of the hard scattering.}
\label{fig:regions}
\end{figure}

 The choice of subdiagrams and hence PSs can be pictured directly in coordinate space.  In Fig.\ \ref{fig:regions},  each point represents the projection of the position of an interaction vertex in some very high-order diagram onto the plane defined by two noncollinear Wilson lines, for example. The closed curves represent the jets and hard scatterings in a transparent fashion.   The normal variables for vertices in either jet are given simply by their distances to the corresponding lines in this diagram, and normal variables for vertices in the hard function are their distances from the origin, as in Eq.\ (\ref{eq:normal-list}).   We denote these subdiagrams by $S^{(\rho)}$, $J_I^{(\rho)}$ and $H^{(\rho)}$, respectively.   We suppress  their explicit orders, which are implicit in the choice of PS $\rho$.   It is clear from the figure that assignments of vertices to jet, hard and soft subdiagrams are shared by many diagrams, that is, all the perturbative diagrams that are found by connecting the points in the figure.
 
  We can now quantify the identification of leading regions, as derived in Ref.\ \cite{Erdogan:2013bga} and illustrated in Fig.\ \ref{fig:shjets-first}.   It was shown in Ref.\ \cite{Erdogan:2013bga} that in massless gauge theories, 
the scaling power $p$ of Eq.\ (\ref{eq:scale-normal}) associated with an arbitrary pinch surface is independent of the order in perturbation theory, and depends only on the numbers of lines connecting the hard, jet and soft subdiagrams associated with the PS in question, and on the polarizations carried by gauge lines that connect the jet subdiagrams with the soft and hard subidagram.   To be specific, we adopt the following notation:

\begin{itemize}
 
\item Let $j_I^{f}$ and $j_I^A$  be respectively the numbers of fermion jet lines and gauge jet lines that connect jet subdiagram $J^{(\rho)}_I$ to the hard subdiagram $H^{(\rho)}$, and let $j_I^{A+}\le j_I^{A}$ be the number of these gauge lines that carry scalar polarizations, proportional to  $\beta_I$, the velocity vector associated with jet $I$.  In these numbers, we suppress the PS label ``$\rho$'', because the result will hold for all PSs \cite{Erdogan:2013bga}.   

\item Let $s_I^{f}$ and $s_I^{A}$ be, respectively the number of fermion and gauge soft lines attached to jet subdiagram $J_I^{(\rho)}$, and $s_I^{A+}\le s_I^A$ the number of these soft gauge lines that are coupled to the velocity vector associated with jet $I$, $\beta_I$.

\item Let $s_H^f$ and $s_H^A$ be, respectively the number of fermion soft and gauge soft lines that are attached to the hard subdiagram $H^{(\rho)}$.

\end{itemize}

In this notation, in Ref.\ \cite{Erdogan:2013bga} it was shown that the {\it minimum} scaling power $p$ in Eq.\ (\ref{eq:scale-normal}) for an arbitrary PS can be expressed as a sum over contributions from each jet subdiagram, plus a contribution when the soft subdiagram is attached to the hard subdiagram directly,
\bea
p_{\rm min}\ =\ \sum_{{\rm jets}\ I}\, \frac{1}{2} \left[  \left (j_I^A - j_I^{A+}\right) + j_I^f -1 +s_I^f+\left( s_I^A-s_I^{A+}\right)  \right]\ +\ \frac{3}{2} s_H^f\ +\ s_H^A\, .
\label{eq:p-general}
\eea
Since divergences are associated only with $p\le 0$, a PS is nonleading unless there are no direct connections between the soft and hard subdiagrams,
\bea
s_H^f\ =\ s_H^A \ =\ 0\, ,
\label{eq:conditions-soft}
\eea
and unless  for each jet $I$, 
\bea
j_I^A \ -\  j_I^{A+}\ +\  j_I^f \ =\ 1\, , \quad {\rm and}\quad  s_I^f\ =\ 0\ =\ \left( s_I^A-s_I^{A+}\right)\, .
\label{eq:conditions-jet}
\eea
The first relation in Eq.~(\ref{eq:conditions-jet}) reflects the Ward identities of the theory, which eliminate the case when all the lines of a jet that attach to the hard subdiagram are unphysically polarized gauge propagators ($j_I^A=j_I^{A+}$, $j_I^f=0$).   As a result, $p_{\rm min}\ge 0$ after gauge-invariant sets of diagrams are combined (in the hard subdiagram, specifically).  
Each jet is coupled to the hard subdiagram by at most a single physically polarized gauge vector or a single jet fermion in addition to an arbitrary number of scalar-polarized gauge vectors, and the coupling of all jets to the soft subdiagram is entirely through soft gauge lines with polarizations in the jet direction.

Equation (\ref{eq:p-general}) holds for PSs with multiple, disconnected as well as simply connected hard subdiagrams.   This confirms that, as noted above, any PS with more than one connected hard part is power suppressed relative to leading behavior.   It is also worth noting, however, that in elastic amplitudes for bound-state scattering, where there is more than one physically polarized parton in each incoming and outgoing particle, PSs with disconnected hard-scattering subdiagrams actually dominate leading behavior \cite{Landshoff:1974ew,Botts:1989kf} because of the tree-level power-counting enhancement noted just after Eq.~(\ref{eq:zeta-def}).

By Eq.\ (\ref{eq:conditions-jet}), no soft fermions attach the soft subdiagram to jet subdiagrams at leading PSs.  This implies that at leading PSs, jet functions are diagonal in the flavor of the external partons; that is, for each jet the same quantum numbers entering the diagram appear at the hard subdiagram.      

Taken together, these considerations justify restricting our considerations to PSs with the structure illustrated in Fig.\ \ref{fig:shjets-first}.  We note that for the purposes of this discussion, we have varied the notation of Ref.\ \cite{Erdogan:2013bga} slightly, and include several terms that are discussed in Ref.~\cite{Erdogan:2013bga}, but not included in the relation analogous to Eq.\ (\ref{eq:p-general}) given there, which is an inequality, rather than an equality.

To organize integrals in the presence of this large but finite number of pinch surfaces, we define neighborhoods $n[\rho]$ of pinch surfaces $\rho$ by requirements on normal variables, $s_i^{(\rho)}$, given in Eq.\ (\ref{eq:normal-list}), and intrinsic variables $r_j^{(\rho)}$ from Eq.~(\ref{eq:intrinsic-list}), 
\bea
\sum_i |s_i^{(\rho)}|^2\ &\le& \   d_0^2\, ,
\nn\\
|r_j^{(\rho)}|^2\ &\ge& \ \left ( \sum_i |s_i^{(\rho)}|^2 \right)^{\delta_j} d_0^{2-2\delta_j}
\nn\\
 &\ge& \lambda^{2\delta_j} \left ( \sum_i |s'_i{}^{(\rho)}|^2 \right)^{\delta_j} d_0^{2-2\delta_j}
\, ,
\label{eq:neigh-def}
\eea
for some finite distance scale $d_0$.  A power $0<\delta_j < 1/2$ is chosen for each intrinsic variable $r^{(\rho)}_j$, where the $s_i'^{(\rho)}$ are rescaled normal variables, Eq.\ (\ref{eq:sprime}).  The inequalities for power $\delta_j$ ensure that the leading terms involving normal variables in the soft-collinear and hard-collinear approximations, Eqs.\ (\ref{eq:soft-co}) and (\ref{eq:co-hard}), remain dominant by a power over the first corrections to these approximations, which are relatively suppressed by $\lambda^{1/2}$ in both cases.
With this definition, the soft-collinear and hard-collinear approximations associated with pinch surface $\rho$ remain accurate for $\lambda\rightarrow 0$ in Eq.\ (\ref{eq:scale-normal}) throughout neighborhood $n(\rho)$, in the absence of cancellations between leading terms at pinches of the homogeneous integral.    We can think of Eq.\ (\ref{eq:neigh-def}) as specifying the closed curves of Fig.\ \ref{fig:regions}.

We close this subsection by noting that the homogeneous integrals (\ref{eq:scale-normal}) for PSs with normal variables identified as in Eq.\ (\ref{eq:normal-list}), have lower-order pinches that are precisely the same structure as those in the original diagrams.   This can be seen by considering vertices in each of the subdiagrams associated with an arbitrary PS, $\rho$.  For vertices $x^\mu$ in the soft subdiagram, the only approximations are for denominators attached to the jets, for which jet vertices are set on the light cones, $\beta_I$.   In neighborhood $n[\rho]$, the $x^\mu$ stay away from all of the light cones, and the physical picture correspondence eliminates PSs involving vertices in $S^{(\rho)}$, just as in the original integral.   For vertices $z_\mu^{(K)}$ in jet $K$, the integrals are unchanged, except for lines attached directly to the hard scattering, where terms that are nonleading in the scaling variable are neglected.  No approximations are made for lines internal to $H^{(\rho)}$.     Pinches of the homogeneous integral are still controlled by the distances of the external vertices $x_K$ of $J^{(\rho)}_K$ to the relevant light cone, and these pinches develop in the same manner in the homogeneous as in the original integral.   In the homogeneous integral, defined as in Eq.\ (\ref{eq:scale-normal}), however, one or more of the rescaled normal variables are always order unity.  Thus, the pinch surfaces of the homogeneous integral will involve fewer vanishing denominators than those of the original PSs.   We will use this observation in our construction of nested subtractions.

\subsection{Approximation operators and region-by-region finiteness}

We will now employ the approximations identified above to define a new set of approximation operators, denoted by $t_\rho$, one for each leading pinch surface $\rho$.    Each operator $t_\rho$ is defined to act on any diagram $\gamma^{(n)}$ that possesses the corresponding PS and to give an expression that corresponds to the leading, singular behavior of $\gamma^{(n)}$ in the neighborhood of PS $\rho$.    Of course, this condition defines the operator $t_\rho$ only up to a finite ambiguity.  For our purposes it will be most useful to construct subtractions similar to those employed in proofs of factorization in Ref.~\cite{Collinsbook}.    

We define the action of the approximation operator $t_\rho\gamma^{(n)}$ as the imposition of the soft-collinear and hard-collinear approximations given above in Eqs.\ (\ref{eq:soft-co}), (\ref{eq:co-hard}) and (\ref{eq:co-hard-parton}) on all lines to which they apply at PS $\rho$ of diagram $\gamma^{(n)}$.  This action can be represented schematically by
\bea
t_\rho \gamma^{(n)}\
&\equiv&\ 
\prod_I
\int d\tau^{(I)} \ S^{(\rho)}_{\{\mu_I\}}(\{\tau^{(I)}\})\ \beta_I^{\mu_I}\bbeta_{I,\mu'_I}
\nn\\
&\ & \hspace{1mm} \times\ \int d\eta^{(I)}\;
\int d^{D-1} z^{(I)} \ J_I^{(\rho)\mu_I'\nu_I'}(z^{(I)},\eta^{(I)})\ \bbeta_{I,\nu'_I}\beta_I^{\nu_I}\ \int d^{D-1} y^{(I)} \ H^{(\rho)}_{\{\nu_I\}}(y^{(I)})\, .
\label{eq:soft-approx}
\eea
In this expression, each vector index or vertex position, for example, $\mu_I$ or $z^{(I)}$, respectively, represents arbitrary numbers of such indices and positions for gluons connecting the subdiagrams specific to this leading region, $S^{(\rho)}$, $J_I^{(\rho)}$ and $H^{(\rho)}$.   
The net effect of $t_\rho$ is to replace the full integral of diagram $\gamma^{(n)}$ by the homogeneous integral that corresponds to PS $\rho$.

As mentioned above, the soft-collinear and hard-collinear approximations defined for coordinate-space integrals in Ref.~\cite{Erdogan:2013bga} are equivalent to approximations with similar names in discussions of factorization in momentum space~\cite{Collins:1981uk,Collins:1989gx}\@. In this case, the approximation isolates ultraviolet divergences in the neighborhood of the PS in coordinate space, so long as the soft-collinear and hard-collinear approximations apply. We represent this result by
\bea
\left. t_\rho\, \gamma^{(n)} \right |_{\mathrm{div}\ n[\rho]} \ = \ 
  \left.\gamma^{(n)}\right|_{\mathrm{div}\ n[\rho]} \ ,
  \label{eq:t-rho-gamma}
   \eea 
where the subscript ``div $n[\rho]$'' represents the divergent UV behavior, from short-distance and/or collinear configurations of PS~$\rho$.  This relation is not guaranteed to apply on subsurfaces where the homogeneous integral in $t_\rho\gamma^{(n)}$ develops pinches of its own.   In the following we will generalize Eq.~(\ref{eq:t-rho-gamma}) by introducing a system of nested subtractions.
   We emphasize first, however, that although the relation (\ref{eq:t-rho-gamma})  refers to the result of an integral over the neighborhood, $n[\rho]$ of PS  $\rho$, where the approximation is accurate, the definition $t_\rho\gamma^{(n)}$ refers to the full integral, extended over the full integration region in coordinate space, including other PSs and regions where $t_\rho$ no longer gives a good approximation to the integrand in general \cite{Collinsbook}.

 In any multiloop diagram,  multiple ultraviolet  divergences can arise from sets of vertices that approach the hard scattering or the collinear directions in partonic amplitudes, or the cusp and/or the Wilson lines in multieikonal amplitudes, at different rates, just as loop momenta may go to infinity faster in some subdiagrams than in others.    As for the renormalization of Green functions, we can classify sets of divergences as either nested or overlapping, in terms of the limiting process in coordinate space.   

Nesting in coordinate space can be classified directly in terms of pinch surfaces.  We say PS $\rho_1$ is nested in PS~$\rho_2$ when a subset of vertices of $\rho_2$, which defines $\rho_1$,  approaches the light cone and/or the origin faster than other vertices in $\rho_2$.   The smaller nested PS has larger subdiagrams with vertices near the light cone (jets) or the origin (hard subdiagram), but it defines a smaller region in coordinate space.   

To be specific, for two leading pinch surfaces, $\rho_1$ is a nested subsurface of $\rho_2$, denoted by
\bea
\rho_1 \subseteq \rho_2\, ,
\eea
 if and only if
\bea
H^{(\rho_2)} &\subseteq& H^{(\rho_1)}\, ,
\nn\\
H^{(\rho_2)} \cup  J_I^{(\rho_2)} &\subseteq& H^{(\rho_1)} \cup   J_I^{(\rho_1)}\, ,
\label{eq:nest-def}
\eea
for all jets $J_I$.    
That is, the jet and/or hard subdiagrams grow as the dimension of the pinch surface decreases.     The equality holds only when $\rho_1=\rho_2$, in which case all these relations are equalities.  Otherwise, we say that $\rho_1$ is contained in $\rho_2$.  Without specifying their ordering, we say that $\rho_1$ and $\rho_2$ nest.  The subsurface, or nesting, relation of course is transitive,
\bea
\rho_3 \subset \rho_2 \ {\rm and}\ \rho_2 \subset \rho_1\ \Rightarrow\ \rho_3 \subset \rho_1\, .
\label{eq:nest-trans}
\eea
We note that the smaller the pinch surface in the sense of Eq.\ (\ref{eq:nest-def}), the {\it larger} the number of its normal variables, and the smaller the number of its intrinsic variables.    Another way of putting this is that smaller pinch surfaces have larger codimension.   We will denote any fully nested set with $M_N$ pinch surfaces by $N=\{\sigma_1 \subset \sigma_2 \subset \cdots \subset \sigma_{M_N}\}$, and the set of all such nested sets for diagram $\gamma$ as  ${\cal N}[\gamma]$.    

We now use the nesting of pinch surfaces and the definitions of neighborhoods, $n[\rho]$ in Eq.\ (\ref{eq:neigh-def}) to  construct a set of regions in coordinate space that cover all pinch surfaces, and in each of which an operator, $t_\rho$ gives a valid approximation to the singular behavior of the diagram throughout.   Starting from the $n[\rho]$, our choice for these ``reduced neighborhoods'' is
\bea
\hat n[\rho]\ =\ n[\rho]\backslash \cup_{\sigma \supset \rho} \left( n[\rho] \cap n[\sigma] \right) \, .
\label{eq:rhohat-def}
\eea
By construction, region $\hat n[\rho]$ is $n[\rho]$  less its intersections with the neighborhoods $n[\sigma]$ of all larger pinch surfaces, $\sigma \supset \rho$.  As we have argued at the end of the previous subsection, the PSs of the homogeneous integral of region $\rho$ correspond to PSs $\sigma$, at which only a proper subset of the lines that are on the light cone or at the origin on PSs $\rho$ remain on the light cone or at the origin.    Such pinch surfaces~$\sigma$ have more intrinsic (and fewer normal) variables than pinch surface $\rho$, and one or more of the intrinsic variables of each $\sigma$ are normal variables of $\rho$.   In addition, by the construction of Eq.~(\ref{eq:neigh-def}),  in neighborhood $n[\sigma]$  the normal variables of $\rho$ that are intrinsic variables of $\sigma$ do not vanish rapidly enough to produce a divergence. Correspondingly, the homogeneous integral for PS $\rho$, Eq.\ (\ref{eq:scale-normal}) integrated over the reduced region $\hat n[\rho]$ is finite. Note that although the PS $\rho$ itself is a subspace of lower dimension in surface $\sigma$,  the neighborhoods $n[\rho]$ and $n[\sigma]$ are of the same dimension, and $\rho \subset \sigma$ does {\it not} imply that $n[\rho] \subset n[\sigma]$.     The neighborhoods $\hat n[\rho]$ cover all pinch surfaces.   

Not all pairs of regions can satisfy the nesting criterion (\ref{eq:nest-def}).   We say two pinch surfaces are overlapping when $\rho\not\subset\sigma$ and $\sigma\not\subset\rho$, which we denote as
\bea
\rho\ :o:\ \sigma \, .
\label{eq:rho-overlap}
\eea
By definition, if $\rho :o: \sigma$, then $\rho$ and $\sigma$ cannot appear in any set $N$ of nested PSs of $\gamma$.   The overlap relation, ``$:o:$'' has a property analogous to transitivity of nesting, Eq.\ (\ref{eq:nest-trans}), which also follows easily from the defining properties of nesting, Eq.~(\ref{eq:nest-def}),
\bea
{\rm given}:\ &\ & \ \sigma_1\subset \sigma_2 \subset \sigma_3\, , \quad {\rm where} \quad
\sigma_3\ :o:\ \rho \quad {\rm and} \quad \sigma_1 \ :o:\ \rho
\nn\\
{\rm then}: \ &\ &\ \sigma_2 \ :o:\ \rho\, .
\label{eq:over-trans}
\eea
Any pair of PSs is either nested or overlapping.   Note that the pinch surface where all vertices are in the hard subdiagram is nested with all other pinch surfaces, so that no pair of pinch surfaces is fully disjoint. Figures \ref{fig:3ov1} and \ref{fig:3ov2} illustrate two overlapping regions.

As we have seen, each pinch surface, and corresponding neighborhood is associated with a distinct matching of the list of vertices to the jet, hard and soft subdiagrams.  In these terms, we can give an explicit form for the requirement of Eq.\ (\ref{eq:t-rho-gamma}), namely that the divergences from PS $\rho$ are equal for $\gamma^{(n)}$ and $t_\rho\gamma^{(n)}$,
\bea
\gamma^{(n)} \big |_{\mathrm{div}\ \hat n[\rho]} \  \ -\  t_\rho \gamma^{(n)} \big |_{\mathrm{div} \hat n[\rho]}
\ &=&\
 \prod_I
\int d\tau^{(I)} \;  \int d^{D-1} z^{(I)} \; \int d\eta^{(I)}\; \int d^{D-1} y^{(I)}\ \Theta({\hat n[\rho]})
\nn\\
&\ & \hspace{0mm} \times\ \left[
S^{(\rho)}_{\{\mu_I\}}(z^{(I)})\ 
 J_I^{(\rho)\mu_I \nu_I}(z^{(I)},y^{(I)})\  H^{(\rho)}_{\{\nu_I\}}(y^{(I)})
\
 \right. \nn \\
 &\ & \left. \hspace{5mm}    -\ S^{(\rho)}_{\{\mu_I\}}(\tau^{(I)})\ \beta_I^{\mu_I}\bbeta_{I,\mu'_I}\
 J_I^{(\rho)\mu_I'\nu_I'}(z^{(I)},\eta^{(I)})\ \bbeta_{I,\nu'_I}\beta_I^{\nu_I} \ H^{(\rho)}_{\{\nu_I\}}(y^{(I)})     \, 
 \right]\, \Big |_{\rm div\ \hat n[\rho]}
 \nn\\
 &=& \ 0\, ,
\label{eq:soft-approx-diff}
\eea
where $\Theta(\hat n[ \rho])$ restricts the integration to the reduced neighborhood $\hat n[\rho]$ [Eq.\ (\ref{eq:rhohat-def})].   This integral over the reduced neighborhood converges because of the accuracy of the soft-collinear and hard-collinear approximations in the entire reduced neighborhood $\hat n[\rho]$. The PSs internal to the original neighborhoods $n[\rho]$ have been removed by construction.  

Equation (\ref{eq:soft-approx-diff}) is the main result we will use for applications in the following sections, treating the neighborhood of each PS separately.   As a more general result, however, we will show that all divergent contributions to amplitudes can be written without restriction to specific regions, in terms of a construction based on nested subtractions \cite{Collinsbook}, which we now discuss.   

\subsection{Nested subtractions}

The quantities $t_\rho\gamma$ [Eq.\ (\ref{eq:soft-approx})]  can also be thought of as counterterms for ultraviolet divergences associated with the limits $x_I^2 \rightarrow 0$ in the partonic matrix elements [Eq.\ (\ref{eq:fields})]  and with multieikonal amplitudes [Eq.\ (\ref{eq:wivertex})].   We will denote an arbitrary $n$-loop diagram  that is one-particle irreducible in the $x_I$ channel as $\gamma^{(n)}$.     Following the momentum-space procedure of Ref.~\cite{Collinsbook}, we define a regulated version of $\gamma^{(n)}$ by
\bea
R^{(n)}\, \gamma^{(n)}\ &=&\ 
\gamma^{(n)}\ +\  \sum_{N \in {\cal N}[\gamma^{(n)}]}\  \prod_{\rho\in N} \big(-t_{\rho}\big)\, \gamma^{(n)}\, ,
\label{eq:R-n-gamma}
\eea
where $\cal N[\gamma]$ is the set of all nonempty nestings for diagram $\gamma$.  We will refer to $R^{(n)}$ as the subtraction operator at $n$th order.   We may then write for the full $n$th-order $x_I$-irreducible partonic amplitude (\ref{eq:gen-G}),  $\bar G^{(n)} = \sum \gamma^{(n)}$, 
\bea
 \bar G^{(n)} \ &=&\     
\sum_{ \gamma^{(n)}}\; \left[ -\ \sum_{N\in {\cal N}[\gamma^{(n)}]}\ \prod_{\rho\in N} \big(-t_{\rho}\big) 
 \, \gamma^{(n)} \ +\ R^{(n)}\, \gamma^{(n)} \right] \, .
  \label{eq:R-n-1-0}
  \eea
The products in Eqs.\ (\ref{eq:R-n-gamma}) and (\ref{eq:R-n-1-0}) are ordered with the larger PSs to the right of smaller PSs.    Thus, the first approximation operators $t_\rho$ to act on $\gamma^{(n)}$ involve the fewest points on the light cones or at short distances.   As in Eq.\ (\ref{eq:soft-approx}), the approximation operators act on the diagram over the full integration region, and are not restricted to the neighborhood of the corresponding pinch surface.  
 
 Among the approximation operators that appear in $R^{(n)}\gamma^{(n)}$, we may identify the smallest, $\rho_\gamma$, for which 
all vertices approach the origin, that is, for which $H^{(\sigma_\gamma)}=\gamma^{(n)}$.    Now because $\rho_\gamma$ is the smallest PS, it nests with every other pinch surface. Its approximation operator, which we denote by $t_{uv}$ for any diagram, always appears to the left of every other operator in Eq.\ (\ref{eq:R-n-1-0}).  Operator $t_{uv}$ acts only on the external propagators that attach to $\gamma^{(n)}$.  We can thus separate it in the sum over nestings, and we find
\bea
 \bar G^{(n)} \ &=&\      
\sum_{ \gamma^{(n)}}\; \left\{ t_{uv}\gamma^{(n)}\ +\ \left( 1 - t_{uv}\right) \ \left [ -\ \sum_{N\in {\cal N}_P[\gamma^{(n)}]}\ \prod_{\rho\in N} \big(-t_{\rho}\big) 
 \, \gamma^{(n)} \ +\ R_P^{(n)}\, \gamma^{(n)} \right ] \right \}\, ,
  \label{eq:R-n-1-0b}
  \eea
where now ${\cal N}_P$ refers to the set of all proper nestings, not including $t_{uv}$, and $R_P^{(n)}$ is the corresponding ``proper" subtraction operation, defined by Eq.\ (\ref{eq:R-n-gamma}) with ${\cal N}$ replaced by ${\cal N}_P$, 
\bea
R_P^{(n)}\, \gamma^{(n)}\ &=&\ \gamma^{(n)}\ +\  \sum_{N \in {\cal N}_P[\gamma^{(n)}]}\  \prod_{\rho\in N} \big(-t_{\rho}\big)\, \gamma^{(n)}\, .
\label{eq:R-P-def}
\eea
The operator $R_P^{(n)}$ is related to $R^{(n)}$ by
\bea
\nn\\
R^{(n)}\gamma^{(n)}\ &=&\ (1-t_{uv})\; R_P^{(n)}\gamma^{(n)}\, .  
\label{eq:R-RP}
\eea
In the following, we will show that $R_P^{(n)}\gamma^{(n)}$ is free of subdivergences.

Specifically, we will show that the nesting, from regions to subregions, eliminates double counting, allowing the subtractions $t_{\rho}$ for each leading  PS $\rho$ to be extended from $\hat n[\rho]$  to the full space, as in the momentum-space discussion  in Ref.\ \cite{Collinsbook}.   We can also think of individual subtractions acting region by region; the purpose of the nested products is to cancel the action of subtractions outside their corresponding reduced neighborhoods $\hat n[\rho]$.  In summary, we claim that for each diagram $\gamma^{(n)}$, the action of the proper subtraction operation, $R_P^{(n)}$ is to remove divergences from leading pinch surfaces $\rho$, 
   \bea
   R_P^{(n)} \, \gamma^{(n)}\big  |_{{\rm div}\ \hat n[\rho]}\ &=&
\left[ \gamma^{(n)}\ +\  \sum_{N \in {\cal N}_P[\gamma^{(n)}]}\  \prod_{\rho\in N} \big(-t_{\rho}\big)\, \gamma^{(n)}  \right]\, \bigg |_{{\rm div}\ \hat n[\rho]}
\  =  \ 0\, ,
   \label{eq:R-gam-zero}
   \eea
for any PS $\rho$ with $H^{(\rho)} \subset\gamma^{(n)}$.  Assuming this result, the proper-subtracted diagram $R_P^{(n)}\gamma^{(n)}(y_1,\dots, y_a)$ is free of all subdivergences.   In particular, because all collinear singularities have been canceled, it remains finite when any of the $y_I^\mu$ approach the light cone, and because all soft subdiagrams are subtracted, it vanishes on dimensional grounds when the positions of external vertices go to infinity,
\bea
\lim_{\{y_I^2\rightarrow 0\}} R_P^{(n)}\gamma^{(n)}(y_1,\dots, y_a)\ &=&\ f^{(n)}\left(y_I\cdot y_J,\mu^2\right)\, ,
\nn\\
\lim_{\{y_K\cdot y_L\rightarrow\infty\}} f^{(n)}\left(y_I\cdot y_J,\mu^2\right)\ &=&\ 0\, ,
\label{eq:RP-limit}
\eea
in terms of some function $f^{(n)}$ that depends on the inner products $y_I\cdot y_J$ and in general on $\mu^2$, the renormalization scale.   

This result has important consequences for the full nested set of subtractions, including nestings that include $t_{uv}$, acting on the full $n$th-order amplitude, $G^{(n)}(x_1\dots x_a)$, Eq.\ (\ref{eq:gen-G}).   For this ``improper" PS, the soft function in Eq.\ (\ref{eq:soft-approx-diff}) is taken as unity and the jet subdiagrams are truncated propagators.    The approximation associated with $t_{uv}$ is, by Eq.\ (\ref{eq:co-hard-parton}), to replace $y_I^\mu$ by $\bbeta_I^\mu y\cdot \beta_I$ on the external propagators of $G^{(n)}_a$.    In schematic form, we can then represent the action of the full set of nested subtractions as
\bea
R^{(n)}G_a^{(n)} (x_1,\dots, x_a) \ &=&\ \left( 1 - t_{uv}\right)\ \prod_I \int d^4y_I\, G_2(x_I-y_I)\, R_P^{(n)} \,   \bar G_a \left(\{y_I\}\right)
\nn\\
&=& \
\prod_I \int d^4y_I\, \left( 1\ -\ \frac{G_2(x_I- y_I\cdot \beta_I \bbeta_I)} {G_2(x_I - y_I)}\  \right)\, R_P^{(n)} \,  G_a \left(\{y_I\}\right)\, ,
\eea
where the fraction represents the matrix inverse for fields with spin.   We now recall that poles in $x_I^2$ come about only from pinches in the integrals over the internal vertices of $G_a$, at configurations associated with physical processes.   For such configurations, $x_I\cdot y_I\sim x_I\cdot\bbeta_I y_I\cdot \beta_I $, and the right-hand side vanishes when the $x_I^\mu$ approach the light cone.   Thus, the full set of nested subtractions acting on the amplitudes, $R^{(n)}G_a^{(n)}$ lack poles in $x_I^2$, and their Fourier transforms will not contribute to the S-matrix,
\bea
R^{(n)}G_a^{(n)} \big |_{\rm div}\ =\ 0\, ,
\label{eq:R-n-G-n-result}
\eea
where in this case ``div'' refers specifically to the leading light-cone singularity in all external coordinates $x_I$.   Equivalently, from Eq.\ (\ref{eq:R-n-gamma}), we have
\bea
\gamma^{(n)} \big |_{\rm div}\ &=&\ 
 -\  \sum_{N \in {\cal N}[\gamma^{(n)}]}\  \prod_{\rho\in N} \big(-t_{\rho}\big)\, \gamma^{(n)}\big |_{\rm div} \, .
\label{eq:gamma-result}
\eea
 This conclusion is analogous to the result of Collins in Ref.\ \cite{Collinsbook} that the Sudakov form factor is power suppressed when subtracted according to the 
momentum-space procedure on which our approach is based.   Here we extend the reasoning to the general class of multiparton amplitudes.

  Returning to the sum of proper subtractions, we first note that for multieikonal amplitudes, the absence of subdivergences [Eq.\ (\ref{eq:R-gam-zero})] is easy to prove, because the {\it largest} PS for such an amplitude is one in which all noneikonal vertices are in the soft subdiagram.   As usual the approximation operator, $t_{\rm eik}$ for this PS takes the soft-collinear approximation (\ref{eq:soft-co}) for all external lines of the soft subdiagram, and because all such lines are attached to the Wilson lines, in this case, $t_{\rm eik}=1$ when acting on the amplitude.   Thus, since this PS can nest with every other PS, all terms in Eq.\ (\ref{eq:finite-condition}) cancel pairwise.   Indeed, the cancellation is exact, and for multieikonal amplitudes, we have
   \bea
   R_P^{(n)} \gamma^{(n)}_{\rm eikonal}\ =\ 0\, , \quad n>0\, ,
   \label{eq:R-P-zero}
   \eea
   with no remainder, or, equivalently, for $n\ge 1$,
   \bea
   \Gamma^{(n)}\ =\ -\ \sum_{\gamma^{(n)}_{\rm eikonal}}\ \sum_{N \in {\cal N}_P[\gamma^{(n)}]}\  \prod_{\rho\in N} \big(-t_{\rho}\big)\, \gamma^{(n)}_{\rm eikonal}\, .
   \label{eq:Gamma-sum}
   \eea
     For $n=0$, of course, there are no subtractions.   This reasoning does not apply to partonic amplitudes, for which the largest soft approximation is not  accurate in general.    

Before going to the proof of Eq.\ (\ref{eq:R-gam-zero}) for partonic amplitudes, it is worth noting the relationship between the subtraction approach here  and the momentum-space ``strategy of regions''~\cite{Smirnov:1999bza}.  In the latter, approximations tailored to regions of loop momenta that are the sources of leading behavior are also extended to all of loop momentum space.  
  We are doing something very similar here; each of the subtraction terms in each nesting is associated with a particular leading PS,  but we extend each such expression over the full coordinate integration space.  The list of PSs specifies the list of regions 
each of which defines an expansion in kinematic variables.
   By showing that all double counting is eliminated in the sum over all nestings, we will verify  that the sum of subtractions is an acceptable representation of the original amplitude, up to well-defined finite corrections.   There is also a connection to the organizations of the various subtraction methods that underly next-to-next-to-leading-order calculations of amplitudes and cross sections \cite{Binoth:2000ps}.

\subsection{Proof of the cancellation of subdivergences}
\label{sec:proof-cancel}

  To derive Eq.\ (\ref{eq:R-gam-zero}) for an arbitrary PS $\rho$ of diagram $\gamma^{(n)}$, we start by reorganizing the sum over nestings in $R_P^{(n)}\gamma^{(n)}$, Eq.\ (\ref{eq:R-P-def}), to highlight the role of an individual approximation $t_\rho$,
\be\begin{split}
  R^{(n)}_P\ \gamma^{(n)}   \ = 
  \ \gamma^{(n)}\ &+\ (-t_\rho)\gamma^{(n)}\ +\ \sum_{N^\rho\ne \rho} \left(  \prod_{\sigma\in N^\rho} \big(-t_{\sigma}\big)
\ +\  
  \prod_{\sigma\in N^\rho\backslash\rho} \big(-t_{\sigma}\big)\, \right) \gamma^{(n)}  \\
  &
  +\
  \sum_{\bar N^\rho}   \prod_{\sigma\in \bar N^\rho } \big(-t_{\sigma}\big) \, \gamma^{(n)}\, .\end{split}
\label{eq:reorganize-RP}
\ee
In the sum, we have separated those nestings denoted by $N^\rho$ that include $\rho$, along with the set $N^\rho\backslash \rho$, in which region $\rho$ can nest but is excluded, and finally the set of nestings with PSs that overlap with $\rho$, denoted by $\bar N^\rho$,  which cannot include $\rho$ because $\rho\, :o:\, \sigma$ for at least one element $\sigma \in \bar N^\rho$.    
  
 We now look at the contribution to Eq.\ (\ref{eq:reorganize-RP}) from region $\hat n[\rho]$, where we wish to verify Eq.\ (\ref{eq:R-gam-zero}), i.e., that the divergence from this region should vanish.  We already know from Eq.\ (\ref{eq:soft-approx-diff}) that the divergent parts of the first two terms on the right-hand side of Eq.\ (\ref{eq:reorganize-RP}) cancel in $\hat n[\rho]$, so that Eq.~(\ref{eq:R-gam-zero}) implies
  \bea
  \ \sum_{N^\rho\ne \rho} \left(  \prod_{\sigma\in N^\rho} \big(-t_{\sigma}\big)
\ +\  
  \prod_{\sigma\in N^\rho\backslash\rho} \big(-t_{\sigma}\big)\, \right) \gamma^{(n)}  \big |_{{\rm div}\ \hat n[\rho]}
  +\
  \sum_{\bar N^\rho}   \prod_{\sigma\in \bar N^\rho } \big(-t_{\sigma}\big) \, \gamma^{(n)} \big |_{{\rm div}\ \hat n[\rho]}
    \ &=&\ 0\, .
  \label{eq:finite-condition}
  \eea
We see that for Eq.~(\ref{eq:R-gam-zero}) to hold in each neighborhood $\hat n[\rho]$, the divergent parts of  all subtraction terms  {\it except} for $t_\rho \gamma^{(n)}$ alone must cancel (or vanish) in region $\hat n[\rho]$ defined by Eq.\ (\ref{eq:rhohat-def}).  To prove the absence of divergences in $R_P^{(n)}G^{(n)}$ for an arbitrary $\hat n[\rho]$, we must examine all nestings in Eq.~(\ref{eq:finite-condition}).

 We start with those nestings, $N^\rho$ in which $\rho$ appears along with at least one other PS.   For all such nestings, in neighborhood $\hat n[\rho]$, the term corresponding to nesting $N^\rho$ cancels the nesting, $N^\rho \backslash\rho$.   This is  because the action of $t_\rho$ is equivalent to the identity in region $\hat n[\rho]$, so that
\bea 
 \sum_{N^\rho\ne \rho} \left(  \prod_{\sigma\in N^\rho} \big(-t_{\sigma}\big)
\ +\  
  \prod_{\sigma\in N^\rho\backslash\rho} \big(-t_{\sigma}\big)\, \right) \gamma^{(n)} \, \Big |_{{\rm div}\ \hat n[\rho]}\ =\ 0 \ ,
  \label{eq:R-P-zero-2}
   \eea
where, as in Eq.\ (\ref{eq:t-rho-gamma}), the subscript ``div $\hat n[\rho]$'' refers to the sum of all divergent parts from the integral over $\hat n[\rho]$.  This implies that the proof of Eq.\ (\ref{eq:R-gam-zero}) reduces to showing that the sum of all overlapping subtractions cancels independently,
\bea
  \sum_{\bar N^\rho}   \prod_{\sigma\in \bar N^\rho } \big(-t_{\sigma}\big) \, \gamma^{(n)}\, \Big |_{{\rm div}\ \hat n[\rho]}\ =\ 0\, .
\label{eq:left-nests}
\eea
 Again, nestings $\bar N^\rho$ cannot include PS $\rho$ because one or more of its PSs $\sigma$ overlap with $\rho$.   Because we are interested in singular contributions, we need to treat only those nestings, $\bar N^\rho$ that are divergent in region $\rho$, and we will use this condition below.

Consider, then, an arbitrary nesting $\bar N^\rho$ that contains some set of PSs $\sigma$ that overlap with PS $\rho$.      Because of the transitive properties of nesting, Eqs.\ (\ref{eq:nest-trans}) and (\ref{eq:over-trans}),  we can partition the PSs $\sigma_i \in \bar N^\rho$ into three ordered sets
\cite{Collinsbook}: those that are larger than $\rho$, those that are smaller than $\rho$ and those that overlap with $\rho$,
\bea
\bar N^\rho\ &=& N_L \cup N_o \cup N_S\, , 
\nn\\
&\ & N_L[\rho]\ = \ \{ \sigma_j \supset \rho \}\, ,
\nn\\
&\ & N_o[\rho] \ = \ \{ \sigma_k \ :o: \  \rho \}\, ,
\nn\\
&\ & N_S[\rho] \ = \ \{ \sigma_l \subset \rho \}\, ,
\label{eq:rho-prime-sets}
\eea
where all $\sigma_j \supset \sigma_k \supset \sigma_l$. By Eq.\ (\ref{eq:over-trans}), there is only a single subset $N_o[\rho]$.

In the following, we will identify an ``enclosing'' PS ${\tau_\ind}$, which is intermediate between the sets $N_o[\rho]$ and $N_L[\rho]$ in Eq.\ (\ref{eq:rho-prime-sets}).   This PS, $\tau_\ind$ will contain both PS $\rho$ and every element $\sigma_k\in N_o$.   It will at the same time be contained in every element $\sigma_j\in N_L[\rho]$, including the case when it equals the smallest element of $N_L[\rho]$.  Specifically, for any element $\sigma$ of $N_o[\rho]$, the enclosing region, $\tau_\ind[\sigma,\rho]$ will be constructed to act as the identity when combined with $t_{\sigma} \gamma^{(n)}$ in neighborhood $\hat n[\rho]$, up to finite corrections, that is, 
\bea
t_{\sigma}\, \Big( 1 - t_{\tau_\ind}[\sigma,\rho] \Big)\, \gamma^{(n)} \Big |_{{\rm div}\ \hat n(\rho)}& = & 0\, .
\label{eq:ind-verify}
\eea
This is the basic property we will need.

\begin{figure}
\centering
\subfigure[]{
\includegraphics[height=4cm]{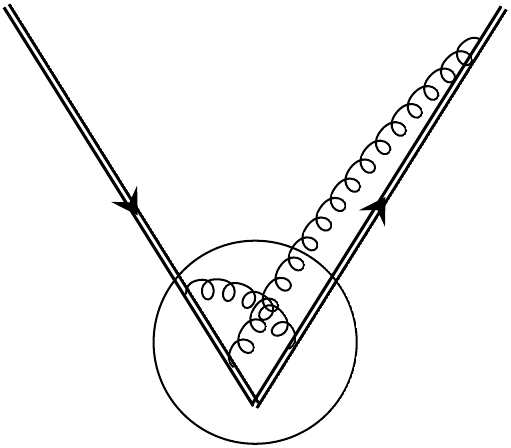}\label{fig:3ov1}} \quad
\subfigure[]{
\includegraphics[height=4cm]{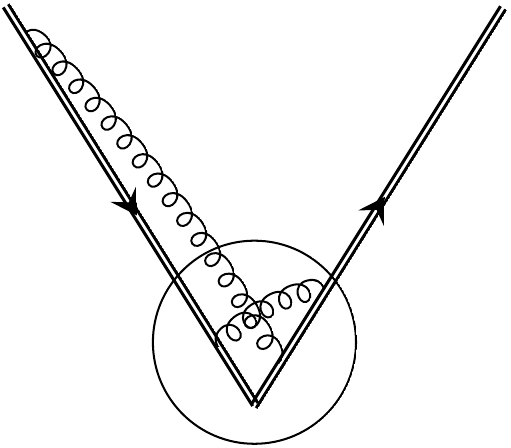}\label{fig:3ov2}} \quad
\subfigure[]{
\includegraphics[height=4cm]{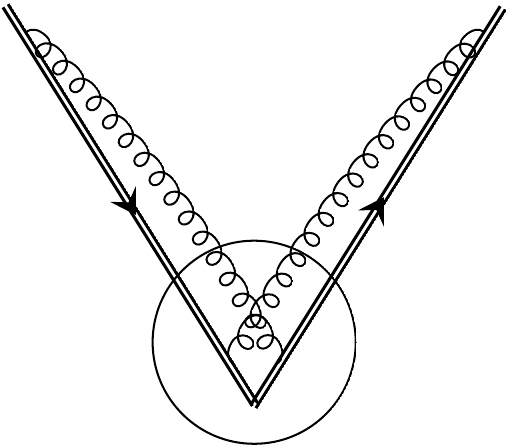}\label{fig:3ov3}}
\caption{\subref{fig:3ov1} and \subref{fig:3ov2} are examples of overlapping regions.  \subref{fig:3ov3} shows the enclosing region specified by Eqs.\ (\ref{eq:induce-1})--(\ref{eq:induce-larger}).}
\label{fig:3overlaps}
\end{figure}

The appropriate enclosing PS, $\tau_\ind[\sigma,\rho]$ is defined as usual by its hard, jet and soft subdiagrams.  These subdiagrams are determined in turn by the subdiagrams of PSs $\sigma$ and $\rho$ in the following manner,
\bea
S^{(\tau_\ind)} \ &=& \ S^{(\sigma)}\cup S^{(\rho)} \, \, \cup \prod_{I,K,\, I\ne K} \left(J^{(\sigma)}_I \cap J^{(\rho)}_K\right) \, , 
\label{eq:induce-1} \\
J^{(\tau_\ind)}_L \ &=& \ J^{(\sigma)}_L \cup J^{(\rho)}_L \, \big \backslash\,  S^{(\tau_\ind)}  \ , 
\label{eq:induce-2} \\
H^{(\tau_\ind)}\  &=&\  H^{(\sigma)}\cap\ H^{(\rho)} \nn \\
  &=& \gamma \, \big \backslash\ \left(  S^{(\tau_\ind)} \, \cup\, \prod_L J_L^{(\tau_\ind)} \right) \ . 
  \label{eq:induce-larger}
\eea
We claim that $\tau_\ind$ constructed in this manner satisfies  Eq.\ (\ref{eq:ind-verify}).   Equation (\ref{eq:ind-verify}) will hold  in region $\rho$ if two sets of conditions are met by $\tau_\ind$.   First, the construction must be self-consistent, which requires that $\tau_\ind$ represents a PS in the class already included in the nestings of Eq.\ (\ref{eq:R-n-gamma}).   This will be the case if the following is true:
\begin{itemize}

\item[]{1)} Whenever $t_\sigma\gamma$ is singular in region $\rho$, the overlap of $H^{(\sigma)}$ and $H^{(\rho)}$ is not empty.

\item[]{2)} Whenever $t_\sigma\gamma$ is singular in region $\rho$, $S^{(\tau_\ind)}$ is not connected to $H^{(\tau_\ind)}$.

\end{itemize}
In addition, for Eq.\ (\ref{eq:ind-verify}) to hold, we must also have the following:

\begin{itemize}

\item[]{3)} The hard-collinear approximations of Eq.\ (\ref{eq:co-hard}), applied by $t_{\tau_\ind}$  are accurate at PS $\rho$.

\item[]{4)} The soft-collinear approximations of Eq.\ (\ref{eq:soft-co}), applied by $t_{\tau_\ind}$  are accurate at PS $\rho$.

\end{itemize}
If all of these conditions are satisfied up to corrections that vanish as a power of one or more of the normal variables of PS $\rho$, then Eq.\ (\ref{eq:ind-verify}) holds, because the overall integral is logarithmically divergent and we have constructed  the reduced neighborhood $\hat n[\rho]$, Eq.\ (\ref{eq:rhohat-def}) to remove its subdivergences. A simple example illustrating Eqs.\ (\ref{eq:induce-1})--(\ref{eq:induce-larger}) is given by Figs. \ref{fig:3ov1}--\ref{fig:3ov3}.

Much of the subtlety in the construction of $\tau_\ind$ involves ``overlapping jets'' in different directions, in which some subsets of lines shift from one light cone in $\sigma$ to another light cone in $\rho$.  Many such subdiagrams, $J^{(\sigma)}_I \cap J_K^{(\rho)}$, are possible, and are defined by the list of PSs of each diagram $\gamma$.   We make two preliminary observations regarding these overlaps.

First, lines that carry physical polarization from each jet to the hard part do not contribute to these overlaps. This is easiest to see for fermionic external lines.   The shift of a fermion line of jet $I$ in region $\sigma$ to jet $K$ in region $\rho$ would require that the line pass through the soft subdiagram of region $\rho$, and we have seen in Eq.\ (\ref{eq:conditions-jet}) that fermion lines cannot connect jet and soft subdiagrams at leading PSs.   Similarly, also by Eq.\ (\ref{eq:conditions-jet}), a physically polarized gauge propagator of jet $I$ in region $\sigma$ cannot pass through $S^{(\rho)}$ to jet $K$ if $\rho$ is to remain a leading PS.

Second, in the coordinate-space integrals of $t_\sigma \gamma$,   certain PSs are modified by $t_\sigma$.   Specifically, PSs  $\rho$ involving overlaps  $J^{(\sigma)}_I \cap J_K^{(\rho)}$ are replaced by PSs  where the vertices of $J^{(\sigma)}_I \cap J_K^{(\rho)}$ are either pinched at the origin, or align only in the direction $\bbeta_I$, complementary to the direction of $J_I^{(\sigma)}$ and independent of the direction $\beta_K$ (that is, of the precise direction of the jet in region $\rho$).    This is because the soft-collinear, Eq.\ (\ref{eq:soft-co}),  and hard-collinear, Eq.\ (\ref{eq:co-hard}), approximations that act on the external lines and vertices of $J_I^{(\sigma)}$ eliminate dependence on all vectors except for $\beta_I$ and $\bbeta_I$.

 With these observations in mind, we can now give proofs of conditions 1) -- 4) above.

\subsubsection{ {\bf Overlap of hard subdiagrams} }

The construction of the enclosing PS, $\tau_\ind$ using Eq.\ (\ref{eq:induce-larger}) requires a nonvanishing overlap between the hard subdiagrams $H^{(\sigma)}$ and $H^{(\rho)}$.   As we have seen in Sec.\ \ref{subset:leading-regions}, for leading regions $\sigma$ and $\rho$,  the hard subdiagrams $H^{(\sigma)}$ and $H^{(\rho)}$ are themselves simply connected. For the cusp or processes initiated by a single external current, the hard subdiagrams of all leading regions overlap at the current, as illustrated by Fig.\ \ref{fig:shjets-cusp}.   For scattering amplitudes, however, there are many cases where regions $\sigma$ and $\rho$ have disjoint hard subdiagrams.  This happens whenever the hard subdiagram in $\sigma$, $H^{(\sigma)}$, is entirely contained in the union of soft and jet subdiagrams in $\rho$, $S^{(\rho)}\cup \prod_L J_L^{(\rho)}$.  We now show that in all such cases, either PS $\rho$ is suppressed, or  $\rho$ is actually not a PS of $t_\sigma\gamma$.   

\begin{figure}
\centering
\includegraphics[height=4.5cm]{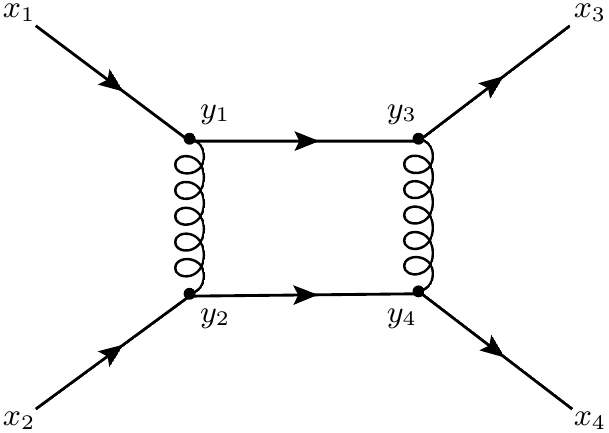}

\caption{An example of  disjoint hard subdiagrams.}
\label{fig:2hseq}
\end{figure}

Let us suppose that $H^{(\sigma)}$ and $H^{(\rho)}$ are disjoint.   We then consider the external lines of $H^{(\sigma)}$, on which the hard-collinear approximations (\ref{eq:co-hard}) and (\ref{eq:co-hard-parton}) have acted.   Because the hard subdiagrams are disjoint, neither these lines nor the vertices of $H^{(\sigma)}$ to which they attach can be in $H^{(\rho)}$.     Then,  each external vertex of the hard subdiagram $H^{(\sigma)}$ either appears as an internal vertex in some jet subdiagram $J_L^{(\rho)}$ of $\rho$, or is an internal vertex of the soft function, $S^{(\rho)}$ of $\rho$.   We will see that when any such vertex is in a jet subdiagram of $\rho$, then $\rho$ is not a PS for  $t_\sigma\gamma$.   We will then show that when all the external vertices of $H^{(\sigma)}$ are in $S^{(\rho)}$, then $\rho$ is not a leading PS of $t_\sigma\gamma$.   The only possibility left is that at least one vertex of $H^{(\sigma)}$ is also a vertex in $H^{(\rho)}$, so that the hard subdiagrams are not disjoint.

Consider first Fig.\ \ref{fig:2hseq}, which illustrates the possibility that an external vertex of $H^{(\sigma)}$ is in a jet subdiagram of PS $\rho$. In the figure, we identify the hard scattering $H^{(\sigma)}$ as the propagator $y_1-y_2$, along with the vertices $y_1$ and $y_2$ which it connects.    In $H^{(\sigma)}$, $y_1\rightarrow y_2 \rightarrow 0$, while $y_3$ and $y_4$ remain at finite distances from each other with $(y_3-y_4)^2\ne0$, and approach the light cones defined by $x_3$ and $x_4$, respectively.   We then let $H^{(\rho)}$ be  defined by $y_3-y_4$ and $\rho$ by the limit $y_3\rightarrow y_4 \rightarrow 0$ while $y_1$ and $y_2$ remain at finite distances with $(y_1-y_2)^2\ne 0$, such that the lines $(y_1,y_3)$ and $(y_2,y_4)$ are in the jet subdiagrams associated with external points $x_1$ and $x_2$, respectively.   These two regions clearly cannot be nested, and their hard subdiagrams are disjoint.         Now in the neighborhood of PS $\rho$, as $y_3$ and $y_4$ approach the origin, $y_1$ should move to the $\beta_1$ light cone, which requires a pinch in the $y_1$ integral.   Since $(y_1- y_2)^2\ne 0$ at $\rho$, this pinch can  come from the denominators $x_1-y_1$ and $y_1-y_3$.      The action of $t_\sigma$, however, partially decouples the internal integrals of the hard subdiagram $(y_1,y_2)$ from the $\sigma$ jet subdiagrams.  For the subtracted diagram, $t_\sigma \gamma$, $t_\sigma$ applies the hard-collinear approximation $hc(1)$ from Eq.\ (\ref{eq:co-hard-parton}) to $(x_1- y_1)^\mu$, which is then replaced by a line $(x_1-y_1\cdot \beta_1\bbeta_1)^\mu$.   Similarly, it applies $hc(3)$ to  $y_1-y_3$, which is replaced by $(y_1\cdot \beta_3 \bbeta_3-y_3)^\mu$.     The two lines thus depend on different components of $y_1^\mu$, and  the $y_1$ integrals cannot be pinched at $y_1\cdot \beta_1=0$.   The lines $x_1-y_1$ and $y_1-y_3$ are therefore not pinched in region $\rho$ after the action of $t_\sigma$.

The feature of the example in Fig.\ 5 that extends to more general cases is that operator $t_\sigma$ acts on the line $y_1-y_3$ with the hard-collinear approximation $hc(3)$, which eliminates the pinch that fixes vertex $y_1$ in $J_1^{(\rho)}$.   In the original integral, the coordinate $y_1\cdot \beta_1$ is pinched at the origin between poles from the propagators of $x_1-y_1$ and $y_1-y_3$ when they are in $J_1^{(\rho)}$.  After the action of $t_\sigma$, however,  the propagator associated with line $y_1-y_3$ no longer has a pole when  $y_1\cdot \beta_1=0$ because $y_1\cdot \beta_3\ne 0$ when $y_1\propto \beta_1$.   To extend this result to higher orders, we must treat general hard-scattering subdiagrams, and allow the possibility that more than one external vertex of $H^{(\sigma)}$ appears in the jet subdiagrams $J_L^{(\rho)}$ of PS $\rho$.

Suppose then, that in the general case more than one external vertex of $H^{(\sigma)}$ attaches to lines in a subdiagram $J_L^{(\rho)}$.   Now, because $H^{(\sigma)}$ is connected, and because in region $\rho$ each jet corresponds to physically propagating lines, one or more lines in jet $J_L^{(\rho)}$ must enter subdiagram $H^{(\sigma)}$ at one of its external vertices, and leave at another external vertex.   Since these lines are external to $H^{(\sigma)}$, they must be included in jet subdiagrams $J_K^{(\sigma)}$ of PS $\sigma$.  At all such external vertices, then, the approximation operator $t_\sigma$ will have applied the hard-collinear approximation [Eq.\ (\ref{eq:co-hard}) or (\ref{eq:co-hard-parton})] that is appropriate for the directions  and polarizations of these external lines of $H^{(\sigma)}$ in PS $\sigma$.   As we have observed above, however, the imposition of the hard-collinear approximation for jet $J^{(\sigma)}_K$, say, by $t_\sigma$ eliminates pinches in that subdiagram except in the $\beta_K$ and $\bbeta_K$ directions.   (We assume for simplicity that no pair of jets satisfies $\beta_I=\bbeta_L$.).   As a result, no pinch that sets these lines to the light cone in $\rho$ is possible, unless the overlap of $H^{(\sigma)}$ with jet $J_L^{(\rho)}$ involves only external lines that are also in $J^{(\sigma)}_L$, {\it i.e.} in $J_L^{(\rho)} \cap J^{(\sigma)}_L$.  This requires some of the lines of $J_L^{(\sigma)}$ to change direction.    At PS $\sigma$, the lines in $J_L^{(\sigma)}$ all flow in or all flow out of $H^{(\sigma)}$, but at PS $\rho$, some would have to flow in and some out.  In the physical picture corresponding to PS $\rho$, the relevant vertices of $H^{(\sigma)}$ are all either before $H^{(\rho)}$, or after.   For definiteness, we assume they are before, so that $J_L^{(\rho)}$ is an incoming jet.

On the other hand, the external lines of jet $J_L^{(\sigma)} \cap J_L^{(\rho)}$, can carry at most one physical polarization from the external point $x_L$ into subdiagram $H^{(\sigma)}$.  All other lines that attach $J_L^{(\sigma)}$ to $H^{(\sigma)}$ must be scalar polarized.     This physical polarization is then eliminated by the net action of the hard-collinear approximation in diagram $H^{(\sigma)}$, because only one external line of $J_L^{(\sigma)}$ can be physically polarized. 
In the case where the vertex at which the physical polarization reaches $H^{(\sigma)}$ is in $J_L^{(\sigma)}\cap J_L^{(\rho)}$, the physical polarization then cannot reach the hard subdiagram $H^{(\rho)}$ (which is by assumption disjoint from $H^{(\sigma)}$) because all the other$J_L^{(\rho)}$ lines are scalar polarized.  But then PS~$\rho$ is nonleading. 
These considerations imply that the jet subdiagrams $J_L^{(\rho)}$ of PS $\rho$ cannot share lines or vertices with the hard subdiagram $H^{(\sigma)}$, unless all the vertices that attach the individual lines that carry physical polarization from  jets $J_K^{(\sigma)}$ to $H^{(\sigma)}$ are in $S^{(\rho)}$.   As a result, if $H^{(\sigma)} \cap H^{(\rho)}$ were to be empty, the ``physical'' vertices of $H^{(\sigma)}$ would all have to be in $S^{(\rho)}$.

We now treat the possibility that the vertices that bring physical polarizations to the hard subdiagram $H^{(\sigma)}$ in region $\sigma$ are entirely in $S^{(\rho)}$, and show that in this case $\rho$ is nonleading.   The reason  is illustrated by the example of Fig.~\ref{fig:2hardsA}, assuming that the PS $\sigma$ describes the scattering of (massless) fermions.   The alternative physical process in the figure, with a hard scattering involving gluons, would require the fermions to be in the soft subdiagram $S^{(\rho)}$, a configuration that is always nonleading by Eq.\ (\ref{eq:conditions-jet}), see Eq.~(\ref{eq:p-general}) and Ref.~\cite{Erdogan:2013bga}.   This reasoning applies to any order and diagram:  restricting ourselves to fermion-fermion scattering to be specific, at any leading PS, the external fermions must only appear as jet lines, and as external lines of both hard subdiagrams $H^{(\sigma)}$ and $H^{(\rho)}$.    But then, since the fermion lines are continuous, the hard subdiagrams must be connected by these jet lines, which must be in different directions in the two PSs.   The definition of Eq.\ (\ref{eq:induce-larger}) is then guaranteed to give a connected hard subdiagram $H^{(\tau_\ind)}$.    In a similar fashion, for external gluons, the role of fermion lines is taken by gluon lines that carry the external physical polarizations of the gluons.   From the general power-counting result (\ref{eq:p-general}), such polarizations cannot be radiated into soft subdiagrams at leading PSs, and the same conclusion as for external fermions applies.   

In summary, $H^{(\sigma)} \cap H^{(\rho)}$ is never empty.

\subsubsection{{\bf Soft and hard disjoint}} 

The external lines of $S^{(\tau_\ind)}$ are either external lines of  the soft subdiagrams $S^{(\sigma)}$ and/or $S^{(\rho)}$ or of the overlaps of jet subdiagrams $\prod_{I,K} J_I^{(\sigma)} \cap J_K^{(\rho)}$.   Now the external lines of $S^{(\sigma)}$  can only attach to the jet subdiagrams of $\sigma$, $J_I^{(\sigma)}$ and are hence are separated from $H^{(\rho)} \cap H^{(\sigma)}$, and similarly for lines in $S^{(\rho)}$.   

 To verify that lines in $J_I^{(\sigma)}\, \cap\,J_K^{(\rho)}$ cannot attach to $H^{(\tau_\ind)}$ at leading PSs, we consider a gauge line in the $I$th jet subdiagram, $J^{(\sigma)}_I$, attached at one end to an arbitrary vertex at a point in $H^{(\sigma)}$, and at the other end to a vertex that is in subdiagram $J_I^{(\sigma)}$.    It is easy to see that if this line is also in $\prod_{I,K} J_I^{(\sigma)} \cap J_K^{(\rho)}$, it cannot attach directly to $H^{(\tau_\ind)}$, because $t_\sigma$ acts by $hc(I)$,  Eq.\ (\ref{eq:co-hard}), on the external lines of $J_I^{(\sigma)}$, and produces a $\bar\beta_I$ polarization at $H^{(\sigma)}$.  This polarization is suppressed when coupled to the lines of $J_I^{(\sigma)}\, \cap\,J_K^{(\rho)}$,  which, as we have observed below conditions \mbox{1) -- 4)} for the consistency of the construction of $\tau_\ind$, can have PSs only in the $\bbeta_I$ direction in region $\rho$.  As $\bbeta^2_I=0$, leading contributions are eliminated when $J_I^{(\sigma)}\, \cap\,J_K^{(\rho)}$ attaches to $H^{(\tau_\ind)}$. Thus, none of the elements of $S^{(\tau_\ind)}$ can attach directly to $H^{(\tau_\ind)}$, and the two subdiagrams are disjoint.

\subsubsection{{\bf Hard-collinear}}  

Any line from $J_L^{(\tau_\ind)}$ that is attached to $H^{(\tau_\ind)}$ either attaches $J_L^{(\rho)}$  to $H^{(\rho)}$ or $J_L^{(\sigma)}$ to $H^{(\sigma)}$ (or possibly both).   If the line is from $J_L^{(\rho)}$, $t_{\tau_\ind}$ will apply the hard-collinear approximation  $hc(L)$ [Eq.\ (\ref{eq:co-hard})], which is   a good approximation in region $\rho$.  If the line is from $J_L^{(\sigma)}$, both $t_\sigma$ and $t_{\tau_\ind}$ apply the hard-collinear approximation, $hc(L)$, whether or not the line is in $J_L^{(\rho)}$, that is, whether or not $hc(L)$ is a good approximation at PS $\rho$.  The result, however, is the same for $t_\sigma t_{\tau_\ind}\gamma$ or $t_\sigma\gamma$ alone because,  as we easily verify from Eq.\ (\ref{eq:co-hard}), $hc(L)^2=hc(L)$.  Thus, all hard-collinear connections are consistent with Eq.\ (\ref{eq:ind-verify}).

\subsubsection{{\bf Soft-collinear}}  

The soft-collinear approximation must work for all external lines of $S^{(\tau_\ind)}$.   By Eq.\ (\ref{eq:induce-larger}), these external lines are either in  $S^{(\sigma)}\cup S^{(\rho)}$ or in $\prod_{I,K,\, I\ne K} \left(J^{(\sigma)}_I \cap J^{(\rho)}_K\right)$.  

The first case, lines in $S^{(\sigma)}\cup S^{(\rho)}$, is relatively straightforward.    The external lines of $S^{(\rho)}$ attach only to jet lines from subdiagrams $J_L^{(\rho)}$ that are also in $J_L^{(\tau_\ind)}$ and those from $S^{(\sigma)}$ attach only to $J_L^{(\sigma)}$ lines that are in $J_L^{(\tau_\ind)}$.   The operator $t_{\tau_\ind}$ applies the soft-collinear approximation to all such lines.   Applied to lines in $S^{(\rho)}$, this is a good approximation in region $\rho$.   For lines in $S^{(\sigma)}$, both $t_{\tau_\ind}$ and $t_\sigma$ apply the soft-collinear approximation $sc(L)$, Eq.\ (\ref{eq:soft-co}), and since $sc(L)=sc(L)^2$, they are consistent with Eq.\ (\ref{eq:ind-verify}).

\begin{figure}[t]
\centering
\subfigure[]{
\includegraphics[height=3.4cm]{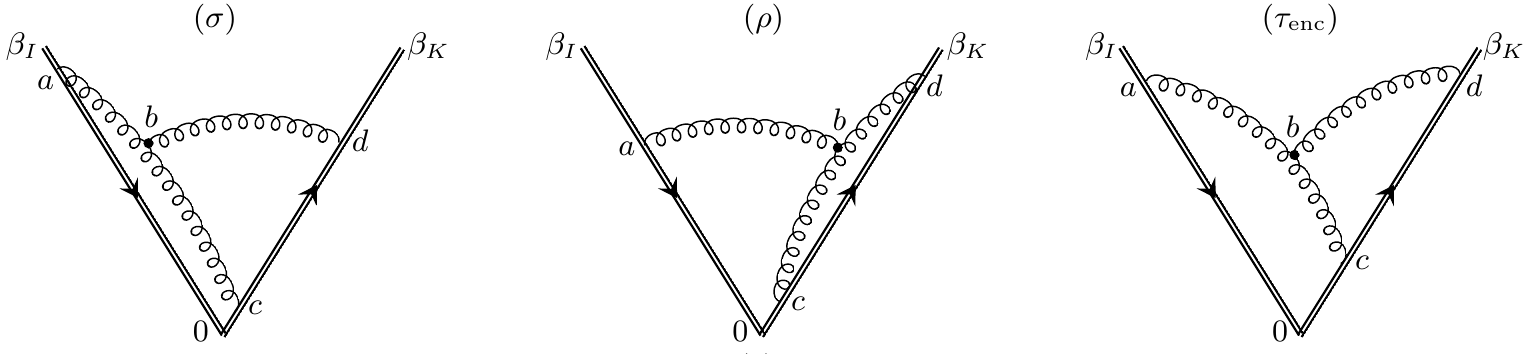}\label{fig:3gex1}}
\vspace{3mm}
\subfigure[]{
\includegraphics[height=3.4cm]{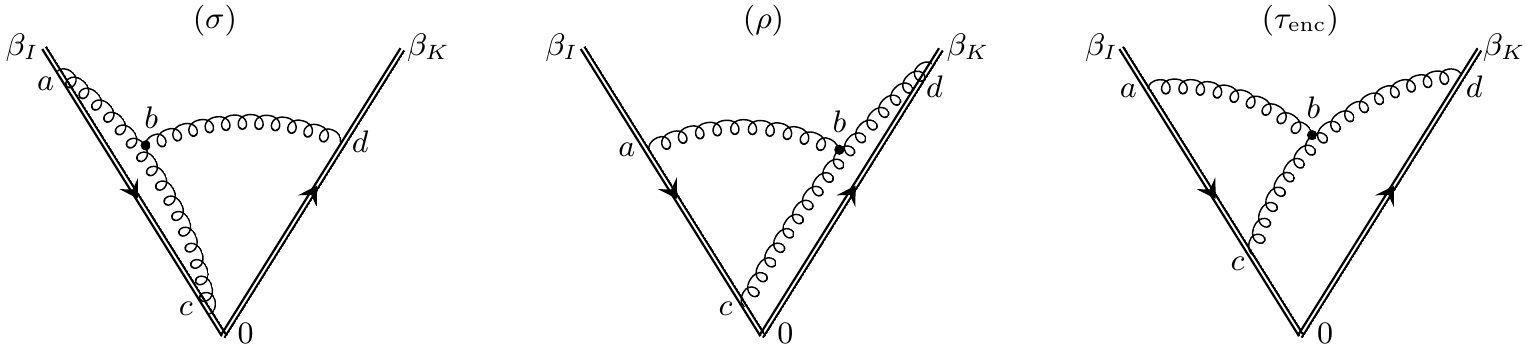}\label{fig:3gex2}}

\caption{Example of nested overlapping subdivergences in PSs $\sigma$ and $\rho$ and their enclosing PS $\tau_\ind[\sigma,\rho]$, where \subref{fig:3gex1} illustrates the case $L\ne I$ and \subref{fig:3gex2} $L= I$\@.}
\label{fig:3gex}
\end{figure}

The case of $S^{(\tau_\ind)}$ lines in $\prod_{I,K,\, I\ne K} \left(J^{(\sigma)}_I \cap J^{(\rho)}_K\right)$ is somewhat more complex.   The external lines of the intersections $J^{(\sigma)}_I \cap J^{(\rho)}_K \in S^{(\tau_\ind)}$, $I\ne K$ are attached to jet lines of subdiagrams $J_L^{(\tau_\ind)}$ of $\tau_\ind$, and will have the soft-collinear approximation $sc(L)$ applied to them by $t_{\tau_\ind}$.    The action of $t_\sigma$, however, depends on whether: (a) $I\ne L$ or (b) $I=L$.   We treat these cases in turn, using the examples of Fig.\ \ref{fig:3gex} to illustrate the method.    The figure represents two pairs of overlapping PSs, $\sigma$ and $\rho$, in a two-loop correction involving two partonic or eikonal lines, labeled $I$ and $K$ in the figure.  We should think of this example as embedded in a larger diagram with any number of external lines, connected at a hard subdiagram denoted by $0$ in the figure.

\begin{itemize}

\item[]({\bf a})  Figure \ref{fig:3gex1} is an example of $L\ne I$.   As indicated by the positions of the vertices in the figure, in PS $\sigma$ of Fig.~\ref{fig:3gex1}, $J^{(\sigma)}_I$ consists of the lines connecting vertices in the set $\{a,b,c,0\}$ except for the line $\{c,0\}$.   In this notation, jet $J_K^{(\sigma)}$ is $\{c,d\}$, $S^{(\sigma)}$ is $\{b,d\}$, and the hard subdiagram $H^{(\sigma)}$ is  $\{0,c\}$. 
On PS $\rho$, $J_I^{(\rho)}$ is $\{0,a\}$, $J_K^{(\rho)}$ is $\{0,c,b,d\}$ and $S^{(\rho)}$ is $\{a,b\}$ and $H^{(\rho)}=\{0\}$.    For this example, the overlap of the jets is  $J^{(\sigma)}_I \cap J^{(\rho)}_K=\{b,c\}$, a single line.    Following Eqs.\ (\ref{eq:induce-1})--(\ref{eq:induce-larger}), PS $\tau_\ind$ has all of its gauge lines in its soft subdiagram, $S^{(\rm \tau_\ind)}=\{a,b,c,d\}$.   Its hard subdiagram is just the vertex at the origin, $H^{(\tau_\ind)}=\{0\}$, and its jets are $J^{(\tau_\ind)}_I=\{0,a\}$ and $J^{(\tau_\ind)}_K$=$\{0,c,d\}$.  Clearly, the intersection, $J^{(\sigma)}_I \cap J^{(\rho)}_K=\{b,c\}$ of jets from $\sigma$ and $\rho$ attaches to the jet line $L=K$.

Turning to the general case, when $L\ne I$, then in PS $\sigma$,  the vertex at which the line in  $J^{(\sigma)}_I \cap J^{(\rho)}_K \in S^{(\tau_\ind)}$ connects to $J_L^{(\tau_\ind)}$ must also connect either a line in $J_L^{(\sigma)}$ and a line in $J_L^{(\rho)}\cap H^{(\sigma)}$ or to two lines in $J_L^{(\rho)}\cap H^{(\sigma)}$, since in PS $\sigma$ a line in $J_I^{(\sigma)}$ can only be attached to a line from the distinct jet $J_L^{(\sigma)}$ at a vertex that is part of the hard subdiagram $H^{(\sigma)}$, as in the example.   In addition, because one or more  lines in $J_L^{(\rho)}\cap H^{(\sigma)}$ attach to the same vertex as the line in $S^{(\tau_\ind)}$, which is in $J_K^{(\rho)}$, we {\it must} have $L=K$, again as in the example.   Otherwise, the vertex would connect lines from different jets $J_K^{(\rho)}$ and $J_L^{(\rho)}$, which would force it to be part of $H^{(\rho)}$.  Such a vertex would then be in $H^{(\sigma)}\cap H^{(\rho)}=H^{(\tau_\ind)}$, contrary to our assumptions.

We are thus restricted to the case $J_I^{(\sigma)}\cap J_L^{(\rho)}$ quite generally. Now, when lines from  $J^{(\sigma)}_I \cap J^{(\rho)}_L$ attach to $J_L^{(\tau_\ind)}$, as at vertex $c$ in Fig.\ \ref{fig:3gex1}, the hard-collinear approximation, $hc(I)$ is applied by $t_\sigma$, after the application of the soft-collinear approximation $sc(L)$ from $t_{\tau_\ind}$ in the product $t_\sigma t_{\tau_\ind}$.   In this case, applying the definitions of the soft-collinear and hard-collinear approximations, Eqs.\ (\ref{eq:soft-co}) and (\ref{eq:co-hard}), we readily verify that the product of operations $hc(I)( 1 - sc(L))$ produces a nonleading contribution, as an identity of the integrand, so that Eq.\ (\ref{eq:ind-verify}) holds.

To be specific, suppose that the vertex where $J^{(\sigma)}_I \cap J^{(\rho)}_L$ attaches to $J_L^{(\tau_\ind)}$  is at point $u$, and the other end of the line in $J^{(\rho)}_L$ is at point $w$. In the example, $u=c$ and $w=b$, but in complete generality,  if we denote these vertices by $v_\mu(u)$ and $v'_\nu(w)$, respectively, we get
\bea
hc^\sigma(I) \left( 1\ -\ sc^{\tau_\ind}(L)\right) \ v'_\mu(w)\frac{-g^{\mu\nu}}{(w-u)^2}v_\nu(u)\ &=&\ 
v'(w) \cdot \bbeta_I \beta_{I\mu}\ 
\left[ \frac{-g^{\mu\nu}}{(w-u\cdot \beta_I \bbeta_I)^2} \bbeta_{I\nu} \beta_I\cdot v(u) \right .
\nn\\
&\ & \hspace{-10mm} -\ 
\ \left. \frac{-g^{\mu\nu}}{(w-u\cdot \bbeta_L \beta_L\cdot \beta_I \bbeta_I)^2} \bbeta_{I\nu} \beta_I\cdot \beta_L \bbeta_L\cdot v(u) \right ]
\label{eq:approx-L-K}
\, ,
\eea
where the superscripts in $hc^\sigma(I)$ and $sc^{\tau_\ind}(L)$ indicate the PS associated with the soft-collinear and hard-collinear approximations. Note that in the special case of ``back-to-back" jets, $\beta_I=\bbeta_L$, the right-hand side of Eq.~(\ref{eq:approx-L-K}) vanishes identically.   This is the case of the Sudakov form factor.     Let us suppose, more generally, that $\beta_I \ne \bbeta_L$, or equivalently, $\beta_L\cdot \bbeta_I \ne 0$.
Since line $w-u$  and vertex $u$ are both in jet $J_L^{(\rho)}$, the point $w$ may be in $J_L^{(\rho)}$ or in the hard-scattering subdiagram $H^{(\rho)}$.  Let us first treat the case when  vertex $v'(w)$ is also in $J_L^{(\rho)}$.   Then, up to terms that vanish as a power of the normal variables of PS $\rho$, we may approximate for both ends of line $w-u$,
\bea
w^\mu &=& \beta_L^\mu \bbeta_L\cdot w\, ,
\nn\\
u^\mu &=& \beta_L^\mu \bbeta_L\cdot u\, .
\label{eq:u-w-jet}
\eea
We consider first the two denominators that represent the line $w-u$ in Eq.~(\ref{eq:approx-L-K}) individually.    At PS $\rho$ this line was originally on the light cone and the denominator $(w-u)^2$ vanishes linearly in the scaling variable $(\lambda)$, in terms of the normal variables introduced in Eq.\ (\ref{eq:normal-list}) of Sec.\ \ref{subsec:power-ctg}.   Also, the presence of two three-point vertices in the jet subdiagram would ensure an additional factor of $\lambda$ in the numerator (see Ref.\ \cite{Erdogan:2013bga}).    After the action of the soft-collinear and hard-collinear approximations, however, both the denominator corresponding to this line, and the numerator factor are order $\lambda^0$ near PS $\rho$.    The net effect is that both terms on the right-hand side of Eq.~(\ref{eq:approx-L-K}) are leading power ($\lambda^0$) at PS $\rho$.    At the same time, using Eq.\ (\ref{eq:u-w-jet}), we find that in neighborhood $\hat n[\rho]$,
\bea
(w-u\cdot \beta_I \bbeta_I)^2\
=\
(w-u\cdot \bbeta_L \beta_L\cdot \beta_I \bbeta_I)^2\ +\ {\cal O}(\lambda^{1/2})\, ,
\eea
so that the right-hand side of Eq.~(\ref{eq:approx-L-K}) is suppressed by a power of the scaling variable at PS $\rho$ when vertex $v(w)$ is in jet $L$.

For the alternative case, the limit that $v(w)$ is in $H^{(\rho)}$, that is, when $w^\mu\rightarrow 0$, the denominators still vanish on PS $\rho$ even after the approximations, while the leading power behavior corresponds to a finite numerator involving a scalar-polarized gauge propagator.    The denominators still cancel to leading power in the scaling variable $\lambda$, however, and the difference is again subleading.  

\item[]{({\bf b})}   Figure \ref{fig:3gex2} represents the case $L=I$, that is, when a line in subdiagram $S^{(\tau_\ind)}$ attaches to a line in subdiagram $J_I^{(\sigma)}$.   The assignment of lines and vertices to subdiagrams is almost the same as in Fig.~\ref{fig:3gex1}, except that vertex $c$ is now part of the $I$ jet in $\sigma$, and is connected to the $K$ jet on PS $\rho$.   

When $L= I$, which implies that $L\ne K$, $t_\sigma$ does not impose the $hc(L)$ approximation because in PS $\sigma$ at least two of the lines [lines $\{a,c\}$ and $\{b,c\}$ in the example of Fig.\ \ref{fig:3gex2}] that meet at the vertex connecting $S^{(\tau_\ind)}$ to $J_I^{(\tau_\ind)}$ are in $J_I^{(\sigma)}$, so that the remaining line must also be in $J_I^{(\sigma)}$. [This is line $\{0,c\}$ in Fig.\ \ref{fig:3gex2}.)  In PS~$\rho$, we use that the line from $S^{(\tau_\ind)}$ is in $J_K^{(\rho)}$.  (This is line $\{b,c\}$ in Fig.\ \ref{fig:3gex2}.] This line attaches to two lines of
$J_I^{(\tau_\ind)}= J^{(\sigma)}_I\cup J^{(\rho)}_I \,\backslash \prod_{I,K,\, I\ne K} (J^{(\sigma)}_I \cap J^{(\rho)}_K$).    Now these lines must be in either $J_I^{(\rho)}$ or $H^{(\rho)}$, and at least one must be in $H^{(\rho)}$.   This is because $I\ne K$, so that in PS $\rho$, lines from $J_I^{(\rho)}$ and $J_K^{(\rho)}$ can join only at the hard subdiagram $H^{(\rho)}$.  [Again, this is $\{0,c\}$ in Fig.\ \ref{fig:3gex2}.] As a result, in PS $\rho$, the hard-collinear approximation $hc^\sigma(K)$ is good, and we may invoke the same analysis as in case ({\bf a}) above for 
$hc^\sigma(K)(1-sc^{\tau_\ind}(I))$, $K\ne I$.  Again the sum of terms is suppressed and all soft-collinear connections are consistent with Eq.\ (\ref{eq:ind-verify}).  

\end{itemize}

This completes our arguments for conditions 1) -- 4) below Eq.\ (\ref{eq:induce-larger}), which ensure the consistency of the construction $\tau_\ind$.

\subsubsection{\bf Cancellation from nesting with the enclosing region}

So far, we have shown how to construct the enclosing PS, $\tau_\ind[\sigma,\rho]$ and have confirmed that $t_{\sigma} t_{\tau_\ind}\gamma$ is a good approximation to $t_\sigma \, \gamma$ in PS~$\rho$, so that Eq.\ (\ref{eq:ind-verify}), $t_\sigma\gamma^{(n)}|_{\rm div}=t_\sigma t_{\tau_\ind}\gamma^{(n)}|_{\rm div}$, is satisfied.    We note that showing Eq.~(\ref{eq:ind-verify}) for $t_\sigma \gamma$ implies the same result for $t_{\sigma'}\, t_{\sigma}\gamma$ for any nested pair, $\sigma' \subset \sigma$, because the approximations of $t_{\sigma'}$ do not modify the list of pinch surfaces or power counting in PS $\tau_\ind$, which was all that was used in the discussion above.   
We are now ready to show that with this definition of $\tau_\ind$, Eq.\  (\ref{eq:left-nests}) is satisfied, that is, that the sum of subtractions cancels for arbitrary overlapping regions.   To proceed, assuming that $N_o[\rho]$ is not empty, we construct the enclosing PS for the pair $\rho$ and the largest PS within $N_o[\rho]$, which we denote by $\sigma_{o \max}[\rho]$.    By construction, both $\rho$ and $\sigma_{o \max}$ are smaller in the sense of nesting  [Eq.\ (\ref{eq:nest-def})] than every element in $N_L[\rho]$.   In fact, $\tau_\ind[\sigma_{o \max},\rho]$ is also smaller than all elements of $N_L[\rho]$, or equal to the smallest, in the sense of Eq.\ (\ref{eq:nest-def}).    To confirm this, consider a PS $\zeta$ in $N_L$.   For $\tau_\ind \subseteq \zeta$, we need
\bea
H^{(\zeta)}\ \subseteq \ H^{(\tau_\ind)}\, ,
\nn\\
S^{(\zeta)}\ \supseteq \ S^{(\tau_\ind)}\, .
\label{eq:nest-reqs}
\eea
The first of these relations follows immediately from the definition of nesting (\ref{eq:nest-def}) and the construction (\ref{eq:induce-1})--(\ref{eq:induce-larger}), since any vertex in $H^{(\zeta)}$ must be in both $H^{(\sigma)}$ and $H^{(\rho)}$, and therefore in $H^{(\tau_\ind)}$.    The second relation requires us to verify that
\bea
S^{(\zeta)} \ \supseteq S^{(\sigma)} \cup S^{(\rho)} \cup \prod_{I\ne K} J^{(\sigma)}_I \cap J^{(\rho)}_K\, .
\label{eq:S-zeta-nest}
\eea
To verify this relation, we note that because PS $\zeta$ is larger than both $\sigma$ and $\rho$, $S^{(\zeta)} \supset S^{(\sigma)}$ and $S^{(\zeta)} \supset S^{(\rho)}$, so that $S^{(\zeta)} \supset S^{(\sigma)} \cup S^{(\rho)}$.   Next, we consider subdiagrams $J^{(\sigma)}_I \cap J^{(\rho)}_K$.   Again, because $\zeta\supset \sigma$, by Eq.\ (\ref{eq:nest-def}), any line in $J_I^{(\sigma)}$ must be in either $J_I^{(\zeta)}$ or $S^{(\zeta)}$,  and similarly, because $\zeta \supset \rho$ as well, $J_K^{(\rho)}$ must be in either $J_K^{(\zeta)}$ or $S^{(\zeta)}$.   The only possibility for a line in $J^{(\sigma)}_I \cap J^{(\rho)}_K$ is then $S^{(\zeta)}$. Equation~(\ref{eq:S-zeta-nest}) then follows, and we have
\bea
\zeta \ \supseteq\ \tau_\ind\, .
\eea
We conclude that the enclosing PS, $\tau_\ind$ is contained by all of the elements of $N_L[\rho]$ or is equal to the smallest element in $N_L[\rho]$.    At the same time, $\tau_\ind$ itself contains PS $\sigma_{o \max}$, the largest of the regions in $N_o[\rho]$.   Therefore, $\tau_\ind[\sigma_{o \max},\rho]$ nests with all the elements of $\bar N^{(\rho)}$, and either $\tau_\ind[\sigma_{o \max},\rho]$ is already contained in $\bar N^\rho$ or the set $\bar N^{\rho,\tau_\ind}\equiv \{\bar N^\rho,\tau_\ind[\sigma_{o \max},\rho]\}$ is an acceptable nesting, and is already included in $R_P^{(n)}G^{(n)}$ [Eq.\ (\ref{eq:R-P-def})].   Also, $t_{\sigma_{o \max}} t_{\tau_\ind[\sigma_{o \max},\rho]} \gamma$ is a good approximation to $t_{\sigma_{o \max}}\,\gamma$ in region $\rho$, 
so that Eq.\ (\ref{eq:ind-verify}) holds.  Then, leading contributions cancel, either between nesting $\bar N^\rho$  and $\bar N^\rho \backslash \tau_\ind[\sigma_{o \max},\rho]$ if $\tau_\ind$ is already in $\bar N^\rho$, or between $\bar N^\rho$ and
$ \bar N^{\rho,\tau_\ind}$ if it is not.    Thus, we have verified the cancellation of  the sum over $\bar N^\rho$ in Eq.~(\ref{eq:left-nests}) and the ultraviolet finiteness of the subtracted diagram (\ref{eq:R-gam-zero}), which is what we were after.

\section{Renormalization of Wilson-Line Amplitudes and Factorization for Partonic Amplitudes}
\label{sec:renorm-fact}

In this section, we apply the regularization procedure of the foregoing section to verify the multiplicative renormalizability of multieikonal vertices involving massless Wilson lines, thus generalizing the results of Brandt, Neri and Sato in Ref.\ \cite{Brandt:1981kf}.  We will also confirm the factorization of partonic amplitudes in coordinate space, corresponding to the momentum-space factorization of S-matrix amplitudes for fixed-angle scattering shown originally by Sen in Ref.~\cite{Sen:1982bt}.   

Our discussion begins by reviewing how the soft-collinear and hard-collinear approximations in Eqs.\ (\ref{eq:soft-co})--(\ref{eq:co-hard-parton}) result in exact scalar polarizations for gauge lines that couple the soft to jet subdiagrams, and for all unphysically polarized gluons coupling jets to the hard subdiagrams \cite{Erdogan:2013bga}.   We conclude from gauge-theory Ward identities that the approximation operators, $t_\rho$ [Eq.\ (\ref{eq:soft-approx})], act to factorize amplitudes into hard, jet and soft subdiagrams at the level of integrands.   We then use these factorization properties and the nested subtractions of Eq.\ (\ref{eq:R-n-1-0}) to renormalize multieikonal vertices coupling massless Wilson lines, and factorize amplitudes for massless partons, when the positions of all external fields define a physical scattering process.

\subsection{Approximations and Ward identities}
\label{subsec:ward-identities}
We recall that the action of the approximation operator $t_{\rho}$ is to perform the soft- and hard-collinear approximations on gauge propagators that attach the soft subdiagram for PS $\rho$ to the jet subdiagrams  and  on gauge propagators that attach the jet subdiagrams to the hard subdiagram of PS $\rho$, such that the leading singularity of $\gamma$ in neighborhood $\hat n[\rho]$ is given by Eq.\ (\ref{eq:soft-approx}).  Reference\ \cite{Erdogan:2013bga} showed how the soft-jet and jet-hard gluon connections, approximated by their dominant polarization states as in Eq.\ (\ref{eq:soft-approx}), may be replaced by scalar polarizations (equivalent to longitudinal polarizations for massless particles).  We begin with a review of the method.  Consider first a soft-jet connection, as specified by Eq.~(\ref{eq:soft-co}).  We  rewrite  the propagator given in Eq.~(\ref{eq:soft-co}) as
\be 
D^{\mu\nu}(x-\tau^{(K)}\beta_K)\ = \ \frac{\partial}{\partial
  \tau^{(K)}}\int^{\tau^{(K)}}_{\infty} d\tau_K
D^{\mu\nu}(x-\tau_K\beta_K) \ =\  
\frac{\partial}{\partial   z^{(K)}\cdot \bbeta_K}\int^{z^{(K)}\cdot \bbeta_K}_{\infty} d\tau_K
D^{\mu\nu}(x-\tau_K\beta_K)\ , 
\ee
where we have used the definition of $\tau^{(K)}$ in terms of $z^{(K)}$, which is the position of the vertex in jet subdiagram~$K$ to which this line attaches.   We then integrate by parts in $\bbeta_K\cdot z^{(K)}$ in Eq.~(\ref{eq:soft-approx}) so that the derivative now acts on the $\beta_K$ component of the jet function in the soft-collinear approximation.  To this, we are free to add derivatives with respect to the other coordinate components of  $z_\mu^{(K)}$ to the integrand, acting on corresponding components of the jet subdiagram, because these terms are total derivatives and vanish after the integration.  The soft-collinear approximation (\ref{eq:soft-co}) then becomes \cite{Erdogan:2013bga}
\bea
sc(K)\ \left[ D^{\mu\nu}(x-z^{(K)}) \right] \ J_\nu(z^{(K)}) & \rightarrow & \int^{z^{(K)}\cdot \bbeta_K}_{\infty} d\tau_K\
D^{\mu\nu'}(x-\tau_K\beta_K)\beta_{K\, \nu'} \ \big(-\partial^{\nu}
J_{\nu}(z^{(K)}) \big) \ , 
\label{eq:soft-collinear-coord}
\eea
where the right-hand side is to be interpreted as the integral over an eikonal line in the jet direction $\beta_K$, and parameter $\tau_K$ is the position of the attachment of the soft line, $x-\tau_K\beta_K$ to this eikonal, multiplied by the divergence of the jet function at vertex $z^{(K)}$. This summarizes the soft-collinear approximation defined in Ref.~\cite{Erdogan:2013bga} for coordinate-space integrals, and can be carried out independently for each gluon to which we apply the soft-collinear approximation.

\begin{figure}[t]
\centering
\includegraphics[height=5cm]{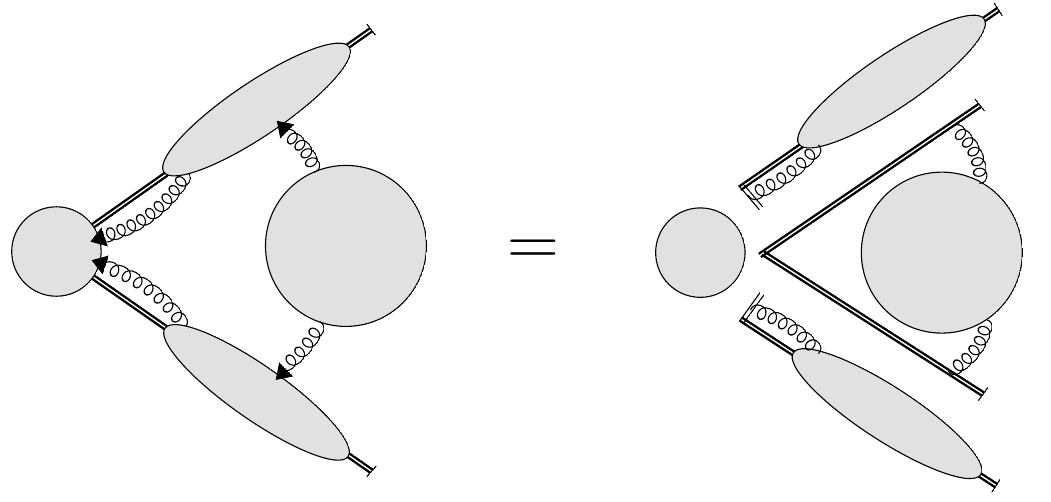}
\caption{Factorization in a leading region of a vertex function, illustrating Eq.\ (\ref{eq:rho-fact}).   Each line ending in an arrow represents arbitrary numbers of gluons in the soft-collinear or hard-collinear approximation defined in the text.   As in Fig.\ \ref{fig:shjets-first}, the double lines passing through the jet functions may represent either Wilson lines or partons.}
\label{fig:shjets}
\end{figure}

Similarly, an unphysical gluon coupling the $I$-jet to the hard-scattering function, as in Eq.\ (\ref{eq:co-hard}) is replaced in the hard-collinear approximation by a convolution in a single component of the gluon propagator with the divergence of the hard-scattering function,
\bea 
hc(I)\ \left[  D^{\mu\nu}(z-y^{(I)}) \right]\ H_\nu(y^{(I)}) \ &\rightarrow&  \int^{y^{(I)} \cdot \beta_I}_{\infty} d\eta_I
D^{\mu\nu'}(z-\eta_I\bbeta_I)\bbeta_{I\, \nu'} \ \big(-\partial^{\nu} H_{\nu}(y^{(I)}) \big) \, . 
\label{eq:hard-co-scalar-pol}
\eea
The right-hand sides of Eqs.\ (\ref{eq:soft-collinear-coord}) and (\ref{eq:hard-co-scalar-pol}) are respectively the Fourier transforms of the soft-collinear and hard-collinear approximations in momentum space. The application of the momentum-space approximations was discussed extensively in Refs.\ \cite{Bodwin:1984hc,Collins:1989gx}, for example.  Replacing the jet-soft connections by scalar-polarized gluon lines that are associated with the scalar operator $\partial_{\mu}A^{\mu}(x)$ allows us to apply the gauge-theory Ward identities. After the sum over all diagrams, the Ward identities then ensure the factorization of the soft lines from jet subdiagrams in coordinate space, in exactly the same way as in momentum space in  Refs.~\cite{Bodwin:1984hc,Collins:1989gx}.    We note that the Ward identity we need for showing the factorization in the case of multieikonal amplitudes was  derived as part of the proof of renormalizability for smooth Wilson lines in Ref.~\cite{Dotsenko:1979wb}. The resulting factorization is illustrated in Fig.\ \ref{fig:shjets}.

The factorized amplitudes illustrated by Fig.\ \ref{fig:shjets} are of course somewhat different in the cases of multieikonal and partonic amplitudes, and we will begin with the multieikonals.  For multieikonal amplitudes the jets are themselves singlet cusp operators,
\bea
j_I^{\, \rm cusp}(\vep)
=
c_I^{\rm cusp}(\beta_I,\bbeta_I,\vep)\ 
\bigg \langle 0\left| T\bigg(  \Phi^{[f_I]}_{\beta_I}(\infty,\tau_I \beta_I)\, \Phi_{\bbeta_I}^{[f_I]}{}^\dagger(\infty,\tau_I\beta_I) \bigg)\right|0  \bigg \rangle\, ,
\label{eq:J-eik-cusp}
\eea
which, as indicated by its argument, may be regulated dimensionally.  The vectors $\tau_I\beta_I$ are the positions of the cusps after the application of the Ward identities, but the matrix element must be independent of $\tau_I$  by translation invariance.     In addition, because of the symmetry between velocities $\beta_I$ and $\bbeta_I$, these jet functions have collinear singularities in both the $\beta_I$ and $\bbeta_I$ directions if $\bbeta_I^2=0$, as in our discussion above.  As shown in the previous section, all $\bbeta_I$ collinear singularities cancel in the sum over nestings, although this is only true for the full amplitude. The prefactor $c_I(\beta_I,\bbeta_I,\vep)$ is chosen to eliminate overall divergences associated with the choice of vector $\bbeta_I$ in the factorized form.   As we shall review in the following section, collinear singularities for $\bbeta_I$ occur additively in the logarithm of $j_I$, and can therefore be compensated by a multiplicative factor in each of the jets, and correspondingly by the inverse factor in the remaining soft and hard factors of the amplitude.  Collinear singularities associated with $\bbeta_I$ correspond to the rapidity divergences discussed  in direct QCD and soft-collinear effective theory in the references cited in Refs.~\cite{Collinsbook,Collins:2008ht}, for example, where specific methods of handling these extra divergences were developed. 
In general, the factorization of jet and soft functions requires an additional renormalization, as we introduce composite operators into the matrix elements for the jet functions, and also in the soft function.   We will show below that the renormalization of the soft function is also multiplicative.

For multieikonal amplitudes, the jet, soft and hard functions are in convolution only with respect to distances $\tau_I$ from the origin along each of the eikonal velocities, $\beta_I$.   The eikonal jet functions, however, are independent of the position of their cusps, and we may write
\be 
t_\rho\, \Gamma =
 \left[  \prod_I \ j_{I,\rho}^{\rm cusp}(\vep)  \int d\tau_I \; \lambda_\rho(\{\tau_I\}) \right]^{(n-m)}  \ \times\ \sum_{\gamma^{(m)}} \gamma^{(m)}\left ( \{ \tau_I\beta_I \}  \right) \, ,
\label{eq:rho-fact}
\ee
where the product $\times$ indicates a product in color space, and where the remaining integrals are over light-cone variables along 
the directions $\beta_I$ of the jets.  After the sum of diagrams necessary for the Ward identities, the dependence on $\rho$ of the right-hand side is all in the order, labeled $m<n$ of the sum over possible hard subdiagrams and in the choices of the individual jet functions $j_I^{\rm cusp}$ and of the soft subdiagram, labeled $\lambda_\rho$.  Their total order is denoted by $n-m$, for a specific PS $\rho$.  In the spirit of the notation of  Eq.\ (\ref{eq:soft-approx}) for the approximation operators, the factorized soft and jet functions
may be represented as
\bea
\lambda_\rho(\{\tau_K\})\ & = & \ \prod_K c_K^{-1} \int_{\tau_K} \overline{d u_K} \ \beta_K^{\mu_K}\
S^{(\rho)}_{\{\mu_K\}}(\{u_K\})\,  ,
\label{eq:lambda_def} \\
j^{{\rm cusp}}_{I,\rho}(\vep)\ & = & \ 
c_I\ \int_0 \overline{d v_I}\,
J_{I}^{(\rho)\, \nu_I'}(\{v_I\},\vep)\ \bbeta_{I,\nu'_I} \ .
\eea
Relative to Eq.\ (\ref{eq:soft-approx}), the integrals over  distances along light-cone directions, $u_K$ and $v_I$, corresponding to $\tau^{(K)}$ in Eq.~(\ref{eq:tau-def}) and $\eta^{(I)}$ in Eq.~(\ref{eq:zeta-def}), act only on the soft and jet functions, and are no longer in convolution with the jet and hard functions, respectively, except through the lower limit $\tau_K$ for the Wilson lines of the soft function, which is set by the position of the outermost vertex of the hard subdiagram on each eikonal line $\beta_K$.  In both functions, these integrals are ordered along the relevant eikonals, which we indicate by an overline.   

The factorized result (\ref{eq:rho-fact}) is by itself suggestive, and also represents the true behavior of the amplitude in region~$\rho$.   We now turn to a more complete derivation,   which starts from the nested subtraction forms of the multieikonal and partonic amplitudes.    We will derive  a single expression that combines the approximations associated with all PSs.   For multieikonal amplitudes, we will use factorization through Ward identities to construct a soft function that incorporates the color coherence properties of the amplitude, and which is renormalized multiplicatively.   We will then go on to use this result to show that partonic amplitudes factorize into a form that involves the same soft function.

\subsection{Factorization and renormalization for multieikonal vertices with massless Wilson lines}

So far, all of our integrals were computed using the renormalized gauge-theory Lagrangian. As a composite operator, the multieikonal vertex itself produces ultraviolet divergences, and requires further renormalization.  The multiplicative renormalizability of such vertices was proved in Ref.\ \cite{Brandt:1981kf} for massive Wilson lines.   In this section we confirm that multiplicative renormalization survives the zero-mass limit in Minkowski space, in spite of the presence of nonlocal ultraviolet collinear singularities.   We will find that the latter factor into universal jet functions, depending only on the color representations of the Wilson lines, which can themselves be renormalized multiplicatively.  All color coherence between different Wilson lines is contained in a standard soft function matrix, which requires multiplicative renormalization, as shown in momentum space in Ref.~\cite{Sen:1982bt}.   The discussion below shows how this renormalization and factorization can be implemented in covariant gauges for massless lines, and in coordinate space.   

Starting from Eq.\ (\ref{eq:R-n-1-0}), we consider the sum over nestings of an arbitrary $n$th-order diagram, $\gamma^{(n)}$, either partonic or multieikonal, with external self-energies removed.   We isolate within each nesting the smallest PS that corresponds to the largest, that is, highest-order, hard subdiagram, and denote this PS by $\sigma_0[N]$.     In general, $\sigma_0$ is not the smallest PS in the nesting, because there may also be pinch surfaces with the same hard subdiagram, but larger jet subdiagrams.   These differ from PS $\sigma_0$ by increasing jet subdiagrams at the expense of soft subdiagrams.   Separating the subtractions smaller and larger than $\sigma_0$, we  rewrite our expression for the $n$-loop amplitude in terms of approximation operators, Eq.\ (\ref{eq:R-n-1-0}) as
\bea
 G^{(n)} \ &=&\     
\sum_{ \gamma^{(n)} }\; \sum_{\sigma_0[\gamma^{(n)}]}\ \sum_{N_{coll\ }[\sigma_0]}\ \prod_{\omega\in N_{coll}[\sigma_0]} (-t_\omega)\ t_{\sigma_0}\, \sum_{N_> [\sigma_0]} \ 
\prod_{\sigma \in N_>[\sigma_0 ]} \big(-t_{\sigma} \big) 
 \, \gamma^{(n)} \ +\ R^{(n)}\, G^{(n)} \, ,
  \label{eq:R-n-1-1}
  \eea
  where $N_{coll}[\sigma_0]$ labels nestings smaller than $\sigma_0$, which share the same hard subdiagram (after the use of Ward identities), while $N_>[\sigma_0]$ represents all nestings that have $\sigma_0$ as their smallest element.  At this stage, the symbol $G^{(n)}$ may refer to a partonic as well as a multieikonal amplitude.   Each $\sigma_0$ divides diagram $\gamma^{(n)}$ into two subdiagrams.    The first, which we will denote by $\lambda^{(n-m)}=S^{(\sigma_0)}\cup \prod_I J_I^{(\sigma_0)}$ is an $n-m$th-order ``outer" subdiagram, consisting of lines in the soft and jet subdiagrams of $\sigma_0$.  We count in order $n-m$ those factors of the coupling associated with vertices where jet lines attach to the hard subdiagram of PS $\sigma_0$.    Subdiagram $\lambda^{(n-m)}$ is connected to the remaining, hard subdiagram,  $H^{(\sigma_0)}$, by jet lines only.  The remaining order of $H^{(\sigma_0)}$ is $m$.  The approximation operators $t_\omega$ in Eq.~(\ref{eq:R-n-1-1}) take into account all nestings involving soft-collinear connections in the outer subdiagram.

  For notational purposes, we now identify a ``reduced" hard subdiagram, which we will denote by $\bar \gamma^{(m)}[\sigma_0]$.    This is the diagram found by deleting unphysically polarized jet gluons from the hard subdiagram.    By construction, $\bar \gamma^{(m)}$ is irreducible under cuts of the external eikonal lines.

We now claim, following Collins \cite{Collinsbook},  that the Ward identities can be applied to the subtracted inner diagrams just as for the unsubtracted case. 
We imagine acting with the approximation operators  $t_\sigma$ in Eq.\ (\ref{eq:R-n-1-1}) one at a time, starting with the right most, that is, the one corresponding to the largest PS in nesting $N$, which we will refer to as $\sigma^{(N)}_{\rm max}$.  Summing over diagrams, the application of Ward identities leads to a factorized form, with a soft subdiagram, $S^{(\sigma^{(N)}_{\rm max})}$ and partonic jets, as in Fig.\ \ref{fig:shjets}.   To this set of diagrams we apply the approximation operator corresponding to the next-largest PS.   By the nesting construction, this approximation operator acts only on lines in the jet and soft subdiagrams of PS $\sigma^{(N)}_{\rm max}$, leading through the Ward identities to a new set of jet and hard subdiagrams.   In this set, the nesting requirement allows lines from the jets of $\sigma^{(N)}_{\rm max}$ to be absorbed into the hard subdiagram, and lines of the soft subdiagram of $\sigma^{(N)}_{\rm max}$ to be absorbed into new, fully eikonal jet subdiagrams, which, however, are now disconnected from the partonic jet subdiagrams that are produced by $t_{\sigma^{(N)}_{\rm max}}$.  This procedure can be repeated as many times as there are approximation operators in any nesting, and at each stage, the Ward identities can be used.    As a result of the nesting, all lines and vertices of the soft subdiagram corresponding to $\sigma_0$, the smallest PS that has the hard subdiagram shared by all smaller PSs in the nesting,  are already in a subdiagram of $S^{(\sigma^{(N)}_{\rm max})}$.  Similarly, all jet lines of $\sigma^{(N)}_{\rm max}$ are in the corresponding jet subdiagrams of $\sigma_0$.  The approximation operator, $t_{\sigma_0}$ acts to separate the lines of 
diagrams, $\lambda^{(n-m)}$ from the remaining diagrams, $\bar \gamma^{(m)}$ as in Fig.\ \ref{fig:shjets} and as in the partonic case [Eq.~(\ref{eq:rho-fact})], and also factorizes the soft and jet subdiagrams within $\lambda^{(n-m)}$.  At each stage in this process, the subdiagrams that are left behind as subdiagram $\lambda^{(n-m)}$ is factorized are independent of the number of scalar-polarized lines to which we have applied the Ward identities.  The series of diagrams that result from this procedure is thus identical to the diagrams that would be found by the proper subtractions of diagram $\bar \gamma^{(m)}$.   Then, once $\lambda^{(m-n)}$ is factored, we may replace the sum over nestings $N_{>[\sigma_0]}$ of $\gamma^{(n)}$ in Eq.\ (\ref{eq:R-n-1-1}) by a sum over the proper nestings of $\bar\gamma^{(m)}$.

All these considerations apply as well to partonic and multieikonal amplitudes, but for now we restrict our discussion to multieikonal  amplitudes, and return to the partonic case in the following subsection.
  Applied to the multieikonal case, the Ward identities  factorize subdiagram $\lambda_{\rm eikonal}$ from the remainder of the $n$th-order diagram, giving
 \bea
 \Gamma^{(n)} \  &=& \ 
 \prod_I \int d\tau_I \sum_{ \lambda_{\rm eikonal}^{(n-m)} }\,  \prod_{\omega\in N_{coll}[\lambda_{\rm eikonal}^{(n-m)}]} (-t_\omega)\ \hat t_{\sigma_0[N_{coll}]}\   \lambda_{\rm eikonal}^{(n-m)}\, \left (\{ \tau_I\beta_I  \} \right)\; \times\; {\cal H}_{\rm eikonal}^{(m)}\left ( \{ \tau_I\beta_I \}\right)
  \ +\ R^{(n)}\Gamma^{(n)}\, ,
  \nn\\
  \label{eq:lambda-sum-eikonal}
  \eea
  where the function ${\cal H}_{\rm eikonal}^{(m)}$  absorbs the action of all proper subtractions on $\bar \gamma_{\rm eikonal}^{(m)}$.   Precisely because $\bar \gamma_{\rm eikonal}^{(m)}$ is eikonal, we have for $m>0$
   \bea
 {\cal H}_{\rm eikonal}^{(m)}\left ( \{ \tau_I\beta_I \}\right)\ &=&  
 \sum_{\bar \gamma^{(m)}}\; \sum_{N_P[\bar \gamma_{\rm eikonal}^{(m)}] }\ \prod_{\sigma\in N_P[\bar \gamma_{\rm eikonal}^{(m)}] }\big(-t_{\sigma}\big)  \, \bar \gamma_{\rm eikonal}^{(m)}(\{\tau_I\beta_I\})
 \nn\\
 &=& \sum_{\bar\gamma^{(n)}} \ R_P^{(n)}\; \bar \gamma_{\rm eikonal}^{(n)}
 \nn\\
 &=&\ 0\, ,
  \label{eq:cal-H-def}
\eea
where we have used the vanishing of the sum of proper subtractions in the eikonal approximation (\ref{eq:R-P-zero}).    
Term by term, however, the variables $\tau_I$ are the positions of the $I$th Wilson line vertices farthest from the multieikonal vertex in each diagram. These are the only integration variables that link the diagrams $\lambda_{\rm eikonal}$ with those of ${\cal H}_{\rm eikonal}={\cal H}_{\rm eikonal}^{(0)}=\prod_I \delta(\tau_I)$.

The operator $\hat t_{\sigma_0}$ in Eq. (\ref{eq:lambda-sum-eikonal}) represents the remaining action of $t_{\sigma_0}$ on diagrams $\lambda^{(n-m)}$, which consists of the union of the soft and collinear subdiagrams of PSs $\sigma_0$. In the case where $\lambda^{(n-m)}$ is entirely soft on PS $\sigma_0$, we define $\hat t_{\sigma_0}=1$.

\begin{figure}[t]
\centering
\includegraphics[height=5cm]{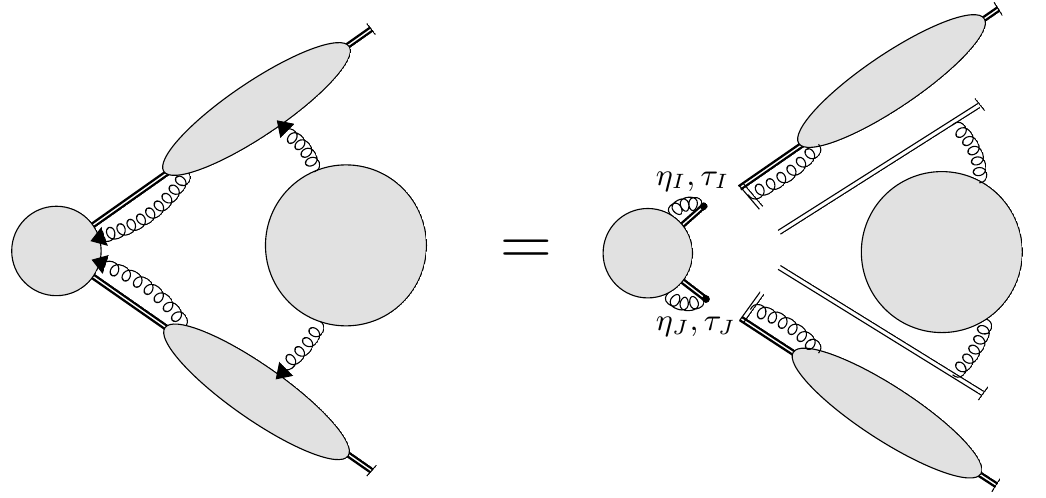}

\caption{Factorization in a leading region of a multieikonal or partonic amplitude, illustrating Eqs.\ (\ref{eq:lambda-sum-eikonal}) and (\ref{eq:full-partonic-2}), at fixed values of $\tau_I$, $\tau_J$ and $\eta_I$, $\eta_J$,  which are coordinates of the outermost vertices of the hard subdiagram along each external line.   In multieikonal amplitudes, the $\eta_I$ are all identically zero.  Only two external lines are shown, but the result generalizes to any number.  The jet functions are given by the perturbative expansions of the matrix element in Eq.~(\ref{eq:J-eik-cusp}).   The hard function on the right contains all proper subtractions and reduces to the lowest-order vertex for multieikonal amplitudes, but retains a full perturbative expansion in partonic amplitudes.}
\label{fig:full-eikonal}
\end{figure}

  Summing Eq.\ (\ref{eq:lambda-sum-eikonal}) over all orders, and using Eq.~(\ref{eq:cal-H-def}), now gives for multieikonal amplitudes,
\bea
\Gamma\ =\ \prod_{I=1}^a\ \int d\tau_I\ \Lambda \left (\{\tau_I\beta_I\} \right)\, ,
\label{eq:full-multieikonal}
\eea
with 
\bea
\Lambda \left (\{\tau_I\beta_I\} \right) \ =
\sum_{i=0}^\infty\  \sum_{\lambda^{(i)}} \sum_{N_{coll}[\lambda^{(i)}]} \prod_{\omega\in N_{coll}} (-t_\omega)\, \hat t_{\sigma_0[N_{coll}]}\, \lambda^{(i)}\left (\{\tau_I\beta_I\} \right)\, ,
\label{eq:Lambda-eik-def}
\eea
where $i=n-m$ in Eq.~(\ref{eq:lambda-sum-eikonal}) and $\lambda^{(i)}$ represents an arbitrary $i$-loop diagram.   This factorization is illustrated by Fig.~\ref{fig:full-eikonal}, with two external lines shown explicitly.   After the sum over all proper subtractions, the eikonal lines of the soft function $\Lambda$  meet at a point as in Fig.~\ref{fig:shjets}.

The action of the Ward identities  on any sum of $i$th-order multieikonal diagrams $\lambda^{(i)}$ by any $t_\omega$, including $\hat t_{\sigma_0}$ is
\bea
- t_\omega \sum_\lambda \lambda^{(i)}\left (\{\lambda_I\beta_I\} \right) &=& \left( -\prod_{I=1}^{a} \hat t_I^{(i_{I\omega})} \right) \sum_\lambda \lambda^{(i)}\left (\{\tau_I\beta_I\} \right)
\nn\\
&=&\  - \prod_{I=1}^{a}\ j^{(i_{I\omega})}(\tau_I\beta_I)\; \sum_\lambda \lambda^{(i-\sum_{I=1}^a i_{I\omega})}\left (\{\tau_I\beta_I\} \right)\, ,
\label{eq:t-omega-resolve}
\eea
where $\sum_\lambda$ applies to all diagrams of the same order, and where in the first equality the approximation operator $t_\omega$ is resolved into independent approximation operators $\hat t_I^{(i_{I\omega})}$, each implementing the soft-collinear approximation on a subdiagram connected to the $I$th Wilson line.   Each approximation operator $\hat t_I$ includes information on the loop order of this subdiagram, which we denote by $i_{I\omega}$.   Operator $\hat t_I^{(i_{I\omega})}$ factorizes jet subdiagram $j^{(i_{I_\omega})}$, and replaces it by a Wilson line attached to the remaining soft subdiagram.   
The jets found this way are, up to a constant, the order-by-order expansion of the matrix elements of Eq.\ (\ref{eq:J-eik-cusp}), and are independent of the positions $\tau_I$ of the cusps for the multieikonal case. This is not, however, the case for the remaining soft subdiagram, which is singular whenever more than one $\tau_I$ vanishes in general.  This procedure can be repeated for each of the jets, and for the action of the nested (smaller) pinch surfaces.

We now define a notation for  products of jet functions, evaluated at fixed loop order, $l$.   These products depend on the vectors that define the jets, their end points $\tau_I$, and the number of loops, but not on the relative ordering (labeled $\omega$ above) of the subtraction within the specific nesting,
\bea
{\cal J}^{(0)}\left (\{\beta_I,\bbeta_I\} \right)\ &=&\ 0\, ,
\nn\\
{\cal J}^{(l)}\left (\{\beta_I,\bbeta_I\} \right)\ &=&\ \sum_{\{l_{I}\}}\ \delta_{l,\sum l_{I}}\ \prod_{I=1}^a j_I^{(l_{I})}( \{\beta_I,\bbeta_I\} )\, , \quad  l>0\, ,
\label{eq:cal-J-defs}
\eea
where the jet functions $j_I^{(l_{I})}$ are the sum of all $l_{I}$-loop-order diagrams for the jet function.   Summing over loop order $l$, we find that $1+\J$ is the product of jet functions, 
\bea
1\ +\ \sum_{l=1}^\infty{\cal J}^{(l)} \ &=&\ \prod_{I=1}^a \left( \sum_{l_I=0}^\infty  j_I^{(l_I)} \right)
\nn\\
&=& \prod_{I=1}^a j_I \, ,
\label{eq:1-plus-cal-J}
\eea
In fact, all massless jet functions are equivalent, differing only in multiplicative color factors that depend on the representation of the Wilson line.

Appling Eqs.\ (\ref{eq:t-omega-resolve}) and (\ref{eq:cal-J-defs})  to the right-hand side of the relation (\ref{eq:Lambda-eik-def})  that defines $\Lambda$, we find
\bea
\Lambda \left (\{\tau_I,\beta_I,\bbeta_I\} \right) \ &=& \ \sum_{i=0}^\infty \sum_{l=0}^i \
\sum_{n_{\cal J}=0}^l \sum_{l_0=0}^l \cdots \sum_{l_{n_{\cal J}}=0}^l \delta_{l-l_0,\sum_{k=1}^{n_{\cal J}} l_{k}} \
\left( \delta_{l_0,0} + {\cal J}^{(l_0)}\left (\{\beta_I,\bbeta_I\} \right)\right) 
\nn\\
&\ & \hspace{20mm} \times \left( \delta_{l-l_0,0}\ +\ \prod_{k=1}^{n_{\cal J}} \left( -{\cal J}^{(l_k)} \left (\{\beta_I,\bbeta_I\} \right) \right) \right) \ \lambda^{(i-l)} \left (\{\tau_I\beta_I\} \right)
\nn\\
&=& E\
\sum_{i'=0}^\infty \lambda^{(i')}\left (\{\tau_I\beta_I\} \right)\, ,
\label{eq:Lambda-E}
\eea
where in the first equality $n_{\cal J}$ is the number of nontrivial approximation operators $\hat t_I$ from Eq.\ (\ref{eq:t-omega-resolve})  that act as we sum over all nestings, and $l_0$ is the number of loops in the jet functions of $\hat \sigma_0$. The case when the operator $\hat t_{\sigma_0}=1$  in Eq.~(\ref{eq:R-n-1-1}) is represented by the product of Kronecker delta terms.  In the second equality, we have changed the summation over the total order, $i$ to one over the order of the function $\lambda$ that remains after the factorization of all jet functions, which is $i'=m-l$.  This factorizes the sums over orders for the jets from the remaining diagram.   The complete sum over jet functions is now represented by
\bea
E\  &=& \sum_{l=0}^\infty \sum_{l_0=0}^l \sum_{n_{\cal J}=0}^{l-l_0}\ 
\sum_{l_1=0}^{l-l_0} \cdots \sum_{l_{n_{\cal J}}=0}^{l-l_0} \delta_{l-l_0,\sum_{k=1}^{n_{\cal J}} l_k}\
\left( \delta_{l_0,0} + {\cal J}^{(l_0)}\right) \ \left( \delta_{l-l_0,0}\ +\ \prod_{k=1}^{n_{\cal J}} \left( -{\cal J}^{(l_k)} \right) \right)\, .
\nn\\
\label{eq:E-def}
\eea
Here and below we drop the arguments of the collective jet functions $\J$, leaving their dependence implicit.   
 
\begin{figure}[t]
\centering
\includegraphics[height=4cm]{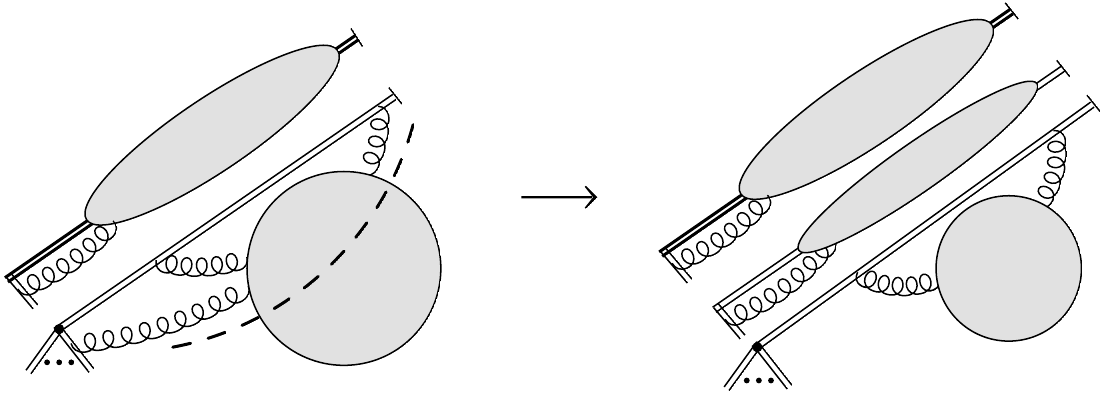}
\caption{The action of a pair of approximation operators in a given nesting. The resulting subsequent jets are eikonal.}
\label{fig:t-w-action}
\end{figure}

 In fact, we easily see that $E=(1+\J)/(1+\J)=1$, where $\J\equiv \sum_{l\ge 1}\J^{(l)}$.  In detail, starting from Eq.\ (\ref{eq:E-def}), the proof is
  \bea
 E\ &=&\ \sum_{l_0=0}^\infty \left( \delta_{l_0,0} + {\cal J}^{(l_0)}\right)\ \left(\, 1\ +\ \sum_{n_\J=1}^\infty\
 \sum_{l_1=0}^\infty \cdots \sum_{l_{n_{\cal J}}=0}^\infty\ \prod_{k=1}^{n_{\cal J}} \left( -{\cal J}^{(l_k)} \right)\
 \sum_{l'=n_\J}^\infty \ \delta_{l',\sum_{k=1}^{n_{\cal J}} l_k}\, \right)
 \nn\\
 &=&\ \sum_{l_0=0}^\infty \left( \delta_{l_0,0} + {\cal J}^{(l_0)}\right)\ \left( 1\ +\ \sum_{n_\J=1}^\infty\ \left( - \sum_{l=0}^\infty \J^{(l)}\right)^{n_\J}\right )
 \nn\\
 &=&\ 1\, .
 \label{eq:E-equals-1}
 \eea
 In the first equality, $l'=l-l_0$ in Eq.\ (\ref{eq:E-def}) is the total order of all the nontrivial jet functions, $\cal J$, and is always greater than or equal to $n_{\cal J}$ by the definitions (\ref{eq:cal-J-defs}).    Equation (\ref{eq:E-equals-1}) is useful because  first, from Eq.\ (\ref{eq:1-plus-cal-J}), $1+\J=\prod_I  j^{\rm cusp}_I$ is the product of jet functions, so that we have shown the factorization of jet functions, and second, $(1+\J)^{-1}\sum_m\lambda^{(m)}$ is precisely the collinear-subtracted multieikonal amplitude when the limit $\tau_I=0,\ I=1,\dots ,a$ is taken.  We illustrate the repeated use of nested approximation operators in Fig.\ \ref{fig:t-w-action}. 
 
   Back in Eq.\ (\ref{eq:full-multieikonal}), we can use Eqs.\ (\ref{eq:1-plus-cal-J}) and (\ref{eq:Lambda-E})  to rewrite the full amplitude as
\bea
\Gamma\ &=&\ \prod_{I=1}^a\ \int d\tau_I\ E\left(\{\tau_I\beta_I\} \right)\
\sum_{i=0}^\infty \lambda^{(i)}\left (\{\tau_I\beta_I\} \right)
\nn\\
&=& \ \prod_{I=1}^a j^{\, \rm cusp}_I\ \int \prod_{I=1}^a\ d\tau_I\ \frac{1}{\prod^a_{I=1} j^{\,\rm cusp}_I}\  
 \lambda\left (\{\tau_I\beta_I\} \right)
 \nn\\
 &\equiv& \prod_{I=1}^a j^{\, \rm cusp}_I\ S\left( \{ \beta_I \} \right)\, ,
\label{eq:full-partonic}
\eea
where $\lambda \equiv \sum_{i}\lambda^{(i)}$, and where   $S$ is a ``soft function'', a matrix in color space that is free of collinear singularities, but which requires renormalization for its purely short-distance UV divergences.     In identifying the products of jet functions in the numerator and denominator, we multiply and divide by the products of normalization constants $c_I$, as in Eq.\ (\ref{eq:J-eik-cusp}) to properly normalize the jet functions.    Here, we have 
again used the independence of the  jet functions $1+\J=\prod j^{\rm cusp}_I$ from the positions of their cusps.

The renormalization of the soft function now follows the standard procedure, as outlined for products of spacelike eikonal lines in Ref.\ \cite{Brandt:1981kf}.    We define ${\cal S}^{(1)}$ as the overall UV divergence of the one-loop soft function, and use the iterative construction 
\bea
{\cal S}^{(n)}\, \big |_{\rm div}\ =\ \sum_{m=1}^n S^{(n-m)}\ {\cal S}^{(m)} \big |_{\rm div}\, ,
\eea 
where, ${\cal S}^{(m)}$ is the $m$th-order soft function after multiplicative renormalization up to $m-1$ loops.   This is possible because the soft function has only local, UV divergences.  The original matrix $S$ can now be renormalized by defining~\cite{Brandt:1981kf} 
\bea
S_{\rm ren}(\{\beta_I\cdot \beta_J\})\ =\ Z^{-1}_S\, S \quad \Rightarrow \quad S_{\rm ren}^{(n)}\ =\ Z^{-1}_S{}^{(n)} + S^{(n)} + \sum_{m=0}^{n-1} S^{(n-m)}\, Z^{-1}_S{}^{(m)}\, .
\label{eq:S-ren}
\eea
The inductive construction of the matrix renormalization constant $Z^{-1}_S$ then follows by choosing $Z^{-1}_S{}^{(m)} =-{\cal S}^{(m)}|_{\rm div}$ starting with $m=1$.

From Eq.\ (\ref{eq:full-partonic}), the full multieikonal amplitude is renormalized by the same matrix $Z^{-1}_S$, and is proportional to a product of jet functions times the renormalized soft matrix, 
\bea
\Gamma_{\rm ren} \ =\ \Gamma\ \times\ Z_S^{-1}\ =\ \prod_I j_I\ S_{\rm ren}\, .
\label{eq:Gamma-ren}
\eea
This relation was the starting point for investigations of color evolution, for example, in Ref.\ \cite{Kidonakis:1998nf}.   In the following subsection, we will apply essentially the same procedures to partonic amplitudes, and will find that the same renormalized soft matrix reappears.

\subsection{Factorization for partonic amplitudes}

 Proceeding as for Wilson-line amplitudes, we have for partonic amplitudes,
\bea
  G^{(n)} \ &=&\     
\sum_{ \gamma^{(n)} }\; \sum_{\sigma_0[\gamma^{(n)}]}\ \sum_{N_{coll}[\sigma_0]}\ \prod_{\omega\in N_{coll}[\sigma_0]} (-t_\omega)\ t_{\sigma_0}\,
\prod_{\sigma \in N_{\sigma_0}[\gamma^{(n)} ]} \big(-t_{\sigma}\big) 
 \, \gamma^{(n)} \ +\ R^{(n)}\, \bar G^{(n)} \, ,
  \label{eq:R-n-1-2}
  \eea
which is the analog of Eq.\ (\ref{eq:lambda-sum-eikonal}),  in terms of partonic diagrams, $\gamma^{(n)}$.   Again, operators $t_\sigma$ are ordered from left to right in increasing size of PS, or equivalently decreasing jet and hard subdiagrams, and PS $\sigma_0$ is the smallest PS with the largest hard subdiagram in the nesting.    As for the multieikonal amplitude, the action of approximation operators $t_{\sigma_0}$ and larger is to factor subdiagrams $\lambda^{(n-m)}$ from an $m$th-order short-distance function ${\cal C}^{(m)}$,
   \bea
  G^{(n)} \  &=& \ 
 \prod_I \int d\eta_I \sum_{ \lambda^{(m)} }\;\prod_{\omega\in N_{coll}[\sigma_0]} (-t_\omega)\ \hat t_{\sigma_0}\  \lambda^{(n-m)}\, \left (\{\tau_I\beta_I \},\{ x_I-\eta_I\bbeta_I  \} \right)\; \times\; {\cal C}^{(m)} \left (\{\tau_I\beta_I \}, \{ \eta_I\bbeta_I \}\right)
  \nn\\
  &\ & \hspace{40mm} +\ R^{(n)} G^{(n)}\, ,
  \label{eq:lambda-sum-partonic}
  \eea
where as in Eq.~(\ref{eq:lambda-sum-eikonal}), $\hat t_{\sigma_0}$ represents the action of $t_{\sigma_0}$ restricted to subdiagram $\lambda^{(n-m)}$.
In this expression, the subdiagrams $\lambda$ now depend on two sets of variables. As in the multieikonal case, the factorized soft subdiagram depends on the longitudinal variables $\tau_I$ along the directions of the $I$th jet.    The factorized jet subdiagrams remain independent of the $\tau_I$, but in the partonic case they depend on distances $\eta_I$ along the complementary directions for each jet, $\eta_I\bbeta_I$.  The $\eta_I$ dependence in partonic jets is due to the variability of the jet functions with the position of the vertex at which the physical parton line attaches the jet to the hard subdiagram (here at PS $\sigma_0$), as in Eq.\ (\ref{eq:co-hard-parton}).   Note that in multieikonal amplitudes this dependence is absent.  In effect the physically polarized parton is given infinite energy and is replaced by a Wilson line, on which $\eta_I=0$ identically.

The  partonic short-distance function ${\cal C}^{(m)}$ in Eq.\ (\ref{eq:lambda-sum-partonic}) is given by
 \bea
 {\cal C}^{(m)}\left (\{\tau_I\beta_I \}, \{ \eta_I\bbeta_I \}\right)\ &=&  \prod_I  \int d^4y_I\, \delta(y_I\cdot \beta_I-\eta_I)\, \delta(y_I\cdot \bbeta_I -\tau_I)
 \nn\\
 &\ & \hspace{10mm} \times\ 
 \sum_{\bar \gamma^{(m)}}\; \sum_{N_P[\bar \gamma^{(m)}] }\ \prod_{\sigma\in N_P[\bar \gamma^{(m)}] }\big(-t_{\sigma}\big) 
 \, \bar \gamma^{(m)}(\{y_I\})\,  \, ,
  \label{eq:cal-H-def-partonic}
  \eea
where $\bar \gamma^{(m)}$ is the set of diagrams of order $m$, with external vertices $y_I$, at which physically polarized partons attach.

   As in the multieikonal case, summing Eq.\ (\ref{eq:lambda-sum-partonic}) over all orders gives a factorized form
\bea
G \ =\ \prod_{I=1}^a\ \int d\eta_I\ \int d\tau_I\ \Lambda^{\rm part} \left (\{\tau_I\beta_I\}, \{ \eta_I\bbeta_I \} \right)\ \times\ {\cal C}\left ( \{\tau_I\beta_I\}, \{ \eta_I\bbeta_I \} \right)\, ,
\label{eq:full-partonic-2}
\eea
with a partonic soft-collinear function given by
\bea
\Lambda^{\rm part} \left (\{\tau_I\beta_I\}, \{ \eta_I\bbeta_I \} \right) \ =
\sum_{i=0}^\infty  \sum_{L^{(i)}} \sum_{N_{coll}[L^{(i)}]} \prod_{\omega\in N_{coll}} (-t_\omega)\ \hat t_{\sigma_0}\ L^{(i)}\left (\{\tau_I\beta_I\}, \{ \eta_I\bbeta_I \} \right)\, ,
\label{eq:Lambda-def-partonic}
\eea
now in terms of partonic diagrams $L^{(i)}$.   The same analysis of the approximation operators in $\hat t_{\sigma_0}$ and the nestings $N_{coll}$ leading to Eq.\ (\ref{eq:Lambda-E}) in the multieikonal case now gives
\bea
\Lambda^{\rm part} \left (\{\tau_I\beta_I\}, \{ \eta_I\bbeta_I \} \right)\
=
 \ \prod_I  
  j_I^{\, \rm part}(\eta_I\bbeta_I)\ \frac{1}{\prod_I j_I^{\, \rm cusp}}\ \sum_m \sum_{\lambda^{(m)}}  \lambda^{(m)}(\{\tau_I\beta_I\})\, .
\label{eq:Lambda-part-def}
\eea
After the factorization of the partonic jet functions (by the operator $t_{\sigma_0}$), the functions $\lambda^{(m)}$ here are again multieikonal diagrams, the same as in Eq.\ (\ref{eq:Lambda-eik-def}).  As in the multieikonal case, the partonic jet functions in the numerator and the eikonal jet functions in the denominator can be normalized by the same constants $c_I$ in Eq.\ (\ref{eq:J-eik-cusp}).    All partonic information has been factorized into overall jet factors by the action of $t_{\sigma_0}$ in Eq.\ (\ref{eq:R-n-1-2}).  These partonic jet functions are given by vacuum expectation values of partonic fields, $\phi$, recoiling against a Wilson line in the conjugate color representation,
\bea
j_I^{\, \rm part [f_\phi]}(x_I,\eta_I\bbeta_I)\
=\ c^{\rm cusp}_I(\beta_I,\bbeta_I)
\bigg \langle 0\left| T\bigg(  \phi(x_I)\, \phi^\dagger(\eta_I\bbeta_I) \Phi_{\bbeta_I}^{[f_\phi]}{}^\dagger(\infty,\eta_I\bbeta_I) \bigg)\right|0  \bigg \rangle\, ,
\label{eq:J-eik-parton}
\eea
where $f_\phi$ is the color representation of parton $\phi$, $\bbeta_I$ is again the complementary lightlike vector defined by $x_I$, and $x_I^2$ serves to regulate collinear singularities in the $\bbeta_I$ direction.   The factorization of Eq.\ (\ref{eq:full-partonic-2}) is illustrated in Fig.\ \ref{fig:full-eikonal}, where now the hard subdiagram is nontrivial. The function ${\cal C}^{(m)}$ is the set of all {\it proper} nested subtractions of the $m$th-order diagrams $\bar \gamma^{(m)}$, which, by Eq.\ (\ref{eq:finite-condition}), cancels all subdivergences.

In Eq.\ (\ref{eq:full-partonic-2}), we can now use Eq.\ (\ref{eq:Lambda-part-def})  to rewrite the full amplitude as
\bea
G\ &=&\    \prod_{I=1}^a\ \int d\eta_I\ j^{\, \rm part}_I(x_I,\eta_I\bbeta_I)\  \int  d\tau_I\  \frac{1}{\prod_I j^{\, \rm cusp}_I}\
 \lambda\left (\{\tau_I\beta_I\}, \{\eta_I\bbeta_I\} \right)\,  {\cal C}\left (\{\tau_I\beta_I\}, \{\eta_I\bbeta_I\} \right)\, ,
\label{eq:G-fact}
\eea
 where  the prefactor is now a product of partonic jet functions, which result from the approximation operator $t_{\sigma_0}$, while the denominator is the same product of eikonal jet functions as in the multieikonal case.

 There is an additional difference between the partonic and multieikonal amplitudes.  For multieikonal amplitudes, the subtraction of subdivergences at each order organizes ultraviolet singularities, which require renormalization, as in Eqs.\ (\ref{eq:S-ren}) and (\ref{eq:Gamma-ren}).   In contrast, before subtractions, the partonic hard-scattering subdiagram is ultraviolet finite after taking into account the counterterms of the gauge-theory Lagrangian.   Correspondingly, at fixed values of the $\eta_I$, the $\tau_I$ integrals of ${\cal C}^{(m)}$ converge, since all collinear and soft regions have been subtracted.
 At the same time, when the $\tau_I$ are much smaller than these scales, the eikonal diagrams $\lambda$ and $\cal C$  in Eq.~(\ref{eq:lambda-sum-partonic}) 
both develop ultraviolet singularities as a result of the subtractions,  which must cancel, since they result from adding and subtracting singular behavior.   This pattern is familiar from momentum-space factorizations \cite{Sen:1982bt,Kidonakis:1998nf}.   These singularities, however, are  removed from the soft matrix $S$ by the multiplicative renormalization of Eq.\ (\ref{eq:S-ren}).    We can therefore regularize both the soft and hard subdiagrams  by introducing $Z_S^{-1}Z_S$ between 
$\lambda$ and $\cal C$ in Eq.\ (\ref{eq:G-fact}).   Once this is done, the soft subdiagram $\lambda$ becomes independent of the $\tau_I$ for $\tau_I\rightarrow 0$, and we can treat it as a constant, while integrating the hard subdiagram over the $\tau_I$ at fixed $\eta_I$.   The result is now the final coordinate-space factorized form,
 \bea
G\ &=&\  \prod_{I=1}^a\ \int d\eta_I\  j^{\rm part}_I(x_I,\eta_I\bbeta_I)\  
S_{\rm ren}(\{\beta_I\cdot \beta_J\})\ \  {\cal H}\left (\{\eta_I\bbeta_I\} \right)\, ,
\label{eq:G-fact-ren}
\eea
with $S_{\rm ren}$ the same function as in Eq.\ (\ref{eq:S-ren}) for the multieikonal amplitudes, and with a short-distance coefficient function given by
\bea
{\cal H}\left (\{\eta_I\bbeta_I\} \right)\ =\ Z_S\ \prod_I \int_0 d\tau_I\ {\cal C}\left (\{\tau_I\beta_I\}, \{\eta_I\bbeta_I\} \right)\, .
\label{eq:calC-def}
\eea
Taken together, the Fourier transforms of Eqs.\ (\ref{eq:G-fact-ren}) and (\ref{eq:calC-def}) specify factorized amplitudes in momentum space \cite{Sen:1982bt,Sterman:2002qn}.

\section{Webs and Regularization}
\label{eqwbsub}

In this section we give a detailed treatment of the simplest of the eikonal amplitudes, the ``cusp'', defined by Eq.~(\ref{eq:J-eik-cusp}), with a gauge-singlet vertex.  Our goal is to relate the regularization procedure developed in Sec.\ \ref{co-subt}, where we exhibited an expression for the cusp and other amplitudes in terms of nested approximation operators [Eq.\ (\ref{eq:R-n-1-0})] to the exponentiation properties of the cusp.   We first recall the graphical interpretation of exponentiation.

\subsection{Cusp webs and exponentiation}

All multieikonal amplitudes, of the type of Eqs.\ (\ref{eq:wivertex}) and (\ref{eq:J-eik-cusp}) may  conveniently be written as
exponentials
\bea
\Gamma = \exp\ W\, ,
\label{webexp1}
\eea
where $W$ is determined by a set of  rules that define the so-called web diagrams, which were first identified and analyzed for the special case of the cusp matrix element (\ref{eq:J-eik-cusp}).  In all cases, the exponent $W$ is a sum of eikonal diagrams with modified color factors.   For the special case of the cusp, these diagrams, which we label by $w$, are  irreducible under cuts of the two Wilson lines \cite{Sterman:1981jc,Gatheral:1983cz,Frenkel:1984pz} (thus the name, ``webs''). Webs can be used to show the exponentiation of double logarithms and double poles, and of power corrections related to singularities in the perturbative running coupling~\cite{Laenen:2000ij,Berger:2003zh,Magnea:1990zb,Sterman:2002qn,Dixon:2008gr}\@. They help organize calculations at two loops and beyond in the cusp and in closed Wilson loops~\cite{Erdogan:2011yc,Korchemsky:1987wg,Drummond:2007cf}\@.  The concept of webs can  be generalized beyond the color-singlet cusp and can also serve as a starting point for a beyond-eikonal expansion \cite{Mitov:2010rp,Laenen:2010uz,Gardi:2010rn,Gardi:2011yz}.

For the cusp, the exponent can be represented as
\bea
W\ &=&\ \sum_{{\rm webs}\ w} \bar C\left(w\right)\, {\cal I}\left(w\right)
\, ,
\label{webexp2}
\eea
where ${\cal I}(w)$ is 
the corresponding diagrammatic integral over the positions of  internal vertices of web $w$.   Each web integral is multiplied by a color factor $\bar C(w)$, modified relative to the factor $C(w)$ that would normally be associated with diagram $w$.   It is possible to give a closed form  for $\bar C(w)$ \cite{Gardi:2010rn}, but in the following discussion, we will use  the recursive definition \cite{Gatheral:1983cz}, given for each diagram by 
\be
\bar C\left(w^{(n)}\right) \ =\   C\left(w^{(n)}\right)\ -\ 
\sum_{d\in D} \; \prod_{w_i^{(n_i)}\in d}\ \bar C\left(w^{(n_i)}_i\right)\ ,
\label{cbar}
\ee
where the $w^{(n_i)}_i$ are lower-order webs, of order $n_i$, in the decompositions $d$ of the original diagrams $w^{(n)}$ into lower-order webs,  with $\sum_i n_i=n$\@.  As usual, we denote the coefficient of $(\as/\pi)^n$ in $W$ as $W^{(n)}$, and similarly for all other functions.

The sum in Eq.\ (\ref{cbar}) is over all ``proper'' web decompositions $D[w^{(n)}]$, 
not including $w^{(n)}$ itself, and the right-hand side
vanishes identically for diagrams $\gamma^{(n)}$ that are not webs, for which we
have \cite{Gatheral:1983cz,Frenkel:1984pz}
\be 
\sum_{D[\gamma^{(n)}]}\ \prod_{w_i\in D[\gamma^{(n)}]}\bar C(w_{i})
\ =\ 
C(\gamma^{(n)})\, .
\label{eq:non-web-C}
\ee
As a result, the $n$th-order contribution, $W^{(n)}$, to the sum of {\it all} diagrams 
that contribute to the cusp at the same order can be written as
\bea
W^{(n)} = \sum_{\gamma^{(n)}} 
\left(\ \gamma^{(n)}\ -\ {\cal I}(\gamma^{(n)})\; \sum_{D[\gamma^{(n)}]}\ 
\prod_{w\in D[\gamma^{(n)}]} \bar C(w)\ \right)\, .
\label{splitsum}
\eea
We will use this form below. From now on, all diagrams are eikonal, and we drop subscripts to identify this. The web prescription for $W$, the logarithm of the cusp,  was originally identified in momentum space~\cite{Gatheral:1983cz,Frenkel:1984pz}, but also has a very simple coordinate-space derivation~\cite{Mitov:2010rp}\@.

Web diagrams for cusps with massive eikonals have only a single, overall ultraviolet (and infrared) divergence \cite{Brandt:1981kf}, up to multiple poles associated with the running of the coupling.  In the massless limit, they develop a double pole times the cusp anomalous dimension, again with higher-order poles that can be predicted by the running of the coupling order by order~\cite{Berger:2003zh,Korchemsky:1987wg,Dixon:2008gr}\@.  The treatment  of vanishing mass in the cusp was developed in Ref.~\cite{Korchemsky:1987wg} in momentum space, employing physical gauges.   

We now study the fully massless case in Feynman gauge.   Each diagram $w$ in Eqs.\ (\ref{webexp2}) or (\ref{splitsum}) can be written as an integral
over its ``leading'' vertices, that is, vertices at the furthest distances from the cusp vertex along each Wilson line~\cite{Erdogan:2011yc}, 
\bea
W\ &=&\ \int _0 \frac{d\tau d\bar{\tau}}{\tau\, \bar{\tau}} \ f_{W}(\alpha_s(\mu^2),\mu^2\tau\bar{\tau},\vep)\, ,
\label{leading}
\eea
where in the absence of masses, the dependence of the integrand reduces to just a few variables. Standard perturbative renormalization introduces dependence on the renormalization scale $\mu^2$ 
as the positions of vertices are integrated over at fixed $\tau$ and $\bar\tau$. On a diagram-by-diagram basis
these integrals  have
many nonlocal subdivergences, involving jet and hard subdiagrams, which show up as logarithmic enhancements,
as analyzed in Ref.\ \cite{Erdogan:2013bga}. 
We may think of these integrals as cut off at some large length scale to avoid explicit infrared singularities.
The resulting integrand $f_{W}$ is a renormalization-scale-independent function that is the result of all the
remaining integrals, as in the two-loop example treated 
in detail in Ref.\ \cite{Erdogan:2011yc}.     
For the sum of web diagrams we can thus write
\bea
W \ &=&\ 
\int_0 \frac{d\tau d\bar{\tau}}{\tau\, \bar{\tau}} \ f_{W}(\alpha_s(1/\tau\bar{\tau}),1,\vep)
\, .
\label{eq:web-int}
\eea
  We will refer to the sum over webs at fixed $\tau$ and $\bar{\tau}$ as the ``web   integrand'', and
we will show that after a sum over all diagrams, the full web integrand $f_W$ is ultraviolet  finite for $\vep\rightarrow 0$. 
Renormalization for the web functions is then manifestly additive, and associated with the singular $\tau$ and/or $\bar\tau\rightarrow 0$ limits of the integral.   The connection between multiplicative renormalizability and the structure of web functions has been reviewed recently for both color-singlet cusps and multieikonal 
vertices in Ref.\ \cite{Gardi:2011yz}.      In Sec.\ \ref{sec:multi-eik}, we will use the exponentiation in terms of webs to revisit factorization for multieikonal amplitudes, and  discuss subdivergences in web integrands for these cases.   First, however, we  discuss the web construction for the cusp in its own terms.   Although the demonstration below of finiteness for  the cusp function is in some ways more elaborate than the general discussion of Sec.\ \ref{sec:multi-eik}, it is more explicit, and gives insight into the manner in which perturbative corrections conspire at each order to produce ultraviolet finiteness.

\subsection{Subtractions, webs and decompositions}
\label{softward12}

Consider the $n$-loop web, $W^{(n)}$ given in Eq. (\ref{splitsum}).
On the right-hand side of this equation, we replace the simple sum over diagrams by the sum over all their nested proper subtractions, as in Eq.\ (\ref{eq:Gamma-sum}),
\bea
W^{(n)}\ \ =\ -\ \sum_{\gamma^{(n)}}\sum_{N_P\in {\cal N}_P[\gamma^{(n)}]}
  \prod_{\rho\in N_P} \big(-t_{\rho}\big)\, \gamma^{(n)}\ 
  -\
    \sum_{\gamma^{(n)}} \F(\gamma^{(n)})\
\left(\  \sum_{D[\gamma^{(n)}]}\ 
\prod_{w\in D[\gamma^{(n)}]} \bar C(w)
\ \right) \, .
\label{eq:Delta-W-def}
\eea
The right-hand side is now the difference between the sum of the proper subtractions for $n$th-order diagrams (equal to the diagrams themselves) and the subtractions in Eq.\ (\ref{splitsum}) that define the webs, also summed over all diagrams.  In the following, we will use this form to show that  in every leading region $\rho$ involving a subdivergence, $W^{(n)}$ is integrable.   This in turn implies that the $n$th-order web, Eq.\ (\ref{splitsum}), is itself integrable over all subspaces where subdiagrams are ultraviolet singular. Ultraviolet divergences can arise only when all the vertices of the web approach the origin or the light cone together, which implies the finiteness of the web integrand $f_W$ in Eq.\ (\ref{eq:web-int}).    

Let us  thus consider $W^{(n)}$ in the form (\ref{eq:Delta-W-def}), restricted to the reduced neighborhood $\hat n[\rho]$ of PS $\rho$, which we denote by $ W^{(n)}_\rho$. As we have seen in Eq.~(\ref{eq:soft-approx-diff}) and the subsequent discussion, in each region $\rho$ the ultraviolet behavior of the vertex is well approximated by the single subtraction term, $t_\rho \Gamma^{(n)}$, while all other nestings cancel.   Then, up to nonsingular corrections, when restricted to the neighborhood of $\rho$, Eq.\ (\ref{eq:Delta-W-def}) becomes
\be
 W^{(n)}_{\rho}  = -\sum_{\gamma^{(n)}} (-t_{\rho})\gamma_\rho^{(n)}\ -\ \sum_{\gamma^{(n)}} \F(\gamma_{\rho}^{(n)})\sum_{D[\gamma^{(n)}]} 
\prod_{w\in D[\gamma^{(n)}]} \bar C(w) \,  ,
 \label{web-end}
\ee
where here and below we restrict ourselves to divergent contributions.
We will now argue that in region $\rho$  the first sum on the right-hand side cancels against those web decompositions 
$(D[\gamma^{(n)}])$ in the second sum that ``match" the structure of the leading region $\rho$, and that other, ``unmatched" contributions to the sum either cancel or are suppressed in region $\rho$.    We begin our argument by recalling the action of Ward identities in the first term, as described in Sec.~\ref{subsec:ward-identities}.   In this discussion, the integration region is indicated by a subscript and the perturbative order by a superscript.

For definiteness, we assume that there is a nontrivial soft subdiagram at PS $\rho$, which  we now denote by $S_\rho$, as in the factorized form (\ref{eq:rho-fact}),
\bea
t_\rho \sum_{\gamma^{(n)}} \gamma_\rho^{(n)} & =& \  S^{(n-m_{\rho})}_{\rho}\times
R^{(m_{\rho})}_{\rho}
\nonumber\\
& =& \ \sum_{s_\rho \in S_\rho} s_\rho^{(n-m_{\rho})} \times
 \sum_{r_\rho \in R_\rho} r_\rho^{(m_\rho)}\, .
   \label{eq:t-rho-fact}
\eea
In this rewriting of Eq.\ (\ref{eq:rho-fact}), the soft function $S_{\rho}$ [$S^{(\rho)}$ in Eq.\ (\ref{eq:lambda_def})] multiplies a ``remainder" function, $R_{\rho}$, which (to avoid clutter) includes sums over the jet and hard subdiagrams at PS $\rho$.   
The function $S_{\rho}=\sum s_\rho$ is the sum of the soft subdiagrams, $s_\rho$, of each  $\gamma^{(n)}$ in region $\rho$, connected directly to $\beta$ 
and $\bbeta$ Wilson lines, and similarly for the remainder subdiagram $R_\rho=\sum r_\rho$.
We let $m_\rho$ be the order of the remainder function in region $\rho$.  
In summary, in each leading region $\rho$, after a sum over all
$\gamma^{(n)}$, Ward identities factorize the subdiagrams that make up
$S_{\rho}$ and $R_{\rho}$.   The sum over all $\gamma^{(n)}$ in region
$\rho$ can then be replaced by independent sums over soft subdiagrams
$s_\rho$ and remainder subdiagrams $r_\rho$, as in Eq.~(\ref{eq:t-rho-fact}).   

Next, we separate color and coordinate factors of each $s^{(n-m_{\rho})}_{\rho}$ and $r^{(m_{\rho})}_{\rho}$ in Eq.\ (\ref{eq:t-rho-fact}), 
\bea
t_\rho \sum_{\gamma^{(n)}} \gamma_\rho^{(n)}\ &=&  S^{(n-m_{\rho})}_{\rho}\times R^{(m_{\rho})}_{\rho}
\nonumber\\
&=&\ \sum_{s_{\rho}\in
  S_{\rho}} C\left(s^{(n-m_{\rho})}_{\rho}\right) \F\left(s^{(n-m_{\rho})}_{\rho}\right) 
  \times
 \sum_{r_{\rho}\in R_{\rho}} C\left(r^{(m_{\rho})}_{\rho}\right) \F\left(r^{(m_{\rho})}_{\rho}\right)\, .
\label{gfactorized}
\eea
This is the form that we will compare to the sum of web subtractions, the second sum in Eq.\ (\ref{eq:Delta-W-def}), which becomes
\bea
 W^{(n)}_{\rho}  &=&  \sum_{s_{\rho}\in 
  S_{\rho}} C\left(s^{(n-m_{\rho})}_{\rho}\right) \F\left(s^{(n-m_{\rho})}_{\rho}\right) 
 \sum_{r_{\rho}\in R_{\rho}} C\left(r^{(m_{\rho})}_{\rho}\right) \F\left(r^{(m_{\rho})}_{\rho}\right)
 -\ \sum_{\gamma^{(n)}} \F(\gamma_{\rho}^{(n)})\sum_{D[\gamma^{(n)}]} 
\prod_{w\in D[\gamma^{(n)}]} \bar C(w) \,  .
\nonumber\\
 \label{web-end-2}
\eea
As mentioned below Eq.\ (\ref{web-end}), it is useful to split the set of decompositions, $D[\gamma^{(n)}]$ into the set of those that match the factorization of soft and remainder functions in the first term of this expression, and those that do not.  More specifically, matched decompositions of a diagram $\gamma^{(n)}$  are those in which no web contains lines in {\it both}  the soft subdiagram $s_\rho[\gamma^{(n)}]$, and the remainder subdiagram, $r_\rho[\gamma^{(n)}]$.    Correspondingly, in unmatched decompositions, at least one web contains lines of both the soft subdiagram and the remainder in region $\rho$.   In these terms, every decomposition of diagram $\gamma^{(n)}$ is either matched or unmatched in region $\rho$.
  We represent this division of decompositions for the second term on the right-hand side of Eq. (\ref{web-end-2})  as
\bea
\sum_{ \gamma^{(n)}}\F_{\rho}(\gamma^{(n)})\sum_{D[\gamma^{(n)}]} 
\ \prod_{w\in D[\gamma^{(n)}]} \bar C(w) &\ &
\nonumber\\
 & \ & \hspace{-30mm} = \ \sum_{ \gamma^{(n)}} \F_{\rho}(\gamma^{(n)})  \left( \
 \sum_{D_{S_{\rho}\otimes R_{\rho}}[\gamma^{(n)}]}\ \prod_{w\in D_{S_{\rho}\otimes R_{\rho}}} \bar C(w) \ + 
  \  \sum_{D_{S_{\rho}\cap R_{\rho}}[\gamma^{(n)}]}\  \prod_{w\in D_{S_{\rho}\cap R_{\rho}}} \bar C(w)  \right)
  \nonumber\\
\nonumber\\
 & \ & \hspace{-30mm} \equiv \
w^{(n)}_{\rho}[S_{\rho}\otimes R_{\rho}]\ +\
w^{(n)}_{\rho}[S_{\rho}\cap R_{\rho}]\ , 
\label{eq:web-sum-separate}
\eea
where the first term on the right of the second equality  represents the sum over the set of matched decompositions,  $D_{S_{\rho}\otimes R_{\rho}}$ and the second is the sum over unmatched decompositions, $D_{S_\rho\cap R_\rho}$.

In the following, we will show that the matched decompositions cancel the factorized subtraction terms of Eqs.\ (\ref{gfactorized}) and (\ref{web-end-2}) in region $\rho$,
\bea
0\ &=&\ t_\rho \sum_{\gamma^{(n)}} \gamma_\rho^{(n)}\ -\ w^{(n)}_{\rho}[S_{\rho}\otimes R_{\rho}] 
\nonumber\\
\nonumber\\
&=& S^{(n-m_{\rho})}_{\rho}\times R^{(m_{\rho})}_{\rho}\ -\ w^{(n)}_{\rho}[S_{\rho}\otimes R_{\rho}] 
\nonumber\\
\nonumber\\
&=&\ \sum_{s_{\rho}\in
  S_{\rho}} C\left(s^{(n-m_{\rho})}_{\rho}\right) \F\left(s^{(n-m_{\rho})}_{\rho}\right) 
 \sum_{r_{\rho}\in R_{\rho}} C\left(r^{(m_{\rho})}_{\rho}\right) \F\left(r^{(m_{\rho})}_{\rho}\right)\ -\ w^{(n)}_{\rho}[S_{\rho}\otimes R_{\rho}] \, ,
\label{matchingD}
\eea
while the unmatched decompositions are suppressed,
\be 
w^{(n)}_{\rho}[S_{\rho}\cap R_{\rho}]\ =\ 0 \, .
\label{unmatchedD}
\ee
Substituted into Eq.\ (\ref{web-end-2}), these two results show that $W^{(n)}_\rho =0$, so that the web integrand is free of ultraviolet subdivergences.

Before giving our arguments for the results (\ref{matchingD}) and (\ref{unmatchedD}), we recall that we have assumed that the leading region $\rho$ has a nontrivial soft subdiagram, $S_\rho^{(n-m_\rho)}$.   For the
special case of a leading region with no soft subdiagram ($m_\rho =n$), and only jet and hard subdiagrams, we may pick either of the jet subdiagrams to take the place of
$S_\rho^{(n-m_\rho)}$, with the same result.    In the following, we shall suppress the orders of $S_\rho^{(n-m_\rho)}$ and $R_\rho^{(m_\rho)}$, since these are in principle fixed by the choice of region $\rho$.

\subsection{Matched decompositions}

It is clear that the sum over matched decompositions of Eq.~(\ref{eq:web-sum-separate}),  \mbox{$D_{S_{\rho}\otimes R_{\rho}}[\gamma^{(n)}]$} for each diagram $\gamma^{(n)}$ separates into two independent sums over the web decompositions of the soft and remainder subdiagrams of $\gamma^{(n)}$.  Among these decompositions are the  choices $s_{\rho}[\gamma^{(n)}]$  and $r_{\rho}[\gamma^{(n)}]$, the soft and remainder subdiagrams themselves, which appear along with all of the webs made of their decompositions.  Using the general form for webs, Eq.\ (\ref{splitsum}), we can thus separate the color factors associated with the soft and the remainder subdiagrams,
\bea
w^{(n)}_{\rho}[S_{\rho}\otimes R_{\rho}]\ &=& \
  \sum_{\gamma^{(n)}} \F_{\rho}(\gamma^{(n)})  \
  \left(\; \bar C\left(s_\rho[\gamma^{(n)}]\right) \ +\ \sum_{D\left[ s_{\rho}[\gamma^{(n)}] \right ]}
\ \prod_{d\in D\left[ s_{\rho}[\gamma^{(n)}] \right ]}\!\!\! \bar C(d)\! \right)
\nonumber\\
&\ &  \hspace{25mm} \times\ \left(\,\; \bar C\left(r_{\rho}[\gamma^{(n)}]\right)\ +\ 
\sum_{D\left[ r_{\rho}[\gamma^{(n)}] \right ]}
\prod_{d'\in  D\left[ r_{\rho}[\gamma^{(n)}] \right ]}\!\!\! \bar C(d')\ \right)
\nonumber\\
&=&  \sum_{\gamma^{(n)}} \F_{\rho}(\gamma^{(n)}) \  C\left(s_\rho[\gamma^{(n)}]\right)\  C\left(r_{\rho}[\gamma^{(n)}]\right)\, ,
  \label{eq:DC-sum}
  \eea 
  where  in the second equality we have used Eq.\ (\ref{cbar}) for web-color factors.   In effect, after the sum over matched decompositions,  the web-color factors of the soft and remainder functions revert to their normal form, the same as in the subtraction terms of Eq.\ (\ref{web-end-2}), that is, the first term on the right-hand-side of that equation.    As usual, the sum over  $D[g]$ of diagram $g$ refers only to its proper web decompositions.   Note that the color identity in Eq.~(\ref{eq:DC-sum}) extends to all diagrams, $g$.  For a nonweb~$g'$, for which $\bar C(g')=0$, we recall Eq.\ (\ref{eq:non-web-C}).
  
   Having factorized the product of color factors in the sum over matched decompositions, we now turn to the coordinate integrals.      We reexpress the sum over diagrams $\gamma^{(n)}$ in Eq.\ (\ref{eq:DC-sum}) as independent sums over soft and remainder subdiagrams $s_\rho$ and $r_\rho$, and then a sum over all possible connections of these subdiagrams to the eikonal lines, respecting relative orderings ${\cal O}(s_\rho,r_\rho)$ along the Wilson lines of all the vertices that connect gauge lines from subdiagram $s_\rho$ and from subdiagram $r_\rho$ to the Wilson lines,
   \bea
    \sum_{\gamma^{(n)}} \F_{\rho}(\gamma^{(n)})\ =\ \sum_{s_{\rho}\in S_{\rho}}\sum_{r_{\rho}\in  R_{\rho}}\ \sum_{\substack{{\rm eikonal}\\{\rm orderings\ O}}}
    \ \F_{\rho}(O[s_\rho,r_\rho])\, .
    \label{eq:split-gamma-sum}
    \eea
In Eq.~(\ref{eq:DC-sum}), this gives
\bea
w^{(n)}_{\rho}[S_{\rho}\otimes R_{\rho}]\  &=& \
\sum_{s_{\rho}\in S_{\rho}}\sum_{r_{\rho}\in  R_{\rho}}\ \sum_{\substack{{\rm eikonal}\\{\rm orderings\ O}}} \
\F (O[s_\rho,r_\rho])\ \  C\left(s_\rho\right)\  C\left(r_{\rho}\right)\, .
\label{eq:w-n-order}
\eea
To this result we apply the coordinate-space eikonal identity  \cite{Mitov:2010rp}, applicable whenever we sum over all connections of a set of web subdiagrams 
that are attached to the eikonal lines, respecting the order of gauge lines within each subdiagram,
 \bea
\sum_{\substack{{\rm eikonal}\\{\rm orderings\ O}}}\ \F(O[s_\rho,r_\rho,\dots ])\ =\  \F({s_\rho})\, \times \F({r_\rho}) \times \cdots \, ,
\label{eq:eik-iden}
 \eea
 a ``shuffle algebra'' identity that generalizes to any numbers of subdiagrams and any number of eikonal lines.   In Eq.~(\ref{eq:w-n-order}), this gives the desired result,
 \bea
 w^{(n)}_{\rho}[S_{\rho}\otimes R_{\rho}]\  &=& \
\sum_{s_{\rho}\in S_{\rho}}\sum_{r_{\rho}\in  R_{\rho}}\ 
\F (s_\rho)\ C\left(s_\rho\right)\ \F (r_\rho)\ \   C\left(r_{\rho}\right)
\nn\\
&=& S_\rho \times R_\rho\, ,
\label{eq:w-n-resolved}
\eea
which shows that Eq.~(\ref{matchingD}) holds for the matched decompositions, that is, that the matched decompositions cancel the subtractions in region $\rho$.

\subsection{Unmatched decompositions}

We now treat the unmatched decompositions of Eq.\ (\ref{eq:web-sum-separate}), whose sum we have denoted by $w^{(n)}_{\rho}[S_{\rho}\cap R_{\rho}]$\@. 
For any diagram $\gamma^{(n)}$, this sum consists of decompositions with at least one
web that includes one or more lines in the soft subdiagram $s_\rho[\gamma^{(n)}]$ and one or more lines in $r_\rho[\gamma^{(n)}]$.  For this discussion, we assume that 
the cancellation of subdivergences has been proven to order  $n-1$.

From Eq.\ (\ref{eq:web-sum-separate}), we have for the unmatched decompositions 
\be 
w^{(n)}_{\rho}[S_{\rho}\cap R_{\rho}]\ = \ \sum_{
  \gamma^{(n)}}\F_{\rho}(\gamma^{(n)})\sum_{D_{S_{\rho}\cap R_{\rho}}[\gamma^{(n)}]} \ 
\prod_{w\in D_{S_{\rho}\cap R_{\rho}}[\gamma^{(n)}]}\!\! \bar C(w) \ .
 \ee 
By analogy to our analysis of the matched distributions, we will exchange the sum over diagrams $\gamma^{(n)}$ for sums  over webs.
In every element of the unmatched decompositions $D[\gamma^{(n)}]\in  \{S_{\rho}\cap R_{\rho}\}$ of diagram $\gamma^{(n)}$ there is a nonempty decomposition that includes a subdiagram $u_\rho[\gamma^{(n)}]$ consisting of (one or more) webs, each of which is not all in the soft subdiagram, and not all in the remainder of $\gamma^{(n)}$.   In general, once subdiagram $u_\rho[\gamma^{(n)}]$ is fixed, there is also a subdiagram, $s'_\rho[\gamma^{(n)}]$ whose webs are fully subdiagrams of  $S'_\rho[\gamma^{(n)} \backslash u_\rho]$, the soft subdiagram found by removing the unmatched webs of $u_\rho$ from $\gamma^{(n)}$, and another subdiagram, $r'_\rho[\gamma^{(n)}]$, which is fully a subdiagram of the remainder $R'_\rho[\gamma^{(n)}\backslash u_\rho]$.   We can then write for any such decomposition,
\bea
\gamma_\rho^{(n)}\ \rightarrow\ s'_\rho[\gamma^{(n)}] \cup r'_\rho[\gamma^{(n)}] \cup u_\rho[\gamma^{(n)}]\, .
\eea
The sum over such unmatched web decompositions of $\gamma^{(n)}$, then, can be reorganized as a sum over the independent decompositions of each of these subdiagrams.   For decompositions of the soft and remainder subdiagrams, $s'_\rho$ and $r'_\rho$, the diagrams themselves appear in these sums, along with  all of their decompositions.  For each unmatched subdiagram, $u_\rho$, however,  only those decompositions are included that leave $u_\rho[\gamma^{(n)}]$ fully unmatched. For each choice of $u_\rho$, we can sum over all allowed $s'_\rho$ and $r'_\rho$, and using the color and eikonal identities, derive the analog of Eq.\ (\ref{eq:w-n-resolved}),
\bea
w^{(n)}_{\rho}[S_{\rho}\cap R_{\rho}]\ &=& \ \sum_{m_s,m_r}\ \sum_{s'{}_\rho^{(m_s)}} \sum_{r'{}_\rho^{(m_r)}}\ \F (s'{}^{(m_s)}_\rho)\ C\left(s'{}^{(m_s)}_\rho\right)\ \F (r'{}^{(m_r)}_\rho)\    C\left(r'{}^{(m_r)}_{\rho}\right)
\nn\\
&\ & \hspace{8mm} \times\  
\ \left( \sum_{u_\rho} \F(u_\rho)\ \left( \bar C(u_\rho)\ +\ \sum_{D_{\rm un}[u_\rho]} \prod_{d\in D_{\rm un}[u_\rho]}\ \bar C(d) \right) \right)^{(n-m_s-m_r)}\, ,
\label{eq:cap-sum-1}
\eea
where we sum over the orders of the soft and remainder diagrams.   In the final sum over diagrams $u_\rho$, we group all fully unmatched decompositions of the unmatched  webs $u_\rho$ of order $n-m_s-m_r$.    The coordinate factors of all these terms are the same.    Their color factors, however, get contributions only from a subset $D_{\rm un}[u_\rho]$ of all decompositions, $D_{\rm un}[u_\rho] \subset  D_{S_{\rho}\cap R_{\rho}}$, i.e., those that are fully unmatched.   We now consider the difference, $ D[u_\rho]\backslash D_{\rm un}[u_\rho] $, between this set and the full set of decompositions of each $u_\rho$ .

The set of missing decompositions, $D[u_\rho]\backslash D_{\rm un}[u_\rho]$, for a given $u_\rho$ includes those  that have matched soft and remainder subdiagrams, which we denote by $w_\rho \left [S[u_\rho]\otimes R[u_\rho] \right ] $, where $S[u_\rho]$ is the soft subdiagram of $u_\rho$, and  $R[u_\rho]$ the corresponding remainder.   The set $ D[u_\rho]\backslash D_{\rm un}[u_\rho] $ also includes many more decompositions, i.e., those that have decompositions involving some matched and some unmatched webs of lower order. 
The inductive hypothesis, however, assumes Eq.\ (\ref{unmatchedD}) for lower orders, so the sums over unmatched decompositions of lower order cancel among themselves.    Therefore, by adding and subtracting matched decompositions  $w_\rho \left [S[u_\rho]\otimes R[u_\rho] \right ]$ only, we can derive a factor that consists of the difference between all decompositions of $u_\rho$ and its matched decompositions,
\bea
\sum_{u_\rho} \F(u_\rho)\  \sum_{D_{\rm un}[u_\rho]} \prod_{d\in D_{\rm un}[u_\rho]}\ \bar C(d)
\ =\
\sum_{u_\rho} \F(u_\rho)\ \sum_{D[u_\rho]} \prod_{d\in D[u_\rho]}\ \bar C(d)\ -\ 
w_\rho \left [S[u_\rho]\otimes R[u_\rho] \right ] \, .
\label{eq:u-unmatched-matched}
\eea
Substituting this into Eq.\ (\ref{eq:cap-sum-1}), we now have the full color factor for each diagram $u_\rho$ in the sum, and we can use the web-color identity (\ref{cbar}) to confirm that the sum of unmatched decompositions vanishes,
\bea
w^{(n)}_{\rho}[S'_{\rho}\cap R'_{\rho}]\ &=& \   \sum_{m_s,m_r}\
\left( \sum_{s'_{\rho}} s'_\rho \right)^{(m_s)}\
\left(\sum_{r'_{\rho}} r'_\rho \right)^{(m_r)}
\left(\sum_{u_{\rho}} \Big \{ u_\rho\ -\ w_\rho \left [S[u_\rho ]\otimes R[u_\rho] \right]  \Big \} \right)^{(n-m_s-m_r)}
\nn\\
&=& \ 0\, .
\eea
Again, the first factors on the right-hand side are factorized soft (order-$m_s$) and remainder (order-$m_r$) subdiagrams, while the third factor is now a sum of all subdiagrams of order $n-m_s-m_r$.  The third factor vanishes by Eq.~(\ref{matchingD}), which states that all subdivergences cancel against those in the sum of matched decompositions, up to order $n$.  Thus, all unmatched decompositions cancel in region $\rho$ to order $n$, and we confirm Eq.\ (\ref{unmatchedD}) and hence the absence of subdivergences [Eq.\  (\ref{matchingD})] in the logarithm of the cusp amplitude \cite{Erdogan:2011yc}.   As discussed above, this result confirms the UV finiteness of the web integrand, $f_W$ in Eq.\ (\ref{eq:web-int}).

\section{Multieikonal  Amplitudes}
\label{sec:multi-eik}

The arguments of the previous section apply specifically to the cusp, where we 
have used the inductive construction of web-color factors, Eq.\ (\ref{cbar}).
We go on now to study how these considerations change for amplitudes with multiple Wilson lines connected at a local vertex, and to explore the relationship of their exponentiation properties to the factorization demonstrated in Sec.\ \ref{sec:renorm-fact}.

\subsection{Cancellation of web subdivergences for large $N_c$}

For a multieikonal vertex, $\Gamma_a$ with $a>3$ Wilson lines, and a consequent mixing of color tensors \cite{Brandt:1981kf}, it will be useful to use an alternative expression for webs, introduced in Ref.\  \cite{Mitov:2010rp}.  We label each web function with an index  $E$, which represents a list of the numbers of gauge lines attached to each Wilson line, $E\equiv\{e_1\dots e_a\}$ for $a$ Wilson lines.   We then express the sum of all webs with the same index $E$, $w_E^{(i)}$, as an integral $\F_E$ of integrand  $\W_E^{(i)}$ , 
\bea
w^{(i)}_{E}
&=&
\prod_{\alpha=1}^a\, 
\prod_{j=1}^{e_\alpha}\int_{\tau^{(a)}_{j-1}}^\infty d\tau^{(\alpha)}_j\,
\W_{E}^{(i)}\left(\{\tau^{(\alpha)}_j\}\right)
\nonumber\\
\ &\equiv&\ \F_{E}\, [\W_{E}^{(i)}]\, ,
\label{eq:calFcalW}
\eea
where the $\tau^{(\alpha)}_j$ label the locations of the vertices coupling gauge lines to 
Wilson line  $\alpha$, ordered as $\tau_1^{(\alpha)}\le \tau_2^{(\alpha)} \le \cdots \le \tau_{e_\alpha}^{(\alpha)}$.    The functions
$\W_{E}^{(i)}$ represent sums over all diagrams with the specified numbers of
eikonal connections, and are symmetric under exchange, including color, of the gauge lines attached at each
vertex $\tau_j^{(\alpha)}$.   Summing over connections, $E$, we find the complete web, $ W_a^{(i)}$ 
as a sum of the $w^{(i)}_E$, and the amplitude is given by
\bea
\Gamma_a\ &=&\ \exp \left[ \sum_i W_a^{(i)} \right]
\nn\\
&=& \ \exp \left[ \sum_i\ \sum_E \F_E\; [\W_E^{(i)}] \right]\, .
\label{eq:web-calW}
\eea
In these terms, we can write an iterative expression for the $n$th-order web function with $a$ Wilson lines as \cite{Mitov:2010rp}
\bea
W_a^{(n)} 
&=& 
\sum_E\, \sum_{\gamma_E^{(n)}}  \Bigg( \,  \gamma_E^{(n)}   - \ \left \{ \exp \left[ \sum_{i=1}^{n-1}\ \sum_E \F_E\; [\W_E^{(i)}]  \right] \right \}^{(n)} \
\Bigg)\, ,
\label{eq:multi-sum}
\eea
where the superscript on the exponential specifies the $n$th order in the expansion of the exponential of webs up to order $n-1$.   In this expression, the functions $\W_E^{(i)}$ are ordered web integrands, whose color factors are matrices that do not commute in general.   In the case of two (or three) Wilson lines, or in the ``planar" limit of large $N_c$, however, these factors do commute \cite{Platzer:2013fha}, and the sum over orderings is equivalent to the modified color factor $\bar C(w_i)$ in Eq.\ (\ref{cbar}) above.
 
 We shall assume that each of the web functions $W^{(i)}_a=\sum_E\F_E[ \W_E^{(i)}]$ for $i < n$ gets finite contributions only from regions where all of its vertices are integrated over finite distances from the light cone, and where all of its vertices move to the light cone together.  This is to say, we assume that that all $W^{(i)}_a$, $i<n$ are free of  subdivergences.  We shall see under what conditions we may infer this result for $W^{(n)}_a$.

The regularization discussion of Sec.\ \ref{co-subt} applies as well to multieikonal vertices as to the cusp.   Similarly, for any neighborhood $\hat n[\rho]$ for the diagrams of $W^{(n)}_a$,  defined as in Eqs.\ (\ref{eq:neigh-def}) and (\ref{eq:rhohat-def}), we may construct  an 
expression for $W_{a,\rho}^{(n)}$, by analogy to Eq.\ (\ref{web-end}) above,
\bea
W_{a,\rho}^{(n)}\ &=& \  \sum_E \sum_{\gamma_E^{(n)}} (-t_\rho)\gamma_E^{(n)}
\  - \ \left \{ \exp \left[ \sum_{i=1}^{n-1}\ \sum_E \F_E\; [\W_E^{(i)}]  \right] \right \}^{(n)}_\rho  
\nn \\
&=& \  \ (-t_\rho)\Gamma_{a,\rho}^{(n)}
\  - \ \left \{ \exp \left[ \sum_{i=1}^{n-1}\, W_a^{(i)}  \right] \right \}^{(n)}_\rho  \
\, ,
\label{eq:multi-Delta}
\eea
where now the subscript $\rho$ on the exponential term denotes the
contribution of the integrals of the expanded exponential to region~$\rho$, which
defines a potential subdivergence of $W^{(n)}_a$.   In any such region
$\rho$, the remainder function is defined by some number  $r_\rho<n$ of vertices
in the union of integrals generated by monomials of webs found from the
expansion of the exponential, which shrink to the origin.   Correspondingly, $n-r_\rho$
vertices are left at finite distances from the origin, and define a soft function.    The webs in
Eq.\ (\ref{eq:multi-Delta}), as defined in Eq.\ (\ref{eq:calFcalW}),
are expressed as integrals over the positions of all vertices,
including those that attach to the eikonal lines.    As a result, we
may separate additively the contribution to each web function in the
exponential from the region where all of its vertices approach the
light cone or the origin.  We denote this contribution, which by assumption  contains the only
divergences in $W^{(i)}_a$, $i<n$, by $W^{(i)}_{a, \rm uv}$.

For now, let us assume that all webs commute, in addition to the assumption of no subdivergences up to order $n-1$.    We may then write the result of this separation as 
\bea
 W_{a,\rho}^{(n)}\ &=& \ (-t_\rho)\Gamma_{a,\rho}^{(n)}
\  - \ \left \{ \exp \left[ \sum_{i=1}^{n-1}\ [W^{(i)}_{a, \rm fin} +
    W^{(i)}_{a, \rm uv} ]  \right] \right \}^{(n)}_\rho  
  \, .
\label{eq:multi-fin-uv-factor}
\eea
where we define the finite part as
\bea
W^{(i)}_{a, \rm fin}\ =\ W^{(i)}_a\ -\ W^{(i)}_{a,\rm uv}\, ,
\label{eq:W-fin-def}
\eea
which in effect is a regulated version of the $i$th-order web.  The factorization of the finite and ultraviolet terms of the web exponent is trivial when the web functions commute (more generally, it requires the application of the Campbell-Baker-Hausdorf theorem).   The situation is equivalent to that in the renormalization of multieikonal webs outlined in Ref.\ \cite{Mitov:2010rp}.    We shall return briefly to this question below, but here we continue with the case in which all web functions commute, and we find simply,
\bea
 W_{a,\rho}^{(n)}\ &=& \  (-t_\rho)\Gamma_{a,\rho}^{(n)}
\  - \ \left \{ \exp \left[ \sum_{i=1}^{n-1}\ W^{(i)}_{a,\rm fin}\right] \, \exp \left[ \sum_{i=1}^{n-1}\ W^{(i)}_{a,\rm uv}   \right] \right \}^{(n)}_\rho  \, .
\label{eq:multi-fin-uv-factor-2}
\eea
The restriction to region $\rho$ now acts entirely on the exponential of the $W^{(i)}_{a, \rm uv}$ and picks out the sum of order-$r_\rho$ remainder contributions to the exponential of webs.   By definition, this is the full set of diagrams $\Gamma_a^{(r_\rho)}$ restricted to the neighborhood of the light cone and the origin.  Similarly, the exponential of finite parts gives the finite integral
of $\Gamma_a^{(n-r_\rho)}$, so that
\bea
W_{a,\rho}^{(n)}\ &=& \  (-t_\rho)\Gamma_a^{(n)}
\  - \  \Gamma^{(n-r_\rho)}_{a,\rm fin}\; \Gamma^{(r_\rho)}_{a,\rm uv}  \
 \, .
\label{eq:multi-fin-uv-factor-3}
\eea
Given the factorization of the full amplitude in region $\rho$, we conclude that the two terms on the right cancel, so that $W_{a,\rho}$ is finite when integrated over the neighborhood $\hat n[\rho]$ of any PS.   For large $N_c$, then, the full multieikonal web is free of subdivergences, just as for the cusp.   As anticipated above, the arguments we have given in this section, relying on exponentiation, are somewhat simpler than those based directly on the web construction itself.   

\subsection{Collinear factorization and web exponentiation for finite $N_c$}

Relaxing the commutativity of the web functions, we can still rederive
an important result for QCD and other theories beyond the planar
limit.    For an arbitrary multieikonal amplitude, the soft-jet-hard factorization derived above ensures that collinear
singularities are color diagonal and enter the web function
additively.  This means that all subdivergences where some, but not
all, vertices approach the light cone are canceled in multieikonal webs quite generally.   The
steps necessary to show this are just the same as when the webs commute; we need
only replace $W^{(i)}_{a, \rm uv}$ with $W^{(i)}_{a, \rm co}$, defined as the contribution where all vertices go to one or more of the light cones, 
\bea
 W_{a,\rho}^{(n)}\ &=& \ (-t_\rho)\Gamma^{(n)}_a
\  - \ \left \{ \exp \left[ \sum_{i=1}^{n-1}\ [W^{(i)}_{a,\rm central} + W^{(i)}_{a,\rm co } ]  \right] \right \}^{(n)}_\rho  \, ,
\label{eq:multi-central-co}
\eea
where $W^{(i)}_{a, \rm central}$ represents the remainder of the web function, where no vertex approaches the light cone, although in this case subsets of vertices may approach the origin.   This additive separation is certainly true for $i=1$, because the collinear  singularities arise from different regions of the same integral.    In addition, the sum of all $i=1$ (one-loop) collinear singularities for any multieikonal vertex is color diagonal (the sum of Casimir invariants, one for each Wilson line).

We now assume  that $W^{(i)}_{a, \rm co }$, $i<n$ is color diagonal
and thus commutes with all $W^{(j)}_{a, \rm central}$.   The
same steps as for the case of $W_{a{\rm uv}}^{(i)}$ for large $N_c$ then lead to a result analogous to Eq.\ (\ref{eq:multi-fin-uv-factor-3}), 
\bea
W_{a,\rho}^{(n)}\ &=& \  (-t_\rho)\Gamma_a^{(n)}
\  - \ \Gamma^{(n-c_\rho)}_{a,\rm central}\; \Gamma^{(c_\rho)}_{a,\rm co}  \
 \, ,
\label{eq:multi-central-co-factor}
\eea
where $c_\rho$ is the order of the collinear subdiagram.
Given this result, all subdivergences involving collinear subdiagrams of order $i<n$ cancel, and the only remaining collinear divergences are those in which all vertices approach any set of the light cones.   Again, these collinear singularities separate into color-diagonal factors, and we conclude that at order $n$ the collinear singularities of the web function are additive.   Thus, to all orders, collinear singularities factor into a product in the amplitude,
\bea
\Gamma_a\ &=&\ \exp \left[ \sum_{i=1}^{\infty}\ \left( W^{(i)}_{a,\rm central} + W^{(i)}_{a,\rm co } \right)\right]
\nonumber\\
&=& \exp \left[ \sum_{i=1}^{\infty}\  W^{(i)}_{a,\rm central} \right]\ \exp \left[ \sum_{i=1}^{\infty}\  W^{(i)}_{a,\rm co } \right]\, ,
\label{eq:full-co-fact}
\eea
where $W^{(i)}_{a, \rm co }$ is the additive part of the $i$th-order
web function that includes its collinear singularities.   In
principle, we could define this function up to a constant by
introducing an appropriate factorization scale.    In the second equality, we use the color-diagonal nature of
the collinear singularities.

We can put the factorized expression (\ref{eq:full-co-fact}) into
a standard form, simply by multiplying and dividing by an appropriate
power of a function whose collinear singularities match those of the
exponential of $W^{(i)}_{a, \rm co }$.   For a jet function corresponding to direction $\beta$, let us denote this function by $J_I(\beta,n_\beta)$, where $n_\beta$ is any other vector introduced in the definition of $J_I$.  As this notation suggests, there is considerable freedom in the choice of $J_I$.  An acceptable choice for $J_I$, however, is to choose $n_\beta=\bbeta$ and the jet function as the square root of the cusp matrix element  \cite{Sterman:2002qn,Aybat:2006mz},
\bea 
J^{\rm eik}_I(\beta,n_\beta)\ \equiv\ \left[ \bigg \langle 0\left| T\bigg(  \Phi^{[f_I]}_{\beta_I}(\infty,\tau_I \beta_I)\, \Phi_{\bbeta_I}^{[f_I]}{}^\dagger(\infty,\tau_I\beta_I) \bigg)\right|0  \bigg \rangle\right]^{1/2}\, ,
\eea
corresponding to a choice 
\bea
c^{\rm cusp}_I = \left[ \bigg \langle 0\left| T\bigg(  \Phi^{[f_I]}_{\beta_I}(\infty,\tau_I \beta_I)\, \Phi_{\bbeta_I}^{[f_I]}{}^\dagger(\infty,\tau_I\beta_I) \bigg)\right|0  \bigg \rangle\right]^{-1/2}
\eea
in the definition of the eikonal jet, Eq.\ (\ref{eq:J-eik-cusp}).
The square root reflects the symmetry between the vectors $\beta$ and $\bbeta$, giving the same collinear singularities associated with both directions in the matrix element, as is manifest in the exponentiated form (\ref{eq:web-int}) for the logarithm of the cusp as a web integral.   

Once we have defined the jet functions, we may reorganize the factorized multieikonal amplitude as
\bea
\Gamma_a \ &=&\ 
 \left( 
 \exp \left[ \sum_{i=1}^{\infty}\  W^{(i)}_{a,\rm central} \right]  
 \frac{ \exp \left[ \sum_{i=1}^{\infty}\  W^{(i)}_{a,\rm co } \right] }{\prod_{I=1}^a J_I^{\rm eik}} 
 \right)
 \ \prod_{I=1}^a J_I^{\rm eik}\ 
 \nn\\
 &=& S_a\ {\prod_{I=1}^a J_I^{\rm eik}}  \, ,
\eea
where $S_a$ is a collinear-finite soft function, just as in Eq.\ (\ref{eq:full-partonic}).  Because the eikonal jets cancel all collinear singularities in the ratio, the ratio may be factorized into soft and hard eikonal subdiagrams, which are renormalized locally, in the same manner as described in Sec.\ \ref{sec:renorm-fact}B, and in the same way as for massive, or other nonlightlike lines~\cite{Brandt:1981kf,Mitov:2010rp,Gardi:2010rn}.

\section{Conclusions}

 We have studied partonic matrix elements along with cusp and multieikonal  amplitudes for massless Wilson lines, in coordinate space and Feynman gauge.   
  In all these amplitudes, ultraviolet collinear and short-distance divergences arise when integrals over the positions of vertices are pinched in configurations set to fixed lightlike directions or short distances.    We have shown that these divergences are well approximated by the series of nested subtractions given in Eq.\ (\ref{eq:R-n-1-0}) for partonic matrix elements, and Eq.\ (\ref{eq:Gamma-sum}) for multieikonal amplitudes. The subtraction procedure allowed us to give very general proofs of the multiplicative renormalizability of multieikonal amplitudes and the factorization of partonic amplitudes in Feynman gauge.   These arguments, although presented in coordinate space, apply as well to the S-matrix in momentum space.

Our discussion confirmed that for the cusp the only sources of divergences are the limits in which all lines approach the light cones or the origin together \cite{Erdogan:2011yc}.   This is the content of Eq.\ (\ref{eq:web-int}), with a function $f_W$ that is finite for finite values of the variables $\tau$ and $\bar \tau$ that define the positions of the eikonal vertices that are furthest from the cusp.   For a conformal theory, this integrand is effectively constant.    For QCD and related renormalizable theories, the running coupling produces nontrivial dependence on the product $(\tau\bar \tau)$, which may be chosen as the inverse of the squared renormalization mass scale.   In the general multieikonal case, due to the nontrivial group structure of the webs the matching between UV subtraction terms, which factorize, and decompositions of the exponent no longer holds in the same fashion.  For the large-$N_c$ limit of gauge theory, however, the arguments go through, and each web becomes a sum of terms involving the two-dimensional integrals found in cusps.  In this case, as for the cusp, a geometrical interpretation of the web function applies \cite{Erdogan:2011yc}.    Further developments along these lines, and a coordinate-space picture for the origin of power corrections in infrared-safe observables \cite{Sveshnikov:1995vi} may be possible.

A coordinate-space program building on the techniques developed here would also include revisiting factorization proofs for electroweak annihilation \cite{Bodwin:1984hc,Collins:1989gx}, jet and single-particle inclusive cross sections in hadron-hadron collisions \cite{Sachrajda:1977mb}, in which we may look for the cancellation of long-distance dynamics directly from a spacetime point of view.   In particular, we may look forward to developing explicit space-time pictures associated with the cancellation and survival of Glauber \cite{Collinsbook,Becher:2014oda,Catani:2011st,Laenen:2014jga},  nonglobal  \cite{Dasgupta:2001sh} and superleading-logarithmic corrections \cite{Forshaw:2012bi}, and to the coordinate-space content of the dynamics to which jet vetos \cite{Berger:2010xi} may be sensitive.  Each of these examples involves the measurement of energy flow, directly or indirectly probing its time development.   In such cases, we may hope that a spacetime description of dynamics will be complementary to momentum-space analyses.

\acknowledgments
This work was supported in part by the National Science Foundation,  grants No.\ PHY-0969739 and No.\ PHY-1316617. The work of O.E. was also supported by the Isaac Newton Trust. We thank Eric Laenen and Kasper Larsen for useful conversations.

\appendix

\section{Uniqueness of the hard scattering}
\label{app:unique}

Here we give a brief discussion of the uniqueness of the position of the hard scattering in amplitudes with four or more external fields.  We suppose that we have already identified a point in spacetime, $y$, which satisfies the Landau equations (see \cite{Erdogan:2013bga}) for all vertices that connect lines that are on the light cone.   We assume that two lines have $y^0>x_i^0$, and the rest have $y^0<x_i^0$.   
For the vertex $y$, connecting four jets in particular the Landau equations are
\bea
\sum_{i=1}^4 \alpha_i \left (x_i^\mu-y^\mu \right )\ =\ 0\, ,
\label{eq:Lan-app-0}
\eea
for each of the external points, $x_i^\mu$, with all $\alpha_i\ge 0$ and $(x_i-y)^2=0$.
  Without loss of generality, we may translate the system so that $y=0$, giving
\bea
\sum_{i=1}^4 \alpha_i x_i^\mu\ =\ 0\, .
\label{eq:Lan-app}
\eea
We now seek another point in spacetime, $y'{}^\mu$, satisfying these same Landau equations.

At such a pinch surface we must have simultaneously,
\bea
x_i^2 \ &=&\ 0
\nonumber\\
(x_i-y')^2\ &=& \ 0\, .
\eea
This implies that
\bea
y'{}^2\ =\ 2x_i\cdot y' \, .
\label{eq:y-prime-condition}
\eea
Because the $x_i$ are all lightlike and noncollinear, it is not possible that all $x_i\cdot y'=0$ unless $y'=0$.   Thus, if $y'\ne 0$, $y'{}^2\ne 0$.

We may now search for a solution to Eq.\ (\ref{eq:y-prime-condition}), in terms of a rescaled vector,
\bea
z^\mu \ \equiv \frac{y'{}^\mu}{y'{}^2}\, ,
\eea
in terms of which Eq.~(\ref{eq:y-prime-condition}) becomes
\bea
1\ =\ 2 x_i\cdot z\, .
\label{eq:z-condition}
\eea
The Landau equations (\ref{eq:Lan-app}), however, ensure that
\bea
\det \left( x^\mu{}_i\right) \ =\ 0\, ,
\eea
and this implies that Eq.\ (\ref{eq:z-condition}) has no solution, other than $y'=0$.    

To go beyond four external points, $x_i^\mu$, we suppose we have another external vector, $x_5^\mu$.    Either $x_5$ is a linear combination of $x_1,\dots, x_4$ or $x_1,\dots, x_4$ are themselves linearly dependent.   In the former case, the Landau equations can be rewritten entirely in terms of the first four $x$'s, and in the latter, the first four $x$'s obey another linear relation that again ensures that $\det x^\mu_i=0$, with the same result.


\end{document}